\documentclass[
	aps,pra,
	superscriptaddress,
	twocolumn,
	10pt,
	floatfix, 
	tightenlines
]{revtex4-2}

\usepackage[toc,page]{appendix}

\usepackage[utf8]{inputenc}
\usepackage[T1]{fontenc}
\usepackage{lmodern}

\usepackage{csquotes}

\usepackage{parskip}
\usepackage{microtype}
\usepackage{bm} 
\usepackage{soul}

\usepackage{dsfont} 
\usepackage{amsmath,amssymb,amsthm,thmtools}
\usepackage{mathtools}
\usepackage{cases}
\usepackage{calc}
\usepackage{mathrsfs} 
\usepackage[normalem]{ulem} 
\usepackage{nameref}
\usepackage[colorlinks=true]{hyperref}
\usepackage[nameinlink]{cleveref}

\usepackage{physics} 

\usepackage[caption=false]{subfig}
\usepackage{graphicx}

\usepackage[dvipsnames]{xcolor}
\usepackage{easyReview}

\usepackage{tikz}
\usetikzlibrary{calc,shapes.geometric}

\usepackage{placeins}
\usepackage{multirow,tabularx,booktabs}
\usepackage{blindtext}

\usepackage[printonlyused,withpage,nohyperlinks,smaller]{acronym}

\graphicspath{{./figures/}}

\newcommand{\RR}{\mathbb{R}}
\newcommand{\CC}{\mathbb{C}}
\newcommand{\NN}{\mathbb{N}}

\newcommand{\PP}{\mathbb{P}}

\newcommand{\calC}{\mathcal{C}}
\newcommand{\calD}{\mathcal{D}}
\newcommand{\calH}{\mathcal{H}}
\newcommand{\calK}{\mathcal{K}}
\newcommand{\calP}{\mathcal{P}}
\newcommand{\calS}{\mathcal{S}}

\newcommand{\on}[1]{\operatorname{#1}}
\newcommand{\bs}[1]{\boldsymbol{#1}}

\newcommand{\bslambda}{{\bs\lambda}}
\newcommand{\bsgamma}{{\bs\gamma}}
\newcommand{\bsP}{{\bs P}}
\newcommand{\bshatn}{{\hat{\bs{n}}}}
\newcommand{\calCcoh}{{\calC_{\on{coh}}}}

\newcommand{\parTitle}[1]{\noindent\textcolor{Mahogany}{(\emph{#1})}}

\begin{document}

\title{Nonclassicality detection from few Fock-state probabilities}

\author{Luca Innocenti}
\affiliation{Department of Optics, Palack{\'y} University, 17. Listopadu 12, 771 46 Olomouc, Czech Republic}
\affiliation{Centre for Theoretical Atomic, Molecular and Optical Physics, Queen's University Belfast, Belfast BT7 1NN, United Kingdom}
\affiliation{Università degli Studi di Palermo, Dipartimento di Fisica e Chimica – Emilio Segrè, via Archirafi 36, I-90123 Palermo, Italy}
\email{luca.innocenti@community.unipa.it}

\author{Lukáš Lachman}
\affiliation{Department of Optics, Palack{\'y} University, 17. Listopadu 12, 771 46 Olomouc, Czech Republic}

\author{Radim Filip}
\affiliation{Department of Optics, Palack{\'y} University, 17. Listopadu 12, 771 46 Olomouc, Czech Republic}

\begin{abstract}\noindent
	We devise a new class of criteria to certify the nonclassicality of photon- and phonon-number statistics. Our criteria extend and strengthen the broadly used Klyshko's criteria, which require knowledge of only a finite set of Fock-state probabilities, making them easier to apply in realistic experimental scenarios.
    Moreover, we prove the completeness of our criteria in some scenarios, showing that \emph{all} finite distributions incompatible with classical states are detected as such by our criterion.
    In particular, we show that our criteria detect a broad class of noisy Fock states as nonclassical, even when Klyshko's criteria do not. The method is directly applicable to trapped-ion, superconducting circuits, and optical and optomechanical experiments with photon-number resolving detectors.
    This work represents a significant milestone towards a complete characterisation of the nonclassicality accessible from limited knowledge of the Fock-state probabilities.
\end{abstract}

\begin{abstract}
    Experimentally certifying the nonclassicality of quantum states in a reliable and efficient way is a challenge that remains both fundamental and daunting. Despite decades of topical research, techniques that can exploit optimally the information available in a given experimental setup are lacking. Here, we propose a novel paradigm to tackle these challenges, that is both directly applicable to experimental realities, and extendible to a wide variety of circumstances. We demonstrate that Klyshko’s criteria, which remained a primary approach to tackle nonclassicality for the past 20 years, is a special case of a much more general class of nonclassicality criteria. We provide both analytical results and numerical evidence for the optimality of our approach in several different scenarios of interest for trapped-ion, superconducting circuits, optical and optomechanical experiments with photon-number resolving detectors. This work represents a significant milestone towards a complete characterisation of the nonclassicality detectable from the limited knowledge scenarios faced in experimental implementations.
\end{abstract}

\maketitle

\section{Introduction}

Many fundamental quantum information protocols rely on the nonclassicality of bosonic systems induced by nonlinear phenomena~\cite{glauber1963coherent,sudarshan1963equivalence}.
Nonclassical statistics are a crucial resource for quantum sensing~\cite{tan2019nonclassical}, as demonstrated by the recent experiments with trapped ions~\cite{zhang2018noon,wolf2019motional} and superconducting qubits~\cite{wang2019heisenberglimited}, and more generally for the advancement of quantum information processing~\cite{gao2019entanglement,gan2020hybrid}.
Most existing nonclassicality criteria require knowledge of the statistical moments of the boson-number distribution~\cite{agarwal1992nonclassical,richter2002nonclassicality,miranowicz2010testing,sperling2012subbinomial,perina2017nonclassicality,perina2017higherorder,perina2019simultaneous,kimble1977photon,grangier1986experimental,grunwald2019effective,grunwald2020nonquantum,chavez-mackay2020estimating},
which are hard to estimate accurately in experimental platforms such as superconducting-circuits and trapped-ions~\cite{um2016phonon,wang2017converting,kienzler2017quantum,chu2018creation,gely2019observation}, as well as with photon-number-resolving detectors~\cite{kardynal2008avalanchephotodiodebased,divochiy2008superconducting,namekata2010nongaussian,endo2021quantum}.
At the same time, not being tailored to the observables that are directly accessible in a given experimental platform, these tests do not make optimal use of the information available.
Nonclassicality criteria relying on photon-click statistics~\cite{rigovacca2016nonclassicality,sperling2017detectorindependent,sperling2017identification,filip2013hierarchy,lachman2019criteria,lachman2019faithful} have similar shortcomings.
We here tackle both issues: on the one hand, we devise improved nonclassicality tests that can be directly applied to finite numbers of estimated boson-number probabilities, which is useful to better resolve the different brands of nonclassicality underlying boson-number distributions~\cite{zavatta2004quantumtoclassical,zavatta2007experimental,slodicka2016deterministic,marek2016deterministic,li2018generation,ding2018quantum,chu2018creation,podhora2020unconditional}.
On the other hand, we investigate the ultimate limits of any such test.

While deciding the nonclassicality of an unknown input state is fundamentally impossible with finitely many measurements, we find that, remarkably, in at least some cases of interest it is nonetheless possible to devise criteria that are optimal \textit{with respect to a given finite amount of information}. Such criteria are very useful in providing definitive answers to precisely which states can be certified as nonclassical in a given experimental scenario.
While we focus on the nonclassicality detectable from few Fock-state probabilities, our methodology, based on rather general geometric ideas, can be extended to tackle the nonclassicality of different types of measurements.
This approach departs considerably from methods relying on quasiprobability phase-space distributions, which typically rely on complete tomographic information to determine the (non)classicality of a state~\cite{kiesel2008experimental,lvovsky2009continuousvariable,kiesel2011nonclassicality,kuhn2018quantum,sperling2018quasiprobability,tan2020negativity,bohmann2020phasespace,sperling2020detectoragnostic}.
The two approaches not only provide incomparable results, but also rely on fundamentally different assumptions.

A pioneering step in this direction was taken by D.\,N.~Klyshko~\cite{klyshko_observable_1996}, who developed nonclassicality criteria --- in the form of inequalities for the Fock-state probabilities ---
satisfied by all classical states.
These criteria found numerous applications in both theoretical and experimental contexts~\cite{lee1997application,waks2006highly,wakui2014ultrabroadband,kono2017nonclassical,marek2016deterministic}. Similar criteria were also independently formulated in~\cite{simon1997nonclassicality}.
For many photon and phonon states, Klyshko's inequalities are however still insufficient to detect nonclassicality, and a thorough analysis of their completeness is lacking.

Here, we strengthen Klyshko's methodology, developing new criteria to certify nonclassicality from few Fock-state probabilities that are well-suited to experimental implementations.
More specifically, we address the following open question: \textit{given a vector $\bs P\equiv(P_0,P_1,...,P_n)$ of Fock-state probabilities, are these probabilities incompatible with classical states?}
It is worth stressing that, because knowing the probabilities in a fixed basis is not sufficient to characterize a quantum state, it is possible for a given $\bs P$ to correspond to both classical and non-classical states. Nonetheless, we can assess compatibility with a classical distribution, thus allowing to certify the nonclassicality of a given state. In other words, our approach allows to answer questions of the form: \textit{is nonclassicality detectable at all, given the information we are given?}
Moreover, we prove that in at least some cases our strengthened criteria are already \emph{complete}, in the sense that \emph{all} finite sets of Fock-state probabilities corresponding to nonclassical states are detected as such.
A significant advantage of our approach over previous endeavours is our working directly on the Fock-state probabilities, rather than using photon-click statistics or statistical moments.
This makes the criteria detector-independent and of broader applicability, in particular in the context of recent optical~\cite{harder2016singlemode}, atomic~\cite{hacker2019deterministic}, circuit quantum electrodynamics~\cite{sokolov2020superconducting}, and optomechanical~\cite{hong2017hanbury} experiments, and in light of the recent progress in photon-number-resolving detection technology~\cite{harder2016singlemode,sperling2017detectorindependent,tiedau2019scalability}.

While the focus of this paper is on \emph{$P$-nonclassicality}~\cite{tan2019nonclassical,glauber1963coherent,sudarshan1963equivalence,albarelli2016nonlinearity}, that is, on detecting states whose $P$ function cannot be interpreted as a probability distribution,
our approach can be extended to different notions of nonclassicality, thus paving the way for a similar characterisation of non-Gaussianity~\cite{lachman2019faithful}, a crucial resource for quantum computing with bosonic systems.

\section{Results}

In this section, we will discuss the extension of Klyshko's criteria~\cite{klyshko_observable_1996}, and how our geometrical approach that allows to provide definitive answers regarding the question of which nonclassical states can be certified as such in a given scenario where a limited number of observables are accessible.

Klyshko showed that, for all $k\ge1$, the condition
$k P_k^2>(k+1)P_{k+1}P_{k-1}$
cannot be fulfilled by classical states~\cite{klyshko_observable_1996}.
Here, we generalize these criteria to make them usable in arbitrary subsets of Fock-state probabilities of the form $\{P_0,P_1,...,P_N\}$.
Furthermore, we will show that the nonclassicality criteria involving the unobserved probabilities $\{P_{N+1},P_{N+2},...\}$ can be expressed in terms of the observable ones, providing a stronger nonclassicality criterion.
We will show that, in the $N=2$ case, these criteria \emph{characterize} the set of nonclassical states. This means that our criterion, at least in such special instances, exhausts the amount of information about nonclassicality that can be pried from Fock-state probabilities.
Finally, we will showcase applications of our criteria to several classes of experimentally relevant states whose nonclassicality is impervious to alternative methods.
The geometrical approach we use to derive our results has several advantages over alternative methods such as those based on the maximisation of \emph{task-dependent functionals}~\cite{filip2013hierarchy}.
In the appendices, we discuss how some of our results could be derived using this approach, which further highlights how the geometric approach might be a preferable venue to tackle the problems studied here.
It is worth stressing that our criteria are not directly comparable with standard phase-space-based approaches which rely on the negativity of Wigner or $P$-function of the full state. Such approaches require full information about the state~\cite{kiesel2008experimental,lvovsky2009continuousvariable,kiesel2011nonclassicality,kuhn2018quantum,sperling2018quasiprobability,tan2020negativity,bohmann2020phasespace,sperling2020detectoragnostic}, and cannot be directly compared with the results in the partial-knowledge scenario we consider. Same argument applies for momentum-based approaches~\cite{agarwal1992nonclassical,richter2002nonclassicality,miranowicz2010testing,sperling2012subbinomial,perina2017nonclassicality,perina2017higherorder,perina2019simultaneous,kimble1977photon,grangier1986experimental,grunwald2019effective,grunwald2020nonquantum,chavez-mackay2020estimating}.
All these methods share the shortcomings of relying on information about a state that is not easily estimated in many practical scenarios.
For the sake of completeness,
we present in the appendices the results obtained using such criteria, in order to highlight the fundamental differences between them.

\subsection{Boundary of nonclassicality}
Let $\calS\subset\calH$ be the set of quantum states on a single-mode Hilbert space $\calH$, and $\calCcoh\subset\calS$ the set of \emph{coherent} states, that is, of trace-1 projectors of the form $\{\ketbra\alpha\}_{\alpha\in\CC}$, where $\ket\alpha$ denotes a coherent states with average boson number $|\alpha|^2$.
Finally, let $\calC\subset\cal S$ denote the set of \emph{classical states}, that is, the \emph{convex hull} of $\calCcoh$.
This is the set of states $\rho$ that can be written as
$\rho=\int d^2\alpha P(\alpha)\ketbra\alpha$ for some probability distribution $P(\alpha)$.
Classical states can always be prepared by a classical external driving force on the quantized linear oscillator~\cite{scully1997quantum}. As such, their features are explainable via a classical formalism~\cite{glauber1963coherent}.

Given $N\ge1$, consider the \emph{reduced probability space}
\begin{equation}
	\calP_N\equiv\Big\{(P_0,P_1,...,P_N) : \sum_{k=0}^N P_k\le 1\text{ and }P_k\ge0 \Big\}.
\end{equation}
This is the set of vectors which can be part of some larger probability distribution.
Denote with
$\pi_N:\calS\to\calP_N$
the natural projection sending each state to its corresponding Fock-state probabilities:
$\pi_N(\rho) \equiv (\rho_{kk})_{k=0}^N\in \calP_N$.
We want to characterize algebraically the projection $\pi_N(\calC)$ of $\calC$ onto the reduced probability space $\calP_N$.
A crucial observation is that $\pi_N$ is linear. This implies that convex regions in $\calS$ are mapped into convex regions in $\calP_N$,
and thus in particular $\pi_N(\calC)$ is convex. Characterizing its boundary $\partial\pi_N(\calC)$ is therefore sufficient to characterize the whole of $\pi_N(\calC)$.

\subsection{Generalizing Klyshko's inequalities}
A first investigation of the $N=2$ case was presented in~\cite{filip2013hierarchy}, where nonclassicality criteria using $(P_0, P_k)$ were derived.
We summarize and extend these results, discussing the nonclassicality in general spaces of the form $(P_n,P_m)$.
We then extend these considerations to bound the possible Fock-state probabilities in arbitrary probability spaces.
In particular, we derive criteria in the form of inequalities relating probability tuples $(P_{I_1},...,P_{I_\ell})$ and $(P_{J_1},...,P_{J_\ell})$ such that $\sum_i I_i=\sum_i J_i$ and $I$ and $J$ are comparable via majorization.
We say that a tuple $I$ is \emph{majorized} by $J$, and write $I\preceq J$, if the sum of the $k$ largest elements of $I$ is smaller than the sum of the $k$ largest elements of $J$, for all $k$. We say that $I$ is comparable to $J$ via majorization if either $I\preceq J$ or $J\preceq I$.
An example of non-comparable tuples is $(2,2,2,0)$ and $(3,1,1,1)$~\cite{hardy1952inequalities,muirhead1902some,marshall1979inequalities,bhatia2013matrix}.
More precisely, if $I\preceq J$ and $\rho$ is classical, then the associated probabilities are bound to satisfy
\begin{equation}\label{eq:generalized_klyshko_main_statement}
    \prod_{i=1}^{s} I_i! P_{I_i} \le
    \prod_{i=1}^{s} J_i! P_{J_i},
\end{equation}
where $s\equiv \lvert I\rvert = \lvert J \rvert$.
Each such criterion corresponds to a nonclassicality criterion which can be used when the experimenter is given the corresponding set of Fock-state probabilities.

To prove~\cref{eq:generalized_klyshko_main_statement}, we start by defining $Q_k\equiv k! P_k$, so that the statement reads $\prod_{i=1}^s Q_{I_i}\le \prod_{i=1}^s Q_{J_i}$. Remembering the general identity for products of sums,
\begin{equation}
    \prod_{i=1}^n \sum_{j=1}^m a_{ij} =
    \sum_J \prod_{i=1}^n a_{i J_i},
\end{equation}
where the last sum ranges over all multi-indices $J$ of length $n$, with $J_i\in\{1,...,m\}$ for all $i=1,...,n$. For any classical state, the probabilities have the form
\begin{equation}
    P_k = \sum_\lambda p_\lambda e^{-\lambda} \frac{\lambda^k}{k!},
\end{equation}
and thus
\begin{equation}
    \prod_i Q_{I_i} = \prod_i \sum_\lambda 
    p_\lambda  e^{-\lambda} \lambda^{I_i}
    =
    \sum_{\bs\lambda} p_{\bs\lambda}e^{|\bs\lambda|}\bs\lambda^I,
\end{equation}
where we used the shorthand notation $p_{\bs\lambda}\equiv\prod_i p_{\lambda_i}$, $|\bs\lambda|\equiv\sum_i\lambda_i$, and $\bs\lambda^I\equiv\prod_i \lambda_i^{I_i}$, and the sum is extended to all possible tuples of values of $\lambda$.
From the above expression, we see that
\begin{equation}
    \prod_i Q_{I_i} - \prod_i Q_{J_i} =
    \sum_{\bs\lambda} p_{\bs\lambda} e^{-|\bs\lambda|}
    (\bs\lambda^I - \bs\lambda^J).
\end{equation}
The conclusion then follows from \textit{Miurhead's inequalities}~\cite{hardy1952inequalities,cvetkovski2012schurs}.
More details are provided in~\cref{app:sec:generalized_klyshkos}.

In particular, when $I,J$ have length $3$, we get inequalities involving triples of probabilities: for all $0\le n\le m\le k$, classical states are bound to satisfy:
\begin{equation}
	(m! P_m)^{k-n} \le (n! P_n)^{k-m} (k! P_k)^{m-n}.
	\label{eq:klyshko_inequalities_for_triples}
\end{equation}
Violation of~\cref{eq:klyshko_inequalities_for_triples} thus certifies nonclassicality.
For $n=N-1, m=N$ and $k\ge N$, defining $Q_k\equiv k!P_k$, we have
$
	Q_N^{k-N+1} \le Q_{N-1}^{k-N} Q_k,
$
and thus
\begin{equation}
	k! P_k \ge
	\frac{Q_{N-1}^N}{Q_N^{N-1}}
	\left(\frac{Q_N}{Q_{N-1}}\right)^{k},\,\,\forall k\ge N-1.
	\label{eq:klyshko_triple_inequality_s}
\end{equation}
Using this in conjunction with the normalisation condition $\sum_k P_k=1$ we get
\begin{equation}
	\sum_{k=0}^{N-2} P_k +
	\frac{Q_{N-1}^N}{Q_N^{N-1}}
	\sum_{k=N-1}^\infty
	\frac{1}{k!}
	\left(
		\frac{Q_N}{Q_{N-1}}
	\right)^k\le1.
\end{equation}
Using the Taylor expansion of the exponential function to write
$\sum_{k=N-1}^\infty \frac{x^k}{k!}=e^x - \sum_{k=0}^{N-2}\frac{x^k}{k!}$, we then conclude that all classical states must satisfy the inequality
\begin{equation}
\begin{aligned}\scalebox{0.95}{$\displaystyle
	\sum_{k=0}^{N-2} P_k +
	\frac{Q_{N-1}^N}{Q_N^{N-1}} \left[
		e^{\frac{Q_N}{Q_{N-1}}} - \sum_{k=0}^{N-2} \frac{(Q_N/Q_{N-1})^k}{k!}
	\right]\le1.$}
	\label{eq:klyshko_infinite_inequality_N}
\end{aligned}
\end{equation}

Together with the standard Klyshko conditions in the form $Q_k^2 \le Q_{k-1}Q_{k+1}$,~\cref{eq:klyshko_infinite_inequality_N}, defines a closed region $\calD_N\subset\calP_N$ containing $\pi_N(\calC)$. Any probability vector $\bs P\notin\mathcal D_N$ is certifiably nonclassical.
In the rest of the paper, we will refer to condition~\eqref{eq:klyshko_infinite_inequality_N} as $\calK_{\infty,N}$, and to the Klyshko condition $Q_k^2\le Q_{k-1}Q_{k+1}$ as $\calK_k$. We will also use $\calK_\infty$ to refer more generally to criteria of the type $\calK_{\infty,N}$ for some $N$.

Let us remark two additional facts:
\begin{enumerate}
    \item Having access to a finite set of probabilities $(P_0,...,P_{N-1})$, there are always nonclassical states that are not detectable by any criterion. For example, consider $\rho^{(N)}\equiv p\rho_{\rm cl}+(1-p)\ketbra N$ for some classical state $\rho_{\rm cl}$.
    Having only access to the first $N$ probabilities amounts to working with the reduced distribution $\pi_{N-1}(\rho^{(N)})=p \pi_{N-1}(\rho_{\rm cl})$.
    Being $\rho_{\rm cl}$ classical, as we discussed previously, $\pi_{N-1}(\rho_{\rm cl})$ satisfies all the inequalities $\calK_\infty$ and $\calK_k$.
    The scaled probability vector $p\pi_{N-1}(\rho_{\rm cl})$ is then also bound to satisfy the Klyshko-like inequalities $\calK_k$, as these are scale-invariant. It is then also easy to verify that if $\calK_{\infty,N-1}$ is satisfied for $\pi_{N-1}(\rho_{\rm cl})$, then it must also be satisfies for $p\pi_{N-1}(\rho_{\rm cl})$ for any $0<p<1$.
    \item The $\calK_\infty$ inequalities are a necessary addition to fully expoit the knowledge encoded in the Fock-number probability distributions. Indeed, there are always nonclassical states undetected by the Klyshko-like inequalities. For example, knowing any set $(P_0,...,P_{N})$, the Fock state $\ket N$ can be seen to satisfy all inequalities of type $\calK_k$, but it violates $\calK_{\infty,N}$.
\end{enumerate}


The fundamental question that remains to be addressed is whether the inequalities of the form $\calK_k$ and $\calK_\infty$ exhaust the information about nonclassicality encoded in Fock-state probabilities.
This amounts to asking whether a probability vector sastisfying all the inequalities (those usable given a finite set of probabilities), implies the existence of a classical state resulting in said probabilities.
In other words, we want to know whether satisfying all relevant inequalities certifies \textit{compatibility} with some classical state, which is the most one can ask for in this scenario.

\subsection{Nonclassicality in \texorpdfstring{$(P_0,P_1,P_2)$}{(P0,P1,P2)}}

To analyze the applicability of the new conditions~$\calK_{\infty,N}$ beyond~\cite{klyshko_observable_1996}, we study what nonclassical states can be detected when only the first three Fock-state probabilities are known.
We will find that, remarkably, \textit{the nonclassicality of a state is completely captured by only two algebraic inequalities}.

Let us denote with $\calK_1$ the region:
\begin{equation}
	\calK_1\equiv\{(P_0,P_1,P_2)\in\calP_2: P_1^2=2P_0 P_2\},
    \label{eq:P012_K1}
\end{equation}
and with $\calK_{\infty,2}$ the set of points satisfying~\cref{eq:klyshko_infinite_inequality_N} with $N=2$, that is, the probability vectors $(P_0,P_1,P_2)\in\calP_2$ such that
$P_0 + \frac{P_1^2}{2P_2} \left[\exp(\frac{2P_2}{P_1}) - 1\right] = 1$.
The associated \emph{nonclassicality criteria} are then
\begin{gather}
    P_1^2 > 2P_0 P_2, \label{eq:P012_criterion_K1} \\
    P_0 + \frac{P_1^2}{2P_2} \left[\exp(\frac{2P_2}{P_1}) - 1\right] > 1
    \label{eq:P012_criterion_Kinf}.
\end{gather}
The notation $\calK_1^\gtrless$ and $\calK_{\infty,2}^\gtrless$ will be used to denote the sets obtained by replacing the equality in these definitions with the corresponding inequality sign (e.g. $\calK_1^\ge$ is the set of points such that $P_1^2\ge 2P_0 P_2$, while $\calK_1^<$ is the set of points such that $P_1^2<2 P_0 P_2$).
We will prove in this section that $\pi_2(\calC)=\calK_1^\le\cap\calK_{\infty,2}^\le$, that is, that~\cref{eq:P012_criterion_K1,eq:P012_criterion_Kinf} are necessary and sufficient conditions for a state being detectable as nonclassical when only knowledge of the first three Fock-state probabilities is given.
It is worth stressing that these nonclassicality criteria are strictly stronger than previously reported criteria using pairs of Fock-state probabilities~\cite{filip2013hierarchy}.

We already showed that all classical states are contained in $\calK_1^\le\cap\calK_{\infty,2}^\le$.
To prove that the inequalities provide a \emph{necessary and sufficient} condition for compatibility with classical states, we need to show that any probability vector in $\calK_1^\le\cap\calK_{\infty,2}^\le$ is compatible with a classical state.
For the purpose, we will show that any $\bs P$ inside this region can be written as convex combination of probability vectors compatible with classical states.
In other words, we want to find the mixture of coherent states corresponding to a given triple of probabilities $\bs P\equiv (P_0,P_1,P_2)\in\calK_1^\le\cap\calK_{\infty,2}^\le$.
To achieve this, we will
(1) show that $\bs P$ is a convex mixture of the origin and a point $\bs P_\infty\in\calK_1^\le\cap\calK_{\infty,2}$; 
(2) show that $\bs P_\infty$ can be written as convex combination of $(1,0,0)$ (the point corresponding to the vacuum state) and an element of $\calK_1\cap\calK_{\infty,2}$;
(3) show that all vectors in $\calK_1\cap\calK_{\infty,2}$ are compatible with coherent states.
This will allow us to conclude that $\bs P$ is compatible with a convex combination of coherent states.

Let $\bsP\in\calK_1^\le\cap\calK_{\infty,2}^\le$ be an arbitrary point \emph{not} satisfying the nonclassicality criteria~(\ref{eq:P012_criterion_K1}, \ref{eq:P012_criterion_Kinf}).
Define the quantities $K_1(\bsP)\equiv P_1^2-2P_0 P_2$ and $K_{\infty,2}(\bsP)\equiv P_0 + \frac{P_1^2}{2P_2}[\exp(2P_2/P_1)-1]$.
Note that, upon rescaling $\bsP\to\epsilon\bsP$, the sign of $K_1$ is invariant, and $K_{\infty,2}\to\epsilon K_{\infty,2}$.
We can therefore always find $\epsilon\ge1$ such that $\bsP'\equiv \epsilon\bsP\in\calK_1^\le\cap\calK_{\infty,2}$.
We can thus write $\bsP$ as a convex combination of $\bsP'$ and the origin in probability space, $\bs 0\equiv(0,0,0)$, as
$\bsP =1/\epsilon\bsP'+(1-1/\epsilon)\bs 0$.
Note that $\bs0$ is the probability vector generated by a coherent state in the limit of infinite average boson number, and is thus classical. It now remains to prove that $\bsP'$ is also classical to conclude that $\bsP$ is.
For the purpose, consider convex combinations of $\bsP'$ and $\bs e_0\equiv (1,0,0)\equiv \pi_2(\ketbra0)$, and notice that
\begin{equation}
\scalebox{0.97}{$\displaystyle\begin{gathered}
	K_{\infty,2}(p\bsP'+(1-p)\bs e_0) = p K_{\infty,2}(\bsP') + (1 - p) = 1, \\
    K_1(p\bsP'+(1-p)\bs e_0) = p^2 K_1(\bsP') -2p(1-p) P'_2.
\end{gathered}$}
\end{equation}
Solving for the $p\neq0$ such that $K_1=0$, we find
\begin{equation}
    p = \frac{2P_2'}{(P_1')^2} e^{-2P_2'/P_1'} \ge 1,
\end{equation}
where we used $K_{\infty,2}(\bsP')=1$ and $K_1(\bsP')\le0$.
This means that there is some
$\bsP''\in\calK_1\cap\calK_{\infty,2}$ such that
$\bsP' = \frac{p-1}{p} \bs e_0 + \frac{1}{p} \bsP''$.
To conclude, we now only need to show that $\bsP''$ is compatible with a coherent state.
By definition of $\calK_1\cap\calK_{\infty,2}$, the elements of $\bsP''$ satisfy
\begin{equation}
    \frac{P_1''^2}{2P_2''} e^{2P_2''/P_1''} = P_0'' e^{P_1''/P_0''} = 1.
\end{equation}
We finally note that, for any value of $P_0''$, these conditions uniquely determine $P_1''$ and $P_2''$, and that a coherent state with average boson number $\mu=-\log P_0$ produces these probabilities.

It is worth remarking that the above reasoning not only proves that the given inequalities characterize the boundary of classical states in $\calP_2$, but also provides a constructive method to find classical states compatible with an observed (not nonclassical) probability distribution.

\begin{figure}[tb]
	\centering
	\includegraphics[width=0.8\linewidth]{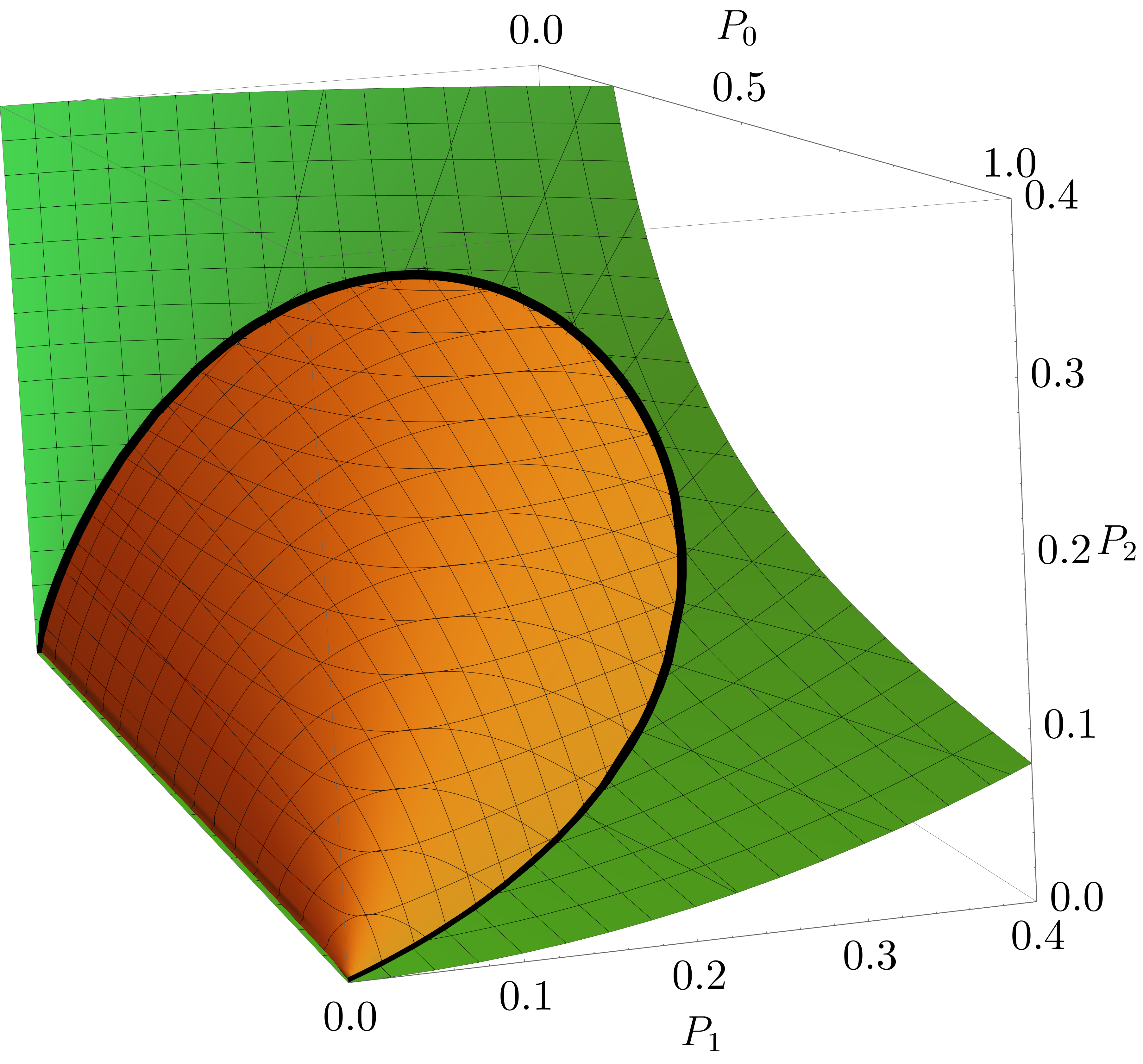}
	\caption{
        \textit{Classical set in the $(P_0,P_1,P_2)$ space.} ---
        The orange (upper) surface is the set of points on $\calK_\infty$, while the green (lower) surface the set of points on $\calK_1$. The black line is the set of coherent states. We notice that the upper surface can be generated as the set of lines going from $(1,0,0)$ to the coherent states, while the lower surface as the set of lines going from $(0,0,0)$ to the coherent states.
        All the states with probabilities lying \emph{above} the upper surface will satisfy Klyshko's inequalities, and can therefore be detected as nonclassical only using~\eqref{eq:P012_criterion_Kinf}.
    }
    \label{fig:P01_boundary}
\end{figure}

\begin{figure}[tb]
	\centering
	\includegraphics[width=0.9\linewidth]{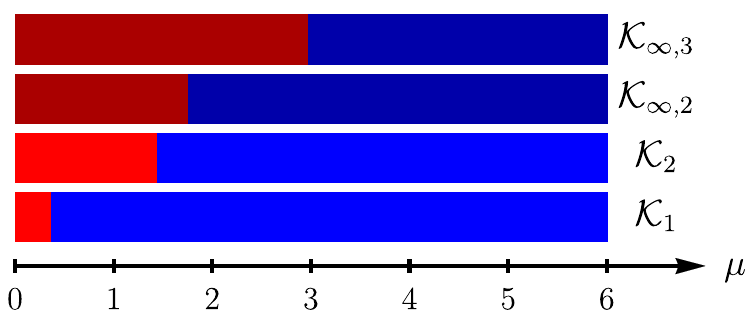}
	\caption{
		\emph{Nonclassicality of boson-added states $p \,a^\dagger \rho_\mu a + (1-p)\rho_\mu$ with $p=0.5$} ---
		For each $\mu$, we highlight whether the different criteria detect the corresponding probability distribution as nonclassical (red) or not (blue).
		Note how restricting to the first three Fock-state probabilities, when only $\calK_1$ and $\calK_{\infty,2}$ are accessible, nonclassicality is certified only up to $\mu\sim1.7$. On the other hand, knowing $P_3$, nonclassicality is certifiable up to $\mu\sim3$, thanks to $\calK_{\infty,3}$.
	}
	\label{fig:nonclassicality_rectangles_dispFockCoh50p}
\end{figure}

\newcommand{\subfigimg}[3][,]{%
  \setbox1=\hbox{\includegraphics[#1]{#3}}
  \leavevmode\rlap{\usebox1}
  \rlap{\hspace*{60pt}\raisebox{\dimexpr\ht1+\baselineskip}{#2}}
  \phantom{\usebox1}
}

\begin{figure}[tb]
	\centering
	\includegraphics[width=0.9\linewidth]{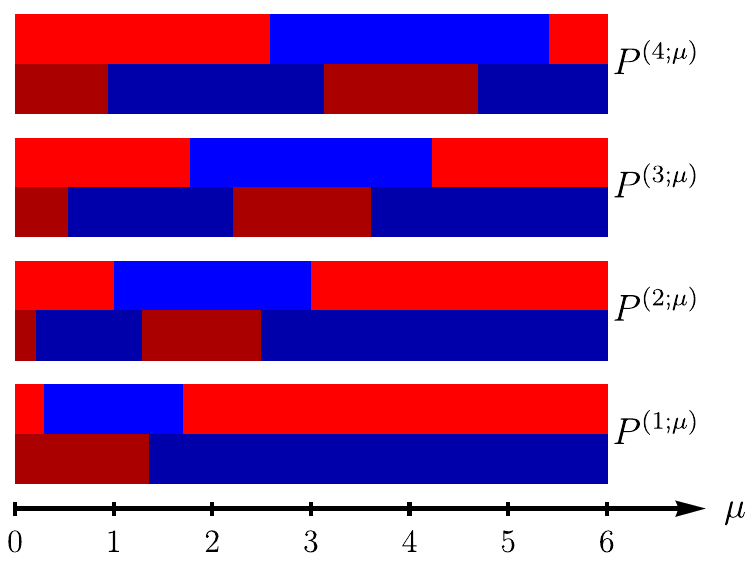}
	\caption{
		\emph{Nonclassicality of noisy Fock states} ---
		Notation is as in~\cref{fig:nonclassicality_rectangles_dispFockCoh50p}.
		We highlight the regions of nonclassicality detected by $\calK_1$ (upper light red) and $\calK_{\infty,2}$ (lower dark red) for noisy Fock states statistics $P^{(k;\mu)}$, for $k=1,2,3,4$.
		For example, we see that $\mu=1.5$ corresponds to a classical statistics for $P^{(1;\mu)}$, a nonclassical one detected by $\calK_{\infty,2}$ for $P^{(2;\mu)}$, and a nonclassical one also for $P^{(3;\mu)}$ and $P^{(4;\mu)}$, now detected by $\calK_1$.
		Further details can be found in~\cref{fig:noisyFock1_nonclassicality_rectangles,fig:noisyFock2_nonclassicality_rectangles}.
	}
	\label{fig:noisyFock_nonclassicality_rectangles}
\end{figure}

\section{Applications}

\subsection{Fock and squeezed states}
\label{sec:fock_and_squeezed_states}
A class of states benefiting from the $\calK_\infty$ criteria are Fock states.
The Fock state $|1\rangle\equiv a^\dagger\ket0$
clearly satisfies $P_1^2>2P_0 P_2$, and is therefore detected as nonclassical by $\calK_1$.
More generally, convex mixtures of $\ket0$ and $\ket1$ are all detected as nonclassical by $\calK_1$ but not by $\calK_{\infty,2}$, as also seen in~\cref{fig:P01_boundary}.
On the other hand, $\ket2= \frac{1}{\sqrt2}a^{\dagger 2}\ket0$ is detected as nonclassical by $\calK_{\infty,2}$ but not by $\calK_1$.

Consider now \textit{attenuated} Fock states, that is, states of the form $\mathcal E_T(\ketbra k)$ with $\ket k$ Fock states and $\mathcal E_T$ the channel corresponding to attenuation through a beamsplitter with transmittivity $T\in[0,1]$ (thus, in particular, $\mathcal E_1(\rho)=\rho$ and $\mathcal E_0(\rho)=\on{Tr}(\rho) \ketbra 0$ for all $\rho$). In these cases, we find that $\mathcal E_T(\ketbra k)$ is, in principle, detected as nonclassical by both $\calK_1$ and $\calK_{\infty,2}$, for all $T\in[0,1]$ and $k\in\NN$. However, the criteria detect this nonclassicality very differently: it is harder to detect the nonclassicality with $\calK_1$ for $T$ closer to $1$, while $\calK_{\infty,2}$ makes it easier in this regime, and viceversa for smaller values of $T$. A Fock state such as $\ket2$ sits on the boundary of the classical region, and is therefore undetectable as nonclassical with finite statistics with $\calK_1$, which is why these results are consistent with our previous statement that $\ket 2$ can only be detected as nonclassical with $\calK_{\infty,2}$.
This hardness for $T$ approaching unity increases for higher Fock states $\ket k$, as shown in~\cref{app:attenuated_fock_states}.
More generally, $\calK_1$ cannot certify the nonclassicality of convex mixtures of $\ket2$ and $\ket0$, which is however revealed by~$\calK_{\infty,2}$.
This suggests squeezed states as another class benefiting from our extended criteria.

In appendix~\ref{app:applications} we show that many squeezed thermal states~\cite{kim1989properties} also require $\calK_{\infty,2}$ to be detected as nonclassical.

\subsection{Boson-added noisy states}
Photon- and phonon-added coherent states~\cite{agarwal1991nonclassical,agarwal1991nonclassical,dominguez-serna2016entangled,gard2016photon} are defined as
$
	\ket{\alpha,\ell}\equiv C_{\alpha,\ell}a^{\dagger \ell}\ket\alpha
$
with $C_{\alpha,\ell}$ normalisation constants.
The associated Fock-state distribution equals that obtained adding single bosons to Poissonian noise with average number $\mu=\abs{\alpha}^2$, here denoted $\rho_\mu$.
The new criteria provide enhanced predictive power also for these highly noisy states $\rho_\mu$.
For example, for $\ell=2$, $\calK_1$ does not predict nonclassicality with $P_0,P_1,P_2$, but $\calK_{\infty,2}$ does.
The same holds for probabilistic boson addition. Consider \emph{e.g.} states of the form $p\, a^\dagger \rho_\mu a+(1-p)\rho_\mu$.
We find that using $\calK_\infty$ criteria allows to detect nonclassicality more efficiently, as shown in~\cref{fig:nonclassicality_rectangles_dispFockCoh50p}.
More details are found in appendix~\ref{app:applications}.

\subsection{Noisy Fock states}
\label{sec:noisy_fock_states}

Consider now \emph{displaced} Fock states, $\ket{\alpha;k}=D(\alpha)\ket k$, obtained applying the displacement operator $D(\alpha)$ to a Fock state $\ket k$~\cite{buvzek1991displaced}.
Averaging over the phases of $\alpha$, these produce the same Fock-state probabilities $(P^{(k;\mu)}_j)_j$ as Fock states with added Poissonian noise, and model states produced in realistic experimental conditions, where the displacement operator causes Fock states higher than $\ket k$ to contribute.
As shown in~\cref{fig:noisyFock_nonclassicality_rectangles}, $\calK_\infty$ criteria increase the predictive power when few Fock-state probabilities are known.
For example, for $k=1$, when $P_0,P_1,P_2$ are known, $\calK_1$ does \emph{not} certify nonclassicality for $0.29\lesssim\mu\lesssim 1.71$, but $\calK_{\infty,2}$ does for $\mu\lesssim1.35$. This means that $\calK_{\infty,2}$ allows to detect nonclassical states in regimes in which $\calK_1$ is not sufficient.
Another striking feature emerging from~\cref{fig:noisyFock_nonclassicality_rectangles} is that adding more noise can make it easier to detect nonclassicality, as highlighted by the presence of bright red regions for large values of $\mu$. This remains the case even if, instead of simply increasing the average boson number of the added Poissonian noise, we add incoherent noise to the state. We find that this can also make the nonclassicality of a distribution easier to detect.
More details are found in appendix~\ref{app:applications}.

\section{Discussion}
We showed that Klyshko's criteria are a special case of a broader class of nonclassicality criteria. Leveraging this result we found that, when only few Fock-state probabilities are known, these new criteria grant additional insight into nonclassical properties of boson statistics, even in realistic experimental conditions. Such criteria are pivotal to deepen our understanding of nonclassical phenomena and uncover new resources for quantum technologies.
Our method is directly applicable to trapped-ion~\cite{zhang2018noon,wolf2019motional}, superconducting-circuit~\cite{wang2019heisenberglimited}, and optical experiments with photon-number resolving detectors~\cite{harder2016singlemode,sperling2017detectorindependent,tiedau2019scalability}.
More specifically, the application of the proposed methodology to a given experimental scenario is completely straightforward, only requiring to verify whether the measured quantities satisfy a finite set of algebraic inequalities.

We proved the optimality of our improved criteria with respect to the first three Fock-state probabilities, and discussed a number of example applications of the criteria for several classes of states of interest, including boson-added thermal states, noisy Fock states, and thermal and Fock states. Others, such as thermal states, are also considered and discussed in the Appendix. We remark that even in cases in which the new $\calK_\infty$ criteria do not provide additional predictive capabilities, as is the case for example for some classes of noisy Fock states, discussed in~\cref{sec:noisy_fock_states,app:noisy_fock_states}, and boson-added thermal states, discussed in~\cref{app:boson_added_thermal_states}.
We stress that, even in these cases, our analysis is useful allowing to conclude that the nonclassicality certification problem, in some circumstances, is simply \textit{unsolvable} with the information given.

The optimality of the proposed criteria in higher-dimensional slices of probability space remains a stimulating open question, which if solved would provide further insight into the nonclassicality of boson statistics.
Another interesting aspect emerging from a combination of this approach with the methodology of~\cite{filip2013hierarchy}, is how adding noise to a state, which is generally an easy operation, can make it easier to detect the nonclassicality of the state from its Fock-state distribution.
Our results, paired with modern optimisation techniques, pave the way to a complete characterisation of the nonclassicality accessible from finite sets of measurable quantities.

\section*{Data availability}
The data that support the findings of this study are available from the corresponding author upon reasonable request.

\section*{Author contribution}
All authors contributed to the conception and development of the idea and to the writing of the manuscript.

\section*{Competing interest}
The authors declare no competing interests.

\section*{Acknowledgments}
The authors acknowledge helpful discussions with Mauro Paternostro and Alessandro Ferraro.
We also gratefully acknowledge support by national funding from MEYS and European Union's Horizon 2020 (2014–2020) research and innovation framework programme under grant agreement No. 731473 (project 8C20002 ShoQC). Project ShoQC has received funding from the QuantERA ERA-NET Cofund in Quantum Technologies implemented within the European Unions Horizon 2020 Programme.
R.F. and L.L. acknowledge grant No. GA19-14988S of the Czech Science Foundation. R.F. also acknowledges support from the MEYS of Czech Republic by the project LTAUSA19099.
L.I. acknowledges  support from \emph{Fondazione Angelo della Riccia}.

\bibliography{draft}

\begin{thebibliography}{77}%
\makeatletter
\providecommand \@ifxundefined [1]{%
 \@ifx{#1\undefined}
}%
\providecommand \@ifnum [1]{%
 \ifnum #1\expandafter \@firstoftwo
 \else \expandafter \@secondoftwo
 \fi
}%
\providecommand \@ifx [1]{%
 \ifx #1\expandafter \@firstoftwo
 \else \expandafter \@secondoftwo
 \fi
}%
\providecommand \natexlab [1]{#1}%
\providecommand \enquote  [1]{``#1''}%
\providecommand \bibnamefont  [1]{#1}%
\providecommand \bibfnamefont [1]{#1}%
\providecommand \citenamefont [1]{#1}%
\providecommand \href@noop [0]{\@secondoftwo}%
\providecommand \href [0]{\begingroup \@sanitize@url \@href}%
\providecommand \@href[1]{\@@startlink{#1}\@@href}%
\providecommand \@@href[1]{\endgroup#1\@@endlink}%
\providecommand \@sanitize@url [0]{\catcode `\\12\catcode `\$12\catcode
  `\&12\catcode `\#12\catcode `\^12\catcode `\_12\catcode `\%12\relax}%
\providecommand \@@startlink[1]{}%
\providecommand \@@endlink[0]{}%
\providecommand \url  [0]{\begingroup\@sanitize@url \@url }%
\providecommand \@url [1]{\endgroup\@href {#1}{\urlprefix }}%
\providecommand \urlprefix  [0]{URL }%
\providecommand \Eprint [0]{\href }%
\providecommand \doibase [0]{https://doi.org/}%
\providecommand \selectlanguage [0]{\@gobble}%
\providecommand \bibinfo  [0]{\@secondoftwo}%
\providecommand \bibfield  [0]{\@secondoftwo}%
\providecommand \translation [1]{[#1]}%
\providecommand \BibitemOpen [0]{}%
\providecommand \bibitemStop [0]{}%
\providecommand \bibitemNoStop [0]{.\EOS\space}%
\providecommand \EOS [0]{\spacefactor3000\relax}%
\providecommand \BibitemShut  [1]{\csname bibitem#1\endcsname}%
\let\auto@bib@innerbib\@empty
\bibitem [{\citenamefont {Glauber}(1963)}]{glauber1963coherent}%
  \BibitemOpen
  \bibfield  {author} {\bibinfo {author} {\bibfnamefont {R.~J.}\ \bibnamefont
  {Glauber}},\ }\bibfield  {title} {\bibinfo {title} {Coherent and incoherent
  states of the radiation field},\ }\href
  {https://doi.org/10.1103/PhysRev.131.2766} {\bibfield  {journal} {\bibinfo
  {journal} {Physical Review}\ }\textbf {\bibinfo {volume} {131}},\ \bibinfo
  {pages} {2766–2788} (\bibinfo {year} {1963})}\BibitemShut {NoStop}%
\bibitem [{\citenamefont {Sudarshan}(1963)}]{sudarshan1963equivalence}%
  \BibitemOpen
  \bibfield  {author} {\bibinfo {author} {\bibfnamefont {E.~C.~G.}\
  \bibnamefont {Sudarshan}},\ }\bibfield  {title} {\bibinfo {title}
  {Equivalence of semiclassical and quantum mechanical descriptions of
  statistical light beams},\ }\href
  {https://doi.org/10.1103/PhysRevLett.10.277} {\bibfield  {journal} {\bibinfo
  {journal} {Physical Review Letters}\ }\textbf {\bibinfo {volume} {10}},\
  \bibinfo {pages} {277–279} (\bibinfo {year} {1963})}\BibitemShut {NoStop}%
\bibitem [{\citenamefont {Tan}\ and\ \citenamefont
  {Jeong}(2019)}]{tan2019nonclassical}%
  \BibitemOpen
  \bibfield  {author} {\bibinfo {author} {\bibfnamefont {K.~C.}\ \bibnamefont
  {Tan}}\ and\ \bibinfo {author} {\bibfnamefont {H.}~\bibnamefont {Jeong}},\
  }\bibfield  {title} {\bibinfo {title} {Nonclassical light and metrological
  power: An introductory review},\ }\href {https://doi.org/10.1116/1.5126696}
  {\bibfield  {journal} {\bibinfo  {journal} {AVS Quantum Science}\ }\textbf
  {\bibinfo {volume} {1}},\ \bibinfo {pages} {014701} (\bibinfo {year}
  {2019})}\BibitemShut {NoStop}%
\bibitem [{\citenamefont {Zhang}\ \emph {et~al.}(2018)\citenamefont {Zhang},
  \citenamefont {Um}, \citenamefont {Lv}, \citenamefont {Zhang}, \citenamefont
  {Duan},\ and\ \citenamefont {Kim}}]{zhang2018noon}%
  \BibitemOpen
  \bibfield  {author} {\bibinfo {author} {\bibfnamefont {J.}~\bibnamefont
  {Zhang}}, \bibinfo {author} {\bibfnamefont {M.}~\bibnamefont {Um}}, \bibinfo
  {author} {\bibfnamefont {D.}~\bibnamefont {Lv}}, \bibinfo {author}
  {\bibfnamefont {J.-N.}\ \bibnamefont {Zhang}}, \bibinfo {author}
  {\bibfnamefont {L.-M.}\ \bibnamefont {Duan}},\ and\ \bibinfo {author}
  {\bibfnamefont {K.}~\bibnamefont {Kim}},\ }\bibfield  {title} {\bibinfo
  {title} {Noon states of nine quantized vibrations in two radial modes of a
  trapped ion},\ }\href {https://doi.org/10.1103/PhysRevLett.121.160502}
  {\bibfield  {journal} {\bibinfo  {journal} {Physical Review Letters}\
  }\textbf {\bibinfo {volume} {121}},\ \bibinfo {pages} {160502} (\bibinfo
  {year} {2018})}\BibitemShut {NoStop}%
\bibitem [{\citenamefont {Wolf}\ \emph {et~al.}(2019)\citenamefont {Wolf},
  \citenamefont {Shi}, \citenamefont {Heip}, \citenamefont {Gessner},
  \citenamefont {Pezzè}, \citenamefont {Smerzi}, \citenamefont {Schulte},
  \citenamefont {Hammerer},\ and\ \citenamefont {Schmidt}}]{wolf2019motional}%
  \BibitemOpen
  \bibfield  {author} {\bibinfo {author} {\bibfnamefont {F.}~\bibnamefont
  {Wolf}}, \bibinfo {author} {\bibfnamefont {C.}~\bibnamefont {Shi}}, \bibinfo
  {author} {\bibfnamefont {J.~C.}\ \bibnamefont {Heip}}, \bibinfo {author}
  {\bibfnamefont {M.}~\bibnamefont {Gessner}}, \bibinfo {author} {\bibfnamefont
  {L.}~\bibnamefont {Pezzè}}, \bibinfo {author} {\bibfnamefont
  {A.}~\bibnamefont {Smerzi}}, \bibinfo {author} {\bibfnamefont
  {M.}~\bibnamefont {Schulte}}, \bibinfo {author} {\bibfnamefont
  {K.}~\bibnamefont {Hammerer}},\ and\ \bibinfo {author} {\bibfnamefont
  {P.~O.}\ \bibnamefont {Schmidt}},\ }\bibfield  {title} {\bibinfo {title}
  {Motional fock states for quantum-enhanced amplitude and phase measurements
  with trapped ions},\ }\href {https://doi.org/10.1038/s41467-019-10576-4}
  {\bibfield  {journal} {\bibinfo  {journal} {Nature Communications}\ }\textbf
  {\bibinfo {volume} {10}},\ \bibinfo {pages} {1} (\bibinfo {year}
  {2019})}\BibitemShut {NoStop}%
\bibitem [{\citenamefont {Wang}\ \emph {et~al.}(2019)\citenamefont {Wang},
  \citenamefont {Wu}, \citenamefont {Ma}, \citenamefont {Cai}, \citenamefont
  {Hu}, \citenamefont {Mu}, \citenamefont {Xu}, \citenamefont {Chen},
  \citenamefont {Wang}, \citenamefont {Song},\ and\ \citenamefont
  {et~al.}}]{wang2019heisenberglimited}%
  \BibitemOpen
  \bibfield  {author} {\bibinfo {author} {\bibfnamefont {W.}~\bibnamefont
  {Wang}}, \bibinfo {author} {\bibfnamefont {Y.}~\bibnamefont {Wu}}, \bibinfo
  {author} {\bibfnamefont {Y.}~\bibnamefont {Ma}}, \bibinfo {author}
  {\bibfnamefont {W.}~\bibnamefont {Cai}}, \bibinfo {author} {\bibfnamefont
  {L.}~\bibnamefont {Hu}}, \bibinfo {author} {\bibfnamefont {X.}~\bibnamefont
  {Mu}}, \bibinfo {author} {\bibfnamefont {Y.}~\bibnamefont {Xu}}, \bibinfo
  {author} {\bibfnamefont {Z.-J.}\ \bibnamefont {Chen}}, \bibinfo {author}
  {\bibfnamefont {H.}~\bibnamefont {Wang}}, \bibinfo {author} {\bibfnamefont
  {Y.~P.}\ \bibnamefont {Song}},\ and\ \bibinfo {author} {\bibnamefont
  {et~al.}},\ }\bibfield  {title} {\bibinfo {title} {Heisenberg-limited
  single-mode quantum metrology in a superconducting circuit},\ }\href
  {https://doi.org/10.1038/s41467-019-12290-7} {\bibfield  {journal} {\bibinfo
  {journal} {Nature Communications}\ }\textbf {\bibinfo {volume} {10}},\
  \bibinfo {pages} {1} (\bibinfo {year} {2019})}\BibitemShut {NoStop}%
\bibitem [{\citenamefont {Gao}\ \emph {et~al.}(2019)\citenamefont {Gao},
  \citenamefont {Lester}, \citenamefont {Chou}, \citenamefont {Frunzio},
  \citenamefont {Devoret}, \citenamefont {Jiang}, \citenamefont {Girvin},\ and\
  \citenamefont {Schoelkopf}}]{gao2019entanglement}%
  \BibitemOpen
  \bibfield  {author} {\bibinfo {author} {\bibfnamefont {Y.~Y.}\ \bibnamefont
  {Gao}}, \bibinfo {author} {\bibfnamefont {B.~J.}\ \bibnamefont {Lester}},
  \bibinfo {author} {\bibfnamefont {K.~S.}\ \bibnamefont {Chou}}, \bibinfo
  {author} {\bibfnamefont {L.}~\bibnamefont {Frunzio}}, \bibinfo {author}
  {\bibfnamefont {M.~H.}\ \bibnamefont {Devoret}}, \bibinfo {author}
  {\bibfnamefont {L.}~\bibnamefont {Jiang}}, \bibinfo {author} {\bibfnamefont
  {S.~M.}\ \bibnamefont {Girvin}},\ and\ \bibinfo {author} {\bibfnamefont
  {R.~J.}\ \bibnamefont {Schoelkopf}},\ }\bibfield  {title} {\bibinfo {title}
  {Entanglement of bosonic modes through an engineered exchange interaction},\
  }\href {https://doi.org/10.1038/s41586-019-0970-4} {\bibfield  {journal}
  {\bibinfo  {journal} {Nature}\ }\textbf {\bibinfo {volume} {566}},\ \bibinfo
  {pages} {509–512} (\bibinfo {year} {2019})}\BibitemShut {NoStop}%
\bibitem [{\citenamefont {Gan}\ \emph {et~al.}(2020)\citenamefont {Gan},
  \citenamefont {Maslennikov}, \citenamefont {Tseng}, \citenamefont {Nguyen},\
  and\ \citenamefont {Matsukevich}}]{gan2020hybrid}%
  \BibitemOpen
  \bibfield  {author} {\bibinfo {author} {\bibfnamefont {H.~C.~J.}\
  \bibnamefont {Gan}}, \bibinfo {author} {\bibfnamefont {G.}~\bibnamefont
  {Maslennikov}}, \bibinfo {author} {\bibfnamefont {K.-W.}\ \bibnamefont
  {Tseng}}, \bibinfo {author} {\bibfnamefont {C.}~\bibnamefont {Nguyen}},\ and\
  \bibinfo {author} {\bibfnamefont {D.}~\bibnamefont {Matsukevich}},\
  }\bibfield  {title} {\bibinfo {title} {Hybrid quantum computing with
  conditional beam splitter gate in trapped ion system},\ }\href
  {https://doi.org/10.1103/PhysRevLett.124.170502} {\bibfield  {journal}
  {\bibinfo  {journal} {Physical Review Letters}\ }\textbf {\bibinfo {volume}
  {124}},\ \bibinfo {pages} {170502} (\bibinfo {year} {2020})}\BibitemShut
  {NoStop}%
\bibitem [{\citenamefont {Agarwal}\ and\ \citenamefont
  {Tara}(1992)}]{agarwal1992nonclassical}%
  \BibitemOpen
  \bibfield  {author} {\bibinfo {author} {\bibfnamefont {G.~S.}\ \bibnamefont
  {Agarwal}}\ and\ \bibinfo {author} {\bibfnamefont {K.}~\bibnamefont {Tara}},\
  }\bibfield  {title} {\bibinfo {title} {Nonclassical character of states
  exhibiting no squeezing or sub-poissonian statistics},\ }\href
  {https://doi.org/10.1103/PhysRevA.46.485} {\bibfield  {journal} {\bibinfo
  {journal} {Physical Review A}\ }\textbf {\bibinfo {volume} {46}},\ \bibinfo
  {pages} {485–488} (\bibinfo {year} {1992})}\BibitemShut {NoStop}%
\bibitem [{\citenamefont {Richter}\ and\ \citenamefont
  {Vogel}(2002)}]{richter2002nonclassicality}%
  \BibitemOpen
  \bibfield  {author} {\bibinfo {author} {\bibfnamefont {T.}~\bibnamefont
  {Richter}}\ and\ \bibinfo {author} {\bibfnamefont {W.}~\bibnamefont
  {Vogel}},\ }\bibfield  {title} {\bibinfo {title} {Nonclassicality of quantum
  states: A hierarchy of observable conditions},\ }\href
  {https://doi.org/10.1103/PhysRevLett.89.283601} {\bibfield  {journal}
  {\bibinfo  {journal} {Physical Review Letters}\ }\textbf {\bibinfo {volume}
  {89}},\ \bibinfo {pages} {283601} (\bibinfo {year} {2002})}\BibitemShut
  {NoStop}%
\bibitem [{\citenamefont {Miranowicz}\ \emph {et~al.}(2010)\citenamefont
  {Miranowicz}, \citenamefont {Bartkowiak}, \citenamefont {Wang}, \citenamefont
  {Liu},\ and\ \citenamefont {Nori}}]{miranowicz2010testing}%
  \BibitemOpen
  \bibfield  {author} {\bibinfo {author} {\bibfnamefont {A.}~\bibnamefont
  {Miranowicz}}, \bibinfo {author} {\bibfnamefont {M.}~\bibnamefont
  {Bartkowiak}}, \bibinfo {author} {\bibfnamefont {X.}~\bibnamefont {Wang}},
  \bibinfo {author} {\bibfnamefont {Y.-x.}\ \bibnamefont {Liu}},\ and\ \bibinfo
  {author} {\bibfnamefont {F.}~\bibnamefont {Nori}},\ }\bibfield  {title}
  {\bibinfo {title} {Testing nonclassicality in multimode fields: A unified
  derivation of classical inequalities},\ }\href
  {https://doi.org/10.1103/PhysRevA.82.013824} {\bibfield  {journal} {\bibinfo
  {journal} {Physical Review A}\ }\textbf {\bibinfo {volume} {82}},\ \bibinfo
  {pages} {013824} (\bibinfo {year} {2010})}\BibitemShut {NoStop}%
\bibitem [{\citenamefont {Sperling}\ \emph {et~al.}(2012)\citenamefont
  {Sperling}, \citenamefont {Vogel},\ and\ \citenamefont
  {Agarwal}}]{sperling2012subbinomial}%
  \BibitemOpen
  \bibfield  {author} {\bibinfo {author} {\bibfnamefont {J.}~\bibnamefont
  {Sperling}}, \bibinfo {author} {\bibfnamefont {W.}~\bibnamefont {Vogel}},\
  and\ \bibinfo {author} {\bibfnamefont {G.~S.}\ \bibnamefont {Agarwal}},\
  }\bibfield  {title} {\bibinfo {title} {Sub-binomial light},\ }\href
  {https://doi.org/10.1103/PhysRevLett.109.093601} {\bibfield  {journal}
  {\bibinfo  {journal} {Physical Review Letters}\ }\textbf {\bibinfo {volume}
  {109}},\ \bibinfo {pages} {093601} (\bibinfo {year} {2012})}\BibitemShut
  {NoStop}%
\bibitem [{\citenamefont {Peřina}\ \emph
  {et~al.}(2017{\natexlab{a}})\citenamefont {Peřina}, \citenamefont
  {Arkhipov}, \citenamefont {Michálek},\ and\ \citenamefont
  {Haderka}}]{perina2017nonclassicality}%
  \BibitemOpen
  \bibfield  {author} {\bibinfo {author} {\bibfnamefont {J.}~\bibnamefont
  {Peřina}}, \bibinfo {author} {\bibfnamefont {I.~I.}\ \bibnamefont
  {Arkhipov}}, \bibinfo {author} {\bibfnamefont {V.}~\bibnamefont
  {Michálek}},\ and\ \bibinfo {author} {\bibfnamefont {O.}~\bibnamefont
  {Haderka}},\ }\bibfield  {title} {\bibinfo {title} {Nonclassicality and
  entanglement criteria for bipartite optical fields characterized by quadratic
  detectors},\ }\href {https://doi.org/10.1103/PhysRevA.96.043845} {\bibfield
  {journal} {\bibinfo  {journal} {Physical Review A}\ }\textbf {\bibinfo
  {volume} {96}},\ \bibinfo {pages} {043845} (\bibinfo {year}
  {2017}{\natexlab{a}})}\BibitemShut {NoStop}%
\bibitem [{\citenamefont {Peřina}\ \emph
  {et~al.}(2017{\natexlab{b}})\citenamefont {Peřina}, \citenamefont
  {Michálek},\ and\ \citenamefont {Haderka}}]{perina2017higherorder}%
  \BibitemOpen
  \bibfield  {author} {\bibinfo {author} {\bibfnamefont {J.}~\bibnamefont
  {Peřina}}, \bibinfo {author} {\bibfnamefont {V.}~\bibnamefont {Michálek}},\
  and\ \bibinfo {author} {\bibfnamefont {O.}~\bibnamefont {Haderka}},\
  }\bibfield  {title} {\bibinfo {title} {Higher-order sub-poissonian-like
  nonclassical fields: Theoretical and experimental comparison},\ }\href
  {https://doi.org/10.1103/PhysRevA.96.033852} {\bibfield  {journal} {\bibinfo
  {journal} {Physical Review A}\ }\textbf {\bibinfo {volume} {96}},\ \bibinfo
  {pages} {033852} (\bibinfo {year} {2017}{\natexlab{b}})}\BibitemShut
  {NoStop}%
\bibitem [{\citenamefont {Peřina}\ \emph {et~al.}(2019)\citenamefont
  {Peřina}, \citenamefont {Haderka},\ and\ \citenamefont
  {Michálek}}]{perina2019simultaneous}%
  \BibitemOpen
  \bibfield  {author} {\bibinfo {author} {\bibfnamefont {J.}~\bibnamefont
  {Peřina}}, \bibinfo {author} {\bibfnamefont {O.}~\bibnamefont {Haderka}},\
  and\ \bibinfo {author} {\bibfnamefont {V.}~\bibnamefont {Michálek}},\
  }\bibfield  {title} {\bibinfo {title} {Simultaneous observation of
  higher-order non-classicalities based on experimental photocount moments and
  probabilities},\ }\href {https://doi.org/10.1038/s41598-019-45215-x}
  {\bibfield  {journal} {\bibinfo  {journal} {Scientific Reports}\ }\textbf
  {\bibinfo {volume} {9}},\ \bibinfo {pages} {1} (\bibinfo {year}
  {2019})}\BibitemShut {NoStop}%
\bibitem [{\citenamefont {Kimble}\ \emph {et~al.}(1977)\citenamefont {Kimble},
  \citenamefont {Dagenais},\ and\ \citenamefont {Mandel}}]{kimble1977photon}%
  \BibitemOpen
  \bibfield  {author} {\bibinfo {author} {\bibfnamefont {H.~J.}\ \bibnamefont
  {Kimble}}, \bibinfo {author} {\bibfnamefont {M.}~\bibnamefont {Dagenais}},\
  and\ \bibinfo {author} {\bibfnamefont {L.}~\bibnamefont {Mandel}},\
  }\bibfield  {title} {\bibinfo {title} {Photon antibunching in resonance
  fluorescence},\ }\href {https://doi.org/10.1103/PhysRevLett.39.691}
  {\bibfield  {journal} {\bibinfo  {journal} {Physical Review Letters}\
  }\textbf {\bibinfo {volume} {39}},\ \bibinfo {pages} {691–695} (\bibinfo
  {year} {1977})}\BibitemShut {NoStop}%
\bibitem [{\citenamefont {Grangier}\ \emph {et~al.}(1986)\citenamefont
  {Grangier}, \citenamefont {Roger},\ and\ \citenamefont
  {Aspect}}]{grangier1986experimental}%
  \BibitemOpen
  \bibfield  {author} {\bibinfo {author} {\bibfnamefont {P.}~\bibnamefont
  {Grangier}}, \bibinfo {author} {\bibfnamefont {G.}~\bibnamefont {Roger}},\
  and\ \bibinfo {author} {\bibfnamefont {A.}~\bibnamefont {Aspect}},\
  }\bibfield  {title} {\bibinfo {title} {Experimental evidence for a photon
  anticorrelation effect on a beam splitter: A new light on single-photon
  interferences},\ }\href {https://doi.org/10.1209/0295-5075/1/4/004}
  {\bibfield  {journal} {\bibinfo  {journal} {Europhysics Letters (EPL)}\
  }\textbf {\bibinfo {volume} {1}},\ \bibinfo {pages} {173–179} (\bibinfo
  {year} {1986})}\BibitemShut {NoStop}%
\bibitem [{\citenamefont {Gr{\"u}nwald}(2019)}]{grunwald2019effective}%
  \BibitemOpen
  \bibfield  {author} {\bibinfo {author} {\bibfnamefont {P.}~\bibnamefont
  {Gr{\"u}nwald}},\ }\bibfield  {title} {\bibinfo {title} {Effective
  second-order correlation function and single-photon detection},\ }\href
  {https://doi.org/10.1088/1367-2630/ab3ae0} {\bibfield  {journal} {\bibinfo
  {journal} {New Journal of Physics}\ }\textbf {\bibinfo {volume} {21}},\
  \bibinfo {pages} {093003} (\bibinfo {year} {2019})}\BibitemShut {NoStop}%
\bibitem [{\citenamefont {Gr{\"u}nwald}(2020)}]{grunwald2020nonquantum}%
  \BibitemOpen
  \bibfield  {author} {\bibinfo {author} {\bibfnamefont {P.}~\bibnamefont
  {Gr{\"u}nwald}},\ }\bibfield  {title} {\bibinfo {title} {Nonquantum
  information gain from higher-order correlation functions},\ }\href
  {https://doi.org/10.1103/PhysRevResearch.2.023147} {\bibfield  {journal}
  {\bibinfo  {journal} {Physical Review Research}\ }\textbf {\bibinfo {volume}
  {2}},\ \bibinfo {pages} {023147} (\bibinfo {year} {2020})}\BibitemShut
  {NoStop}%
\bibitem [{\citenamefont {Chavez-Mackay}\ \emph {et~al.}(2020)\citenamefont
  {Chavez-Mackay}, \citenamefont {Gr{\"u}nwald},\ and\ \citenamefont
  {Rodríguez-Lara}}]{chavez-mackay2020estimating}%
  \BibitemOpen
  \bibfield  {author} {\bibinfo {author} {\bibfnamefont {J.~R.}\ \bibnamefont
  {Chavez-Mackay}}, \bibinfo {author} {\bibfnamefont {P.}~\bibnamefont
  {Gr{\"u}nwald}},\ and\ \bibinfo {author} {\bibfnamefont {B.~M.}\ \bibnamefont
  {Rodríguez-Lara}},\ }\bibfield  {title} {\bibinfo {title} {Estimating the
  single-photon projection of low-intensity light sources},\ }\href
  {https://doi.org/10.1103/PhysRevA.101.053815} {\bibfield  {journal} {\bibinfo
   {journal} {Physical Review A}\ }\textbf {\bibinfo {volume} {101}},\ \bibinfo
  {pages} {053815} (\bibinfo {year} {2020})}\BibitemShut {NoStop}%
\bibitem [{\citenamefont {Um}\ \emph {et~al.}(2016)\citenamefont {Um},
  \citenamefont {Zhang}, \citenamefont {Lv}, \citenamefont {Lu}, \citenamefont
  {An}, \citenamefont {Zhang}, \citenamefont {Nha}, \citenamefont {Kim},\ and\
  \citenamefont {Kim}}]{um2016phonon}%
  \BibitemOpen
  \bibfield  {author} {\bibinfo {author} {\bibfnamefont {M.}~\bibnamefont
  {Um}}, \bibinfo {author} {\bibfnamefont {J.}~\bibnamefont {Zhang}}, \bibinfo
  {author} {\bibfnamefont {D.}~\bibnamefont {Lv}}, \bibinfo {author}
  {\bibfnamefont {Y.}~\bibnamefont {Lu}}, \bibinfo {author} {\bibfnamefont
  {S.}~\bibnamefont {An}}, \bibinfo {author} {\bibfnamefont {J.-N.}\
  \bibnamefont {Zhang}}, \bibinfo {author} {\bibfnamefont {H.}~\bibnamefont
  {Nha}}, \bibinfo {author} {\bibfnamefont {M.~S.}\ \bibnamefont {Kim}},\ and\
  \bibinfo {author} {\bibfnamefont {K.}~\bibnamefont {Kim}},\ }\bibfield
  {title} {\bibinfo {title} {Phonon arithmetic in a trapped ion system},\
  }\href {https://doi.org/10.1038/ncomms11410} {\bibfield  {journal} {\bibinfo
  {journal} {Nature Communications}\ }\textbf {\bibinfo {volume} {7}},\
  \bibinfo {pages} {1} (\bibinfo {year} {2016})}\BibitemShut {NoStop}%
\bibitem [{\citenamefont {Wang}\ \emph {et~al.}(2017)\citenamefont {Wang},
  \citenamefont {Hu}, \citenamefont {Xu}, \citenamefont {Liu}, \citenamefont
  {Ma}, \citenamefont {Zheng}, \citenamefont {Vijay}, \citenamefont {Song},
  \citenamefont {Duan},\ and\ \citenamefont {Sun}}]{wang2017converting}%
  \BibitemOpen
  \bibfield  {author} {\bibinfo {author} {\bibfnamefont {W.}~\bibnamefont
  {Wang}}, \bibinfo {author} {\bibfnamefont {L.}~\bibnamefont {Hu}}, \bibinfo
  {author} {\bibfnamefont {Y.}~\bibnamefont {Xu}}, \bibinfo {author}
  {\bibfnamefont {K.}~\bibnamefont {Liu}}, \bibinfo {author} {\bibfnamefont
  {Y.}~\bibnamefont {Ma}}, \bibinfo {author} {\bibfnamefont {S.-B.}\
  \bibnamefont {Zheng}}, \bibinfo {author} {\bibfnamefont {R.}~\bibnamefont
  {Vijay}}, \bibinfo {author} {\bibfnamefont {Y.~P.}\ \bibnamefont {Song}},
  \bibinfo {author} {\bibfnamefont {L.-M.}\ \bibnamefont {Duan}},\ and\
  \bibinfo {author} {\bibfnamefont {L.}~\bibnamefont {Sun}},\ }\bibfield
  {title} {\bibinfo {title} {Converting quasiclassical states into arbitrary
  fock state superpositions in a superconducting circuit},\ }\href
  {https://doi.org/10.1103/PhysRevLett.118.223604} {\bibfield  {journal}
  {\bibinfo  {journal} {Physical Review Letters}\ }\textbf {\bibinfo {volume}
  {118}},\ \bibinfo {pages} {223604} (\bibinfo {year} {2017})}\BibitemShut
  {NoStop}%
\bibitem [{\citenamefont {Kienzler}\ \emph {et~al.}(2017)\citenamefont
  {Kienzler}, \citenamefont {Lo}, \citenamefont {Negnevitsky}, \citenamefont
  {Flühmann}, \citenamefont {Marinelli},\ and\ \citenamefont
  {Home}}]{kienzler2017quantum}%
  \BibitemOpen
  \bibfield  {author} {\bibinfo {author} {\bibfnamefont {D.}~\bibnamefont
  {Kienzler}}, \bibinfo {author} {\bibfnamefont {H.-Y.}\ \bibnamefont {Lo}},
  \bibinfo {author} {\bibfnamefont {V.}~\bibnamefont {Negnevitsky}}, \bibinfo
  {author} {\bibfnamefont {C.}~\bibnamefont {Flühmann}}, \bibinfo {author}
  {\bibfnamefont {M.}~\bibnamefont {Marinelli}},\ and\ \bibinfo {author}
  {\bibfnamefont {J.~P.}\ \bibnamefont {Home}},\ }\bibfield  {title} {\bibinfo
  {title} {Quantum harmonic oscillator state control in a squeezed fock
  basis},\ }\href {https://doi.org/10.1103/PhysRevLett.119.033602} {\bibfield
  {journal} {\bibinfo  {journal} {Physical Review Letters}\ }\textbf {\bibinfo
  {volume} {119}},\ \bibinfo {pages} {033602} (\bibinfo {year}
  {2017})}\BibitemShut {NoStop}%
\bibitem [{\citenamefont {Chu}\ \emph {et~al.}(2018)\citenamefont {Chu},
  \citenamefont {Kharel}, \citenamefont {Yoon}, \citenamefont {Frunzio},
  \citenamefont {Rakich},\ and\ \citenamefont {Schoelkopf}}]{chu2018creation}%
  \BibitemOpen
  \bibfield  {author} {\bibinfo {author} {\bibfnamefont {Y.}~\bibnamefont
  {Chu}}, \bibinfo {author} {\bibfnamefont {P.}~\bibnamefont {Kharel}},
  \bibinfo {author} {\bibfnamefont {T.}~\bibnamefont {Yoon}}, \bibinfo {author}
  {\bibfnamefont {L.}~\bibnamefont {Frunzio}}, \bibinfo {author} {\bibfnamefont
  {P.~T.}\ \bibnamefont {Rakich}},\ and\ \bibinfo {author} {\bibfnamefont
  {R.~J.}\ \bibnamefont {Schoelkopf}},\ }\bibfield  {title} {\bibinfo {title}
  {Creation and control of multi-phonon fock states in a bulk acoustic-wave
  resonator},\ }\href {https://doi.org/10.1038/s41586-018-0717-7} {\bibfield
  {journal} {\bibinfo  {journal} {Nature}\ }\textbf {\bibinfo {volume} {563}},\
  \bibinfo {pages} {666–670} (\bibinfo {year} {2018})}\BibitemShut {NoStop}%
\bibitem [{\citenamefont {Gely}\ \emph {et~al.}(2019)\citenamefont {Gely},
  \citenamefont {Kounalakis}, \citenamefont {Dickel}, \citenamefont {Dalle},
  \citenamefont {Vatré}, \citenamefont {Baker}, \citenamefont {Jenkins},\ and\
  \citenamefont {Steele}}]{gely2019observation}%
  \BibitemOpen
  \bibfield  {author} {\bibinfo {author} {\bibfnamefont {M.~F.}\ \bibnamefont
  {Gely}}, \bibinfo {author} {\bibfnamefont {M.}~\bibnamefont {Kounalakis}},
  \bibinfo {author} {\bibfnamefont {C.}~\bibnamefont {Dickel}}, \bibinfo
  {author} {\bibfnamefont {J.}~\bibnamefont {Dalle}}, \bibinfo {author}
  {\bibfnamefont {R.}~\bibnamefont {Vatré}}, \bibinfo {author} {\bibfnamefont
  {B.}~\bibnamefont {Baker}}, \bibinfo {author} {\bibfnamefont {M.~D.}\
  \bibnamefont {Jenkins}},\ and\ \bibinfo {author} {\bibfnamefont {G.~A.}\
  \bibnamefont {Steele}},\ }\bibfield  {title} {\bibinfo {title} {Observation
  and stabilization of photonic fock states in a hot radio-frequency
  resonator},\ }\href {https://doi.org/10.1126/science.aaw3101} {\bibfield
  {journal} {\bibinfo  {journal} {Science}\ }\textbf {\bibinfo {volume}
  {363}},\ \bibinfo {pages} {1072–1075} (\bibinfo {year} {2019})}\BibitemShut
  {NoStop}%
\bibitem [{\citenamefont {Kardynał}\ \emph {et~al.}(2008)\citenamefont
  {Kardynał}, \citenamefont {Yuan},\ and\ \citenamefont
  {Shields}}]{kardynal2008avalanchephotodiodebased}%
  \BibitemOpen
  \bibfield  {author} {\bibinfo {author} {\bibfnamefont {B.~E.}\ \bibnamefont
  {Kardynał}}, \bibinfo {author} {\bibfnamefont {Z.~L.}\ \bibnamefont
  {Yuan}},\ and\ \bibinfo {author} {\bibfnamefont {A.~J.}\ \bibnamefont
  {Shields}},\ }\bibfield  {title} {\bibinfo {title} {An
  avalanche‐photodiode-based photon-number-resolving detector},\ }\href
  {https://doi.org/10.1038/nphoton.2008.101} {\bibfield  {journal} {\bibinfo
  {journal} {Nature Photonics}\ }\textbf {\bibinfo {volume} {2}},\ \bibinfo
  {pages} {425–428} (\bibinfo {year} {2008})}\BibitemShut {NoStop}%
\bibitem [{\citenamefont {Divochiy}\ \emph {et~al.}(2008)\citenamefont
  {Divochiy}, \citenamefont {Marsili}, \citenamefont {Bitauld}, \citenamefont
  {Gaggero}, \citenamefont {Leoni}, \citenamefont {Mattioli}, \citenamefont
  {Korneev}, \citenamefont {Seleznev}, \citenamefont {Kaurova}, \citenamefont
  {Minaeva},\ and\ \citenamefont {et~al.}}]{divochiy2008superconducting}%
  \BibitemOpen
  \bibfield  {author} {\bibinfo {author} {\bibfnamefont {A.}~\bibnamefont
  {Divochiy}}, \bibinfo {author} {\bibfnamefont {F.}~\bibnamefont {Marsili}},
  \bibinfo {author} {\bibfnamefont {D.}~\bibnamefont {Bitauld}}, \bibinfo
  {author} {\bibfnamefont {A.}~\bibnamefont {Gaggero}}, \bibinfo {author}
  {\bibfnamefont {R.}~\bibnamefont {Leoni}}, \bibinfo {author} {\bibfnamefont
  {F.}~\bibnamefont {Mattioli}}, \bibinfo {author} {\bibfnamefont
  {A.}~\bibnamefont {Korneev}}, \bibinfo {author} {\bibfnamefont
  {V.}~\bibnamefont {Seleznev}}, \bibinfo {author} {\bibfnamefont
  {N.}~\bibnamefont {Kaurova}}, \bibinfo {author} {\bibfnamefont
  {O.}~\bibnamefont {Minaeva}},\ and\ \bibinfo {author} {\bibnamefont
  {et~al.}},\ }\bibfield  {title} {\bibinfo {title} {Superconducting nanowire
  photon-number-resolving detector at telecommunication wavelengths},\ }\href
  {https://doi.org/10.1038/nphoton.2008.51} {\bibfield  {journal} {\bibinfo
  {journal} {Nature Photonics}\ }\textbf {\bibinfo {volume} {2}},\ \bibinfo
  {pages} {302–306} (\bibinfo {year} {2008})}\BibitemShut {NoStop}%
\bibitem [{\citenamefont {Namekata}\ \emph {et~al.}(2010)\citenamefont
  {Namekata}, \citenamefont {Takahashi}, \citenamefont {Fujii}, \citenamefont
  {Fukuda}, \citenamefont {Kurimura},\ and\ \citenamefont
  {Inoue}}]{namekata2010nongaussian}%
  \BibitemOpen
  \bibfield  {author} {\bibinfo {author} {\bibfnamefont {N.}~\bibnamefont
  {Namekata}}, \bibinfo {author} {\bibfnamefont {Y.}~\bibnamefont {Takahashi}},
  \bibinfo {author} {\bibfnamefont {G.}~\bibnamefont {Fujii}}, \bibinfo
  {author} {\bibfnamefont {D.}~\bibnamefont {Fukuda}}, \bibinfo {author}
  {\bibfnamefont {S.}~\bibnamefont {Kurimura}},\ and\ \bibinfo {author}
  {\bibfnamefont {S.}~\bibnamefont {Inoue}},\ }\bibfield  {title} {\bibinfo
  {title} {Non-gaussian operation based on photon subtraction using a
  photon-number-resolving detector at a telecommunications wavelength},\ }\href
  {https://doi.org/10.1038/NPHOTON.2010.158} {\bibfield  {journal} {\bibinfo
  {journal} {Nature Photonics}\ }\textbf {\bibinfo {volume} {4}},\ \bibinfo
  {pages} {655–660} (\bibinfo {year} {2010})}\BibitemShut {NoStop}%
\bibitem [{\citenamefont {Endo}\ \emph {et~al.}(2021)\citenamefont {Endo},
  \citenamefont {Sonoyama}, \citenamefont {Matsuyama}, \citenamefont {Okamoto},
  \citenamefont {Miki}, \citenamefont {Yabuno}, \citenamefont {China},
  \citenamefont {Terai},\ and\ \citenamefont {Furusawa}}]{endo2021quantum}%
  \BibitemOpen
  \bibfield  {author} {\bibinfo {author} {\bibfnamefont {M.}~\bibnamefont
  {Endo}}, \bibinfo {author} {\bibfnamefont {T.}~\bibnamefont {Sonoyama}},
  \bibinfo {author} {\bibfnamefont {M.}~\bibnamefont {Matsuyama}}, \bibinfo
  {author} {\bibfnamefont {F.}~\bibnamefont {Okamoto}}, \bibinfo {author}
  {\bibfnamefont {S.}~\bibnamefont {Miki}}, \bibinfo {author} {\bibfnamefont
  {M.}~\bibnamefont {Yabuno}}, \bibinfo {author} {\bibfnamefont
  {F.}~\bibnamefont {China}}, \bibinfo {author} {\bibfnamefont
  {H.}~\bibnamefont {Terai}},\ and\ \bibinfo {author} {\bibfnamefont
  {A.}~\bibnamefont {Furusawa}},\ }\bibfield  {title} {\bibinfo {title}
  {Quantum detector tomography of a superconducting nanostrip
  photon-number-resolving detector},\ }\href
  {https://doi.org/10.1364/oe.423142} {\bibfield  {journal} {\bibinfo
  {journal} {Optics Express}\ }\textbf {\bibinfo {volume} {29}},\ \bibinfo
  {pages} {11728} (\bibinfo {year} {2021})}\BibitemShut {NoStop}%
\bibitem [{\citenamefont {Rigovacca}\ \emph {et~al.}(2016)\citenamefont
  {Rigovacca}, \citenamefont {Di~Franco}, \citenamefont {Metcalf},
  \citenamefont {Walmsley},\ and\ \citenamefont
  {Kim}}]{rigovacca2016nonclassicality}%
  \BibitemOpen
  \bibfield  {author} {\bibinfo {author} {\bibfnamefont {L.}~\bibnamefont
  {Rigovacca}}, \bibinfo {author} {\bibfnamefont {C.}~\bibnamefont
  {Di~Franco}}, \bibinfo {author} {\bibfnamefont {B.~J.}\ \bibnamefont
  {Metcalf}}, \bibinfo {author} {\bibfnamefont {I.~A.}\ \bibnamefont
  {Walmsley}},\ and\ \bibinfo {author} {\bibfnamefont {M.~S.}\ \bibnamefont
  {Kim}},\ }\bibfield  {title} {\bibinfo {title} {Nonclassicality criteria in
  multiport interferometry},\ }\href
  {https://doi.org/10.1103/PhysRevLett.117.213602} {\bibfield  {journal}
  {\bibinfo  {journal} {Physical Review Letters}\ }\textbf {\bibinfo {volume}
  {117}},\ \bibinfo {pages} {213602} (\bibinfo {year} {2016})}\BibitemShut
  {NoStop}%
\bibitem [{\citenamefont {Sperling}\ \emph
  {et~al.}(2017{\natexlab{a}})\citenamefont {Sperling}, \citenamefont
  {Clements}, \citenamefont {Eckstein}, \citenamefont {Moore}, \citenamefont
  {Renema}, \citenamefont {Kolthammer}, \citenamefont {Nam}, \citenamefont
  {Lita}, \citenamefont {Gerrits}, \citenamefont {Vogel},\ and\ \citenamefont
  {et~al.}}]{sperling2017detectorindependent}%
  \BibitemOpen
  \bibfield  {author} {\bibinfo {author} {\bibfnamefont {J.}~\bibnamefont
  {Sperling}}, \bibinfo {author} {\bibfnamefont {W.~R.}\ \bibnamefont
  {Clements}}, \bibinfo {author} {\bibfnamefont {A.}~\bibnamefont {Eckstein}},
  \bibinfo {author} {\bibfnamefont {M.}~\bibnamefont {Moore}}, \bibinfo
  {author} {\bibfnamefont {J.~J.}\ \bibnamefont {Renema}}, \bibinfo {author}
  {\bibfnamefont {W.~S.}\ \bibnamefont {Kolthammer}}, \bibinfo {author}
  {\bibfnamefont {S.~W.}\ \bibnamefont {Nam}}, \bibinfo {author} {\bibfnamefont
  {A.}~\bibnamefont {Lita}}, \bibinfo {author} {\bibfnamefont {T.}~\bibnamefont
  {Gerrits}}, \bibinfo {author} {\bibfnamefont {W.}~\bibnamefont {Vogel}},\
  and\ \bibinfo {author} {\bibnamefont {et~al.}},\ }\bibfield  {title}
  {\bibinfo {title} {Detector-independent verification of quantum light},\
  }\href {https://doi.org/10.1103/PhysRevLett.118.163602} {\bibfield  {journal}
  {\bibinfo  {journal} {Physical Review Letters}\ }\textbf {\bibinfo {volume}
  {118}},\ \bibinfo {pages} {163602} (\bibinfo {year}
  {2017}{\natexlab{a}})}\BibitemShut {NoStop}%
\bibitem [{\citenamefont {Sperling}\ \emph
  {et~al.}(2017{\natexlab{b}})\citenamefont {Sperling}, \citenamefont
  {Eckstein}, \citenamefont {Clements}, \citenamefont {Moore}, \citenamefont
  {Renema}, \citenamefont {Kolthammer}, \citenamefont {Nam}, \citenamefont
  {Lita}, \citenamefont {Gerrits}, \citenamefont {Walmsley},\ and\
  \citenamefont {et~al.}}]{sperling2017identification}%
  \BibitemOpen
  \bibfield  {author} {\bibinfo {author} {\bibfnamefont {J.}~\bibnamefont
  {Sperling}}, \bibinfo {author} {\bibfnamefont {A.}~\bibnamefont {Eckstein}},
  \bibinfo {author} {\bibfnamefont {W.~R.}\ \bibnamefont {Clements}}, \bibinfo
  {author} {\bibfnamefont {M.}~\bibnamefont {Moore}}, \bibinfo {author}
  {\bibfnamefont {J.~J.}\ \bibnamefont {Renema}}, \bibinfo {author}
  {\bibfnamefont {W.~S.}\ \bibnamefont {Kolthammer}}, \bibinfo {author}
  {\bibfnamefont {S.~W.}\ \bibnamefont {Nam}}, \bibinfo {author} {\bibfnamefont
  {A.}~\bibnamefont {Lita}}, \bibinfo {author} {\bibfnamefont {T.}~\bibnamefont
  {Gerrits}}, \bibinfo {author} {\bibfnamefont {I.~A.}\ \bibnamefont
  {Walmsley}},\ and\ \bibinfo {author} {\bibnamefont {et~al.}},\ }\bibfield
  {title} {\bibinfo {title} {Identification of nonclassical properties of light
  with multiplexing layouts},\ }\href
  {https://doi.org/10.1103/PhysRevA.96.013804} {\bibfield  {journal} {\bibinfo
  {journal} {Physical Review A}\ }\textbf {\bibinfo {volume} {96}},\ \bibinfo
  {pages} {013804} (\bibinfo {year} {2017}{\natexlab{b}})}\BibitemShut
  {NoStop}%
\bibitem [{\citenamefont {Filip}\ and\ \citenamefont
  {Lachman}(2013)}]{filip2013hierarchy}%
  \BibitemOpen
  \bibfield  {author} {\bibinfo {author} {\bibfnamefont {R.}~\bibnamefont
  {Filip}}\ and\ \bibinfo {author} {\bibfnamefont {L.}~\bibnamefont
  {Lachman}},\ }\bibfield  {title} {\bibinfo {title} {Hierarchy of feasible
  nonclassicality criteria for sources of photons},\ }\bibfield  {journal}
  {\bibinfo  {journal} {Physical Review A}\ }\textbf {\bibinfo {volume} {88}},\
  \href {https://doi.org/10.1103/PhysRevA.88.043827}
  {10.1103/PhysRevA.88.043827} (\bibinfo {year} {2013})\BibitemShut {NoStop}%
\bibitem [{\citenamefont {Lachman}\ and\ \citenamefont
  {Filip}(2019)}]{lachman2019criteria}%
  \BibitemOpen
  \bibfield  {author} {\bibinfo {author} {\bibfnamefont {L.}~\bibnamefont
  {Lachman}}\ and\ \bibinfo {author} {\bibfnamefont {R.}~\bibnamefont
  {Filip}},\ }\bibfield  {title} {\bibinfo {title} {Criteria for single photon
  sources with variable nonclassicality threshold},\ }\href
  {https://doi.org/10.1088/1367-2630/ab34b0} {\bibfield  {journal} {\bibinfo
  {journal} {New Journal of Physics}\ }\textbf {\bibinfo {volume} {21}},\
  \bibinfo {pages} {083012} (\bibinfo {year} {2019})}\BibitemShut {NoStop}%
\bibitem [{\citenamefont {Lachman}\ \emph {et~al.}(2019)\citenamefont
  {Lachman}, \citenamefont {Straka}, \citenamefont {Hloušek}, \citenamefont
  {Ježek},\ and\ \citenamefont {Filip}}]{lachman2019faithful}%
  \BibitemOpen
  \bibfield  {author} {\bibinfo {author} {\bibfnamefont {L.}~\bibnamefont
  {Lachman}}, \bibinfo {author} {\bibfnamefont {I.}~\bibnamefont {Straka}},
  \bibinfo {author} {\bibfnamefont {J.}~\bibnamefont {Hloušek}}, \bibinfo
  {author} {\bibfnamefont {M.}~\bibnamefont {Ježek}},\ and\ \bibinfo {author}
  {\bibfnamefont {R.}~\bibnamefont {Filip}},\ }\bibfield  {title} {\bibinfo
  {title} {Faithful hierarchy of genuine n -photon quantum non-gaussian
  light},\ }\href {https://doi.org/10.1103/PhysRevLett.123.043601} {\bibfield
  {journal} {\bibinfo  {journal} {Physical Review Letters}\ }\textbf {\bibinfo
  {volume} {123}},\ \bibinfo {pages} {043601} (\bibinfo {year}
  {2019})}\BibitemShut {NoStop}%
\bibitem [{\citenamefont {Zavatta}\ \emph {et~al.}(2004)\citenamefont
  {Zavatta}, \citenamefont {Viciani},\ and\ \citenamefont
  {Bellini}}]{zavatta2004quantumtoclassical}%
  \BibitemOpen
  \bibfield  {author} {\bibinfo {author} {\bibfnamefont {A.}~\bibnamefont
  {Zavatta}}, \bibinfo {author} {\bibfnamefont {S.}~\bibnamefont {Viciani}},\
  and\ \bibinfo {author} {\bibfnamefont {M.}~\bibnamefont {Bellini}},\
  }\bibfield  {title} {\bibinfo {title} {Quantum-to-classical transition with
  single-photon-added coherent states of light},\ }\href@noop {} {\bibfield
  {journal} {\bibinfo  {journal} {Science}\ }\textbf {\bibinfo {volume}
  {306}},\ \bibinfo {pages} {660} (\bibinfo {year} {2004})}\BibitemShut
  {NoStop}%
\bibitem [{\citenamefont {Zavatta}\ \emph {et~al.}(2007)\citenamefont
  {Zavatta}, \citenamefont {Parigi},\ and\ \citenamefont
  {Bellini}}]{zavatta2007experimental}%
  \BibitemOpen
  \bibfield  {author} {\bibinfo {author} {\bibfnamefont {A.}~\bibnamefont
  {Zavatta}}, \bibinfo {author} {\bibfnamefont {V.}~\bibnamefont {Parigi}},\
  and\ \bibinfo {author} {\bibfnamefont {M.}~\bibnamefont {Bellini}},\
  }\bibfield  {title} {\bibinfo {title} {Experimental nonclassicality of
  single-photon-added thermal light states},\ }\href
  {https://doi.org/10.1103/PhysRevA.75.052106} {\bibfield  {journal} {\bibinfo
  {journal} {Physical Review A}\ }\textbf {\bibinfo {volume} {75}},\ \bibinfo
  {pages} {052106} (\bibinfo {year} {2007})}\BibitemShut {NoStop}%
\bibitem [{\citenamefont {Slodička}\ \emph {et~al.}(2016)\citenamefont
  {Slodička}, \citenamefont {Marek},\ and\ \citenamefont
  {Filip}}]{slodicka2016deterministic}%
  \BibitemOpen
  \bibfield  {author} {\bibinfo {author} {\bibfnamefont {L.}~\bibnamefont
  {Slodička}}, \bibinfo {author} {\bibfnamefont {P.}~\bibnamefont {Marek}},\
  and\ \bibinfo {author} {\bibfnamefont {R.}~\bibnamefont {Filip}},\ }\bibfield
   {title} {\bibinfo {title} {Deterministic nonclassicality from thermal
  states},\ }\href {https://doi.org/10.1364/OE.24.007858} {\bibfield  {journal}
  {\bibinfo  {journal} {Optics Express}\ }\textbf {\bibinfo {volume} {24}},\
  \bibinfo {pages} {7858} (\bibinfo {year} {2016})}\BibitemShut {NoStop}%
\bibitem [{\citenamefont {Marek}\ \emph {et~al.}(2016)\citenamefont {Marek},
  \citenamefont {Lachman}, \citenamefont {Slodička},\ and\ \citenamefont
  {Filip}}]{marek2016deterministic}%
  \BibitemOpen
  \bibfield  {author} {\bibinfo {author} {\bibfnamefont {P.}~\bibnamefont
  {Marek}}, \bibinfo {author} {\bibfnamefont {L.}~\bibnamefont {Lachman}},
  \bibinfo {author} {\bibfnamefont {L.}~\bibnamefont {Slodička}},\ and\
  \bibinfo {author} {\bibfnamefont {R.}~\bibnamefont {Filip}},\ }\bibfield
  {title} {\bibinfo {title} {Deterministic nonclassicality for
  quantum-mechanical oscillators in thermal states},\ }\href
  {https://doi.org/10.1103/PhysRevA.94.013850} {\bibfield  {journal} {\bibinfo
  {journal} {Physical Review A}\ }\textbf {\bibinfo {volume} {94}},\ \bibinfo
  {pages} {013850} (\bibinfo {year} {2016})}\BibitemShut {NoStop}%
\bibitem [{\citenamefont {Li}\ \emph {et~al.}(2018)\citenamefont {Li},
  \citenamefont {Gröblacher}, \citenamefont {Zhu},\ and\ \citenamefont
  {Agarwal}}]{li2018generation}%
  \BibitemOpen
  \bibfield  {author} {\bibinfo {author} {\bibfnamefont {J.}~\bibnamefont
  {Li}}, \bibinfo {author} {\bibfnamefont {S.}~\bibnamefont {Gröblacher}},
  \bibinfo {author} {\bibfnamefont {S.-Y.}\ \bibnamefont {Zhu}},\ and\ \bibinfo
  {author} {\bibfnamefont {G.~S.}\ \bibnamefont {Agarwal}},\ }\bibfield
  {title} {\bibinfo {title} {Generation and detection of non-gaussian
  phonon-added coherent states in optomechanical systems},\ }\href
  {https://doi.org/10.1103/PhysRevA.98.011801} {\bibfield  {journal} {\bibinfo
  {journal} {Physical Review A}\ }\textbf {\bibinfo {volume} {98}},\ \bibinfo
  {pages} {011801} (\bibinfo {year} {2018})}\BibitemShut {NoStop}%
\bibitem [{\citenamefont {Ding}\ \emph {et~al.}(2018)\citenamefont {Ding},
  \citenamefont {Maslennikov}, \citenamefont {Hablützel},\ and\ \citenamefont
  {Matsukevich}}]{ding2018quantum}%
  \BibitemOpen
  \bibfield  {author} {\bibinfo {author} {\bibfnamefont {S.}~\bibnamefont
  {Ding}}, \bibinfo {author} {\bibfnamefont {G.}~\bibnamefont {Maslennikov}},
  \bibinfo {author} {\bibfnamefont {R.}~\bibnamefont {Hablützel}},\ and\
  \bibinfo {author} {\bibfnamefont {D.}~\bibnamefont {Matsukevich}},\
  }\bibfield  {title} {\bibinfo {title} {Quantum simulation with a trilinear
  hamiltonian},\ }\href {https://doi.org/10.1103/PhysRevLett.121.130502}
  {\bibfield  {journal} {\bibinfo  {journal} {Physical Review Letters}\
  }\textbf {\bibinfo {volume} {121}},\ \bibinfo {pages} {130502} (\bibinfo
  {year} {2018})}\BibitemShut {NoStop}%
\bibitem [{\citenamefont {Podhora}\ \emph {et~al.}(2020)\citenamefont
  {Podhora}, \citenamefont {Pham}, \citenamefont {Le\v{s}und\'{a}k},
  \citenamefont {Ob\v{s}il}, \citenamefont {\v{C}\'{\i}\v{z}ek}, \citenamefont
  {\v{C}\'{\i}p}, \citenamefont {Marek}, \citenamefont {Slodi\v{c}ka},\ and\
  \citenamefont {Filip}}]{podhora2020unconditional}%
  \BibitemOpen
  \bibfield  {author} {\bibinfo {author} {\bibfnamefont {L.}~\bibnamefont
  {Podhora}}, \bibinfo {author} {\bibfnamefont {T.}~\bibnamefont {Pham}},
  \bibinfo {author} {\bibfnamefont {A.}~\bibnamefont {Le\v{s}und\'{a}k}},
  \bibinfo {author} {\bibfnamefont {P.}~\bibnamefont {Ob\v{s}il}}, \bibinfo
  {author} {\bibfnamefont {M.}~\bibnamefont {\v{C}\'{\i}\v{z}ek}}, \bibinfo
  {author} {\bibfnamefont {O.}~\bibnamefont {\v{C}\'{\i}p}}, \bibinfo {author}
  {\bibfnamefont {P.}~\bibnamefont {Marek}}, \bibinfo {author} {\bibfnamefont
  {L.}~\bibnamefont {Slodi\v{c}ka}},\ and\ \bibinfo {author} {\bibfnamefont
  {R.}~\bibnamefont {Filip}},\ }\bibfield  {title} {\bibinfo {title}
  {Unconditional accumulation of nonclassicality in a single-atom mechanical
  oscillator},\ }\href@noop {} {\bibfield  {journal} {\bibinfo  {journal}
  {arXiv preprint arXiv:2004.12863}\ } (\bibinfo {year} {2020})},\ \Eprint
  {https://arxiv.org/abs/2004.12863} {arXiv:2004.12863 [quant-ph]} \BibitemShut
  {NoStop}%
\bibitem [{\citenamefont {Kiesel}\ \emph {et~al.}(2008)\citenamefont {Kiesel},
  \citenamefont {Vogel}, \citenamefont {Parigi}, \citenamefont {Zavatta},\ and\
  \citenamefont {Bellini}}]{kiesel2008experimental}%
  \BibitemOpen
  \bibfield  {author} {\bibinfo {author} {\bibfnamefont {T.}~\bibnamefont
  {Kiesel}}, \bibinfo {author} {\bibfnamefont {W.}~\bibnamefont {Vogel}},
  \bibinfo {author} {\bibfnamefont {V.}~\bibnamefont {Parigi}}, \bibinfo
  {author} {\bibfnamefont {A.}~\bibnamefont {Zavatta}},\ and\ \bibinfo {author}
  {\bibfnamefont {M.}~\bibnamefont {Bellini}},\ }\bibfield  {title} {\bibinfo
  {title} {Experimental determination of a nonclassical
  glauber-sudarshanpfunction},\ }\href
  {https://doi.org/10.1103/PhysRevA.78.021804} {\bibfield  {journal} {\bibinfo
  {journal} {Physical Review A}\ }\textbf {\bibinfo {volume} {78}},\ \bibinfo
  {pages} {021804} (\bibinfo {year} {2008})}\BibitemShut {NoStop}%
\bibitem [{\citenamefont {Lvovsky}\ and\ \citenamefont
  {Raymer}(2009)}]{lvovsky2009continuousvariable}%
  \BibitemOpen
  \bibfield  {author} {\bibinfo {author} {\bibfnamefont {A.~I.}\ \bibnamefont
  {Lvovsky}}\ and\ \bibinfo {author} {\bibfnamefont {M.~G.}\ \bibnamefont
  {Raymer}},\ }\bibfield  {title} {\bibinfo {title} {Continuous-variable
  optical quantum-state tomography},\ }\href
  {https://doi.org/10.1103/RevModPhys.81.299} {\bibfield  {journal} {\bibinfo
  {journal} {Reviews of Modern Physics}\ }\textbf {\bibinfo {volume} {81}},\
  \bibinfo {pages} {299–332} (\bibinfo {year} {2009})}\BibitemShut {NoStop}%
\bibitem [{\citenamefont {Kiesel}\ \emph {et~al.}(2011)\citenamefont {Kiesel},
  \citenamefont {Vogel}, \citenamefont {Bellini},\ and\ \citenamefont
  {Zavatta}}]{kiesel2011nonclassicality}%
  \BibitemOpen
  \bibfield  {author} {\bibinfo {author} {\bibfnamefont {T.}~\bibnamefont
  {Kiesel}}, \bibinfo {author} {\bibfnamefont {W.}~\bibnamefont {Vogel}},
  \bibinfo {author} {\bibfnamefont {M.}~\bibnamefont {Bellini}},\ and\ \bibinfo
  {author} {\bibfnamefont {A.}~\bibnamefont {Zavatta}},\ }\bibfield  {title}
  {\bibinfo {title} {Nonclassicality quasiprobability of single-photon-added
  thermal states},\ }\href {https://doi.org/10.1103/PhysRevA.83.032116}
  {\bibfield  {journal} {\bibinfo  {journal} {Physical Review A}\ }\textbf
  {\bibinfo {volume} {83}},\ \bibinfo {pages} {032116} (\bibinfo {year}
  {2011})}\BibitemShut {NoStop}%
\bibitem [{\citenamefont {Kühn}\ and\ \citenamefont
  {Vogel}(2018)}]{kuhn2018quantum}%
  \BibitemOpen
  \bibfield  {author} {\bibinfo {author} {\bibfnamefont {B.}~\bibnamefont
  {Kühn}}\ and\ \bibinfo {author} {\bibfnamefont {W.}~\bibnamefont {Vogel}},\
  }\bibfield  {title} {\bibinfo {title} {Quantum non-gaussianity and
  quantification of nonclassicality},\ }\href
  {https://doi.org/10.1103/PhysRevA.97.053823} {\bibfield  {journal} {\bibinfo
  {journal} {Physical Review A}\ }\textbf {\bibinfo {volume} {97}},\ \bibinfo
  {pages} {053823} (\bibinfo {year} {2018})}\BibitemShut {NoStop}%
\bibitem [{\citenamefont {Sperling}\ and\ \citenamefont
  {Walmsley}(2018)}]{sperling2018quasiprobability}%
  \BibitemOpen
  \bibfield  {author} {\bibinfo {author} {\bibfnamefont {J.}~\bibnamefont
  {Sperling}}\ and\ \bibinfo {author} {\bibfnamefont {I.~A.}\ \bibnamefont
  {Walmsley}},\ }\bibfield  {title} {\bibinfo {title} {Quasiprobability
  representation of quantum coherence},\ }\href
  {https://doi.org/10.1103/PhysRevA.97.062327} {\bibfield  {journal} {\bibinfo
  {journal} {Physical Review A}\ }\textbf {\bibinfo {volume} {97}},\ \bibinfo
  {pages} {062327} (\bibinfo {year} {2018})}\BibitemShut {NoStop}%
\bibitem [{\citenamefont {Tan}\ \emph {et~al.}(2020)\citenamefont {Tan},
  \citenamefont {Choi},\ and\ \citenamefont {Jeong}}]{tan2020negativity}%
  \BibitemOpen
  \bibfield  {author} {\bibinfo {author} {\bibfnamefont {K.~C.}\ \bibnamefont
  {Tan}}, \bibinfo {author} {\bibfnamefont {S.}~\bibnamefont {Choi}},\ and\
  \bibinfo {author} {\bibfnamefont {H.}~\bibnamefont {Jeong}},\ }\bibfield
  {title} {\bibinfo {title} {Negativity of quasiprobability distributions as a
  measure of nonclassicality},\ }\href
  {https://doi.org/10.1103/PhysRevLett.124.110404} {\bibfield  {journal}
  {\bibinfo  {journal} {Physical Review Letters}\ }\textbf {\bibinfo {volume}
  {124}},\ \bibinfo {pages} {110404} (\bibinfo {year} {2020})}\BibitemShut
  {NoStop}%
\bibitem [{\citenamefont {Bohmann}\ and\ \citenamefont
  {Agudelo}(2020)}]{bohmann2020phasespace}%
  \BibitemOpen
  \bibfield  {author} {\bibinfo {author} {\bibfnamefont {M.}~\bibnamefont
  {Bohmann}}\ and\ \bibinfo {author} {\bibfnamefont {E.}~\bibnamefont
  {Agudelo}},\ }\bibfield  {title} {\bibinfo {title} {Phase-space inequalities
  beyond negativities},\ }\href
  {https://doi.org/10.1103/PhysRevLett.124.133601} {\bibfield  {journal}
  {\bibinfo  {journal} {Physical Review Letters}\ }\textbf {\bibinfo {volume}
  {124}},\ \bibinfo {pages} {133601} (\bibinfo {year} {2020})}\BibitemShut
  {NoStop}%
\bibitem [{\citenamefont {Sperling}\ \emph {et~al.}(2020)\citenamefont
  {Sperling}, \citenamefont {Phillips}, \citenamefont {Bulmer}, \citenamefont
  {Thekkadath}, \citenamefont {Eckstein}, \citenamefont {Wolterink},
  \citenamefont {Lugani}, \citenamefont {Nam}, \citenamefont {Lita},
  \citenamefont {Gerrits},\ and\ \citenamefont
  {et~al.}}]{sperling2020detectoragnostic}%
  \BibitemOpen
  \bibfield  {author} {\bibinfo {author} {\bibfnamefont {J.}~\bibnamefont
  {Sperling}}, \bibinfo {author} {\bibfnamefont {D.~S.}\ \bibnamefont
  {Phillips}}, \bibinfo {author} {\bibfnamefont {J.~F.~F.}\ \bibnamefont
  {Bulmer}}, \bibinfo {author} {\bibfnamefont {G.~S.}\ \bibnamefont
  {Thekkadath}}, \bibinfo {author} {\bibfnamefont {A.}~\bibnamefont
  {Eckstein}}, \bibinfo {author} {\bibfnamefont {T.~A.~W.}\ \bibnamefont
  {Wolterink}}, \bibinfo {author} {\bibfnamefont {J.}~\bibnamefont {Lugani}},
  \bibinfo {author} {\bibfnamefont {S.~W.}\ \bibnamefont {Nam}}, \bibinfo
  {author} {\bibfnamefont {A.}~\bibnamefont {Lita}}, \bibinfo {author}
  {\bibfnamefont {T.}~\bibnamefont {Gerrits}},\ and\ \bibinfo {author}
  {\bibnamefont {et~al.}},\ }\bibfield  {title} {\bibinfo {title}
  {Detector-agnostic phase-space distributions},\ }\href
  {https://doi.org/10.1103/PhysRevLett.124.013605} {\bibfield  {journal}
  {\bibinfo  {journal} {Physical Review Letters}\ }\textbf {\bibinfo {volume}
  {124}},\ \bibinfo {pages} {013605} (\bibinfo {year} {2020})}\BibitemShut
  {NoStop}%
\bibitem [{\citenamefont {Klyshko}(1996)}]{klyshko_observable_1996}%
  \BibitemOpen
  \bibfield  {author} {\bibinfo {author} {\bibfnamefont {D.~N.}\ \bibnamefont
  {Klyshko}},\ }\bibfield  {title} {\bibinfo {title} {Observable signs of
  nonclassical light},\ }\href {https://doi.org/10.1016/0375-9601(96)00091-6}
  {\bibfield  {journal} {\bibinfo  {journal} {Physics Letters A}\ }\textbf
  {\bibinfo {volume} {213}},\ \bibinfo {pages} {7} (\bibinfo {year}
  {1996})}\BibitemShut {NoStop}%
\bibitem [{\citenamefont {Lee}(1997)}]{lee1997application}%
  \BibitemOpen
  \bibfield  {author} {\bibinfo {author} {\bibfnamefont {C.~T.}\ \bibnamefont
  {Lee}},\ }\bibfield  {title} {\bibinfo {title} {Application of klyshko’s
  criterion for nonclassical states to the micromaser pumped by ultracold
  atoms},\ }\href {https://doi.org/10.1103/PhysRevA.55.4449} {\bibfield
  {journal} {\bibinfo  {journal} {Physical Review A}\ }\textbf {\bibinfo
  {volume} {55}},\ \bibinfo {pages} {4449–4453} (\bibinfo {year}
  {1997})}\BibitemShut {NoStop}%
\bibitem [{\citenamefont {Waks}\ \emph {et~al.}(2006)\citenamefont {Waks},
  \citenamefont {Sanders}, \citenamefont {Diamanti},\ and\ \citenamefont
  {Yamamoto}}]{waks2006highly}%
  \BibitemOpen
  \bibfield  {author} {\bibinfo {author} {\bibfnamefont {E.}~\bibnamefont
  {Waks}}, \bibinfo {author} {\bibfnamefont {B.~C.}\ \bibnamefont {Sanders}},
  \bibinfo {author} {\bibfnamefont {E.}~\bibnamefont {Diamanti}},\ and\
  \bibinfo {author} {\bibfnamefont {Y.}~\bibnamefont {Yamamoto}},\ }\bibfield
  {title} {\bibinfo {title} {Highly nonclassical photon statistics in
  parametric down-conversion},\ }\href
  {https://doi.org/10.1103/PhysRevA.73.033814} {\bibfield  {journal} {\bibinfo
  {journal} {Physical Review A}\ }\textbf {\bibinfo {volume} {73}},\ \bibinfo
  {pages} {033814} (\bibinfo {year} {2006})}\BibitemShut {NoStop}%
\bibitem [{\citenamefont {Wakui}\ \emph {et~al.}(2014)\citenamefont {Wakui},
  \citenamefont {Eto}, \citenamefont {Benichi}, \citenamefont {Izumi},
  \citenamefont {Yanagida}, \citenamefont {Ema}, \citenamefont {Numata},
  \citenamefont {Fukuda}, \citenamefont {Takeoka},\ and\ \citenamefont
  {Sasaki}}]{wakui2014ultrabroadband}%
  \BibitemOpen
  \bibfield  {author} {\bibinfo {author} {\bibfnamefont {K.}~\bibnamefont
  {Wakui}}, \bibinfo {author} {\bibfnamefont {Y.}~\bibnamefont {Eto}}, \bibinfo
  {author} {\bibfnamefont {H.}~\bibnamefont {Benichi}}, \bibinfo {author}
  {\bibfnamefont {S.}~\bibnamefont {Izumi}}, \bibinfo {author} {\bibfnamefont
  {T.}~\bibnamefont {Yanagida}}, \bibinfo {author} {\bibfnamefont
  {K.}~\bibnamefont {Ema}}, \bibinfo {author} {\bibfnamefont {T.}~\bibnamefont
  {Numata}}, \bibinfo {author} {\bibfnamefont {D.}~\bibnamefont {Fukuda}},
  \bibinfo {author} {\bibfnamefont {M.}~\bibnamefont {Takeoka}},\ and\ \bibinfo
  {author} {\bibfnamefont {M.}~\bibnamefont {Sasaki}},\ }\bibfield  {title}
  {\bibinfo {title} {Ultrabroadband direct detection of nonclassical photon
  statistics at telecom wavelength},\ }\href
  {https://doi.org/10.1038/srep04535} {\bibfield  {journal} {\bibinfo
  {journal} {Scientific Reports}\ }\textbf {\bibinfo {volume} {4}},\ \bibinfo
  {pages} {1} (\bibinfo {year} {2014})}\BibitemShut {NoStop}%
\bibitem [{\citenamefont {Kono}\ \emph {et~al.}(2017)\citenamefont {Kono},
  \citenamefont {Masuyama}, \citenamefont {Ishikawa}, \citenamefont {Tabuchi},
  \citenamefont {Yamazaki}, \citenamefont {Usami}, \citenamefont {Koshino},\
  and\ \citenamefont {Nakamura}}]{kono2017nonclassical}%
  \BibitemOpen
  \bibfield  {author} {\bibinfo {author} {\bibfnamefont {S.}~\bibnamefont
  {Kono}}, \bibinfo {author} {\bibfnamefont {Y.}~\bibnamefont {Masuyama}},
  \bibinfo {author} {\bibfnamefont {T.}~\bibnamefont {Ishikawa}}, \bibinfo
  {author} {\bibfnamefont {Y.}~\bibnamefont {Tabuchi}}, \bibinfo {author}
  {\bibfnamefont {R.}~\bibnamefont {Yamazaki}}, \bibinfo {author}
  {\bibfnamefont {K.}~\bibnamefont {Usami}}, \bibinfo {author} {\bibfnamefont
  {K.}~\bibnamefont {Koshino}},\ and\ \bibinfo {author} {\bibfnamefont
  {Y.}~\bibnamefont {Nakamura}},\ }\bibfield  {title} {\bibinfo {title}
  {Nonclassical photon number distribution in a superconducting cavity under a
  squeezed drive},\ }\href {https://doi.org/10.1103/PhysRevLett.119.023602}
  {\bibfield  {journal} {\bibinfo  {journal} {Physical Review Letters}\
  }\textbf {\bibinfo {volume} {119}},\ \bibinfo {pages} {023602} (\bibinfo
  {year} {2017})}\BibitemShut {NoStop}%
\bibitem [{\citenamefont {Simon}\ \emph {et~al.}(1997)\citenamefont {Simon},
  \citenamefont {Selvadoray}, \citenamefont {Arvind},\ and\ \citenamefont
  {Mukunda}}]{simon1997nonclassicality}%
  \BibitemOpen
  \bibfield  {author} {\bibinfo {author} {\bibfnamefont {R.}~\bibnamefont
  {Simon}}, \bibinfo {author} {\bibfnamefont {M.}~\bibnamefont {Selvadoray}},
  \bibinfo {author} {\bibnamefont {Arvind}},\ and\ \bibinfo {author}
  {\bibfnamefont {N.}~\bibnamefont {Mukunda}},\ }\bibfield  {title} {\bibinfo
  {title} {Nonclassicality and the concept of local constraints on the photon
  number distribution},\ }\href@noop {} {\bibfield  {journal} {\bibinfo
  {journal} {arXiv preprint arXiv:quant-ph/9708038}\ } (\bibinfo {year}
  {1997})},\ \Eprint {https://arxiv.org/abs/quant-ph/9708038}
  {quant-ph/9708038} \BibitemShut {NoStop}%
\bibitem [{\citenamefont {Harder}\ \emph {et~al.}(2016)\citenamefont {Harder},
  \citenamefont {Bartley}, \citenamefont {Lita}, \citenamefont {Nam},
  \citenamefont {Gerrits},\ and\ \citenamefont
  {Silberhorn}}]{harder2016singlemode}%
  \BibitemOpen
  \bibfield  {author} {\bibinfo {author} {\bibfnamefont {G.}~\bibnamefont
  {Harder}}, \bibinfo {author} {\bibfnamefont {T.~J.}\ \bibnamefont {Bartley}},
  \bibinfo {author} {\bibfnamefont {A.~E.}\ \bibnamefont {Lita}}, \bibinfo
  {author} {\bibfnamefont {S.~W.}\ \bibnamefont {Nam}}, \bibinfo {author}
  {\bibfnamefont {T.}~\bibnamefont {Gerrits}},\ and\ \bibinfo {author}
  {\bibfnamefont {C.}~\bibnamefont {Silberhorn}},\ }\bibfield  {title}
  {\bibinfo {title} {Single-mode parametric-down-conversion states with 50
  photons as a source for mesoscopic quantum optics},\ }\href
  {https://doi.org/10.1103/PhysRevLett.116.143601} {\bibfield  {journal}
  {\bibinfo  {journal} {Physical Review Letters}\ }\textbf {\bibinfo {volume}
  {116}},\ \bibinfo {pages} {143601} (\bibinfo {year} {2016})}\BibitemShut
  {NoStop}%
\bibitem [{\citenamefont {Hacker}\ \emph {et~al.}(2019)\citenamefont {Hacker},
  \citenamefont {Welte}, \citenamefont {Daiss}, \citenamefont {Shaukat},
  \citenamefont {Ritter}, \citenamefont {Li},\ and\ \citenamefont
  {Rempe}}]{hacker2019deterministic}%
  \BibitemOpen
  \bibfield  {author} {\bibinfo {author} {\bibfnamefont {B.}~\bibnamefont
  {Hacker}}, \bibinfo {author} {\bibfnamefont {S.}~\bibnamefont {Welte}},
  \bibinfo {author} {\bibfnamefont {S.}~\bibnamefont {Daiss}}, \bibinfo
  {author} {\bibfnamefont {A.}~\bibnamefont {Shaukat}}, \bibinfo {author}
  {\bibfnamefont {S.}~\bibnamefont {Ritter}}, \bibinfo {author} {\bibfnamefont
  {L.}~\bibnamefont {Li}},\ and\ \bibinfo {author} {\bibfnamefont
  {G.}~\bibnamefont {Rempe}},\ }\bibfield  {title} {\bibinfo {title}
  {Deterministic creation of entangled atom–light schrödinger-cat states},\
  }\href {https://doi.org/10.1038/s41566-018-0339-5} {\bibfield  {journal}
  {\bibinfo  {journal} {Nature Photonics}\ }\textbf {\bibinfo {volume} {13}},\
  \bibinfo {pages} {110–115} (\bibinfo {year} {2019})}\BibitemShut {NoStop}%
\bibitem [{\citenamefont {Sokolov}\ and\ \citenamefont
  {Wilhelm}(2020)}]{sokolov2020superconducting}%
  \BibitemOpen
  \bibfield  {author} {\bibinfo {author} {\bibfnamefont {A.~M.}\ \bibnamefont
  {Sokolov}}\ and\ \bibinfo {author} {\bibfnamefont {F.~K.}\ \bibnamefont
  {Wilhelm}},\ }\bibfield  {title} {\bibinfo {title} {Superconducting detector
  that counts microwave photons up to two},\ }\href
  {https://doi.org/10.1103/PhysRevApplied.14.064063} {\bibfield  {journal}
  {\bibinfo  {journal} {Phys. Rev. Applied}\ }\textbf {\bibinfo {volume}
  {14}},\ \bibinfo {pages} {064063} (\bibinfo {year} {2020})}\BibitemShut
  {NoStop}%
\bibitem [{\citenamefont {Hong}\ \emph {et~al.}(2017)\citenamefont {Hong},
  \citenamefont {Riedinger}, \citenamefont {Marinković}, \citenamefont
  {Wallucks}, \citenamefont {Hofer}, \citenamefont {Norte}, \citenamefont
  {Aspelmeyer},\ and\ \citenamefont {Gröblacher}}]{hong2017hanbury}%
  \BibitemOpen
  \bibfield  {author} {\bibinfo {author} {\bibfnamefont {S.}~\bibnamefont
  {Hong}}, \bibinfo {author} {\bibfnamefont {R.}~\bibnamefont {Riedinger}},
  \bibinfo {author} {\bibfnamefont {I.}~\bibnamefont {Marinković}}, \bibinfo
  {author} {\bibfnamefont {A.}~\bibnamefont {Wallucks}}, \bibinfo {author}
  {\bibfnamefont {S.~G.}\ \bibnamefont {Hofer}}, \bibinfo {author}
  {\bibfnamefont {R.~A.}\ \bibnamefont {Norte}}, \bibinfo {author}
  {\bibfnamefont {M.}~\bibnamefont {Aspelmeyer}},\ and\ \bibinfo {author}
  {\bibfnamefont {S.}~\bibnamefont {Gröblacher}},\ }\bibfield  {title}
  {\bibinfo {title} {Hanbury brown and twiss interferometry of single phonons
  from an optomechanical resonator},\ }\href
  {https://doi.org/10.1126/science.aan7939} {\bibfield  {journal} {\bibinfo
  {journal} {Science}\ }\textbf {\bibinfo {volume} {358}},\ \bibinfo {pages}
  {203–206} (\bibinfo {year} {2017})}\BibitemShut {NoStop}%
\bibitem [{\citenamefont {Tiedau}\ \emph {et~al.}(2019)\citenamefont {Tiedau},
  \citenamefont {Bartley}, \citenamefont {Harder}, \citenamefont {Lita},
  \citenamefont {Nam}, \citenamefont {Gerrits},\ and\ \citenamefont
  {Silberhorn}}]{tiedau2019scalability}%
  \BibitemOpen
  \bibfield  {author} {\bibinfo {author} {\bibfnamefont {J.}~\bibnamefont
  {Tiedau}}, \bibinfo {author} {\bibfnamefont {T.~J.}\ \bibnamefont {Bartley}},
  \bibinfo {author} {\bibfnamefont {G.}~\bibnamefont {Harder}}, \bibinfo
  {author} {\bibfnamefont {A.~E.}\ \bibnamefont {Lita}}, \bibinfo {author}
  {\bibfnamefont {S.~W.}\ \bibnamefont {Nam}}, \bibinfo {author} {\bibfnamefont
  {T.}~\bibnamefont {Gerrits}},\ and\ \bibinfo {author} {\bibfnamefont
  {C.}~\bibnamefont {Silberhorn}},\ }\bibfield  {title} {\bibinfo {title}
  {Scalability of parametric down-conversion for generating higher-order fock
  states},\ }\href {https://doi.org/10.1103/PhysRevA.100.041802} {\bibfield
  {journal} {\bibinfo  {journal} {Physical Review A}\ }\textbf {\bibinfo
  {volume} {100}},\ \bibinfo {pages} {041802} (\bibinfo {year}
  {2019})}\BibitemShut {NoStop}%
\bibitem [{\citenamefont {Albarelli}\ \emph {et~al.}(2016)\citenamefont
  {Albarelli}, \citenamefont {Ferraro}, \citenamefont {Paternostro},\ and\
  \citenamefont {Paris}}]{albarelli2016nonlinearity}%
  \BibitemOpen
  \bibfield  {author} {\bibinfo {author} {\bibfnamefont {F.}~\bibnamefont
  {Albarelli}}, \bibinfo {author} {\bibfnamefont {A.}~\bibnamefont {Ferraro}},
  \bibinfo {author} {\bibfnamefont {M.}~\bibnamefont {Paternostro}},\ and\
  \bibinfo {author} {\bibfnamefont {M.~G.~A.}\ \bibnamefont {Paris}},\
  }\bibfield  {title} {\bibinfo {title} {Nonlinearity as a resource for
  nonclassicality in anharmonic systems},\ }\href
  {https://doi.org/10.1103/PhysRevA.93.032112} {\bibfield  {journal} {\bibinfo
  {journal} {Physical Review A}\ }\textbf {\bibinfo {volume} {93}},\ \bibinfo
  {pages} {032112} (\bibinfo {year} {2016})}\BibitemShut {NoStop}%
\bibitem [{\citenamefont {Scully}\ and\ \citenamefont
  {Zubairy}(1997)}]{scully1997quantum}%
  \BibitemOpen
  \bibfield  {author} {\bibinfo {author} {\bibfnamefont {M.~O.}\ \bibnamefont
  {Scully}}\ and\ \bibinfo {author} {\bibfnamefont {M.~S.}\ \bibnamefont
  {Zubairy}},\ }\href {https://doi.org/10.1017/CBO9780511813993} {\emph
  {\bibinfo {title} {Quantum Optics}}}\ (\bibinfo  {publisher} {Cambridge
  University Press},\ \bibinfo {year} {1997})\BibitemShut {NoStop}%
\bibitem [{\citenamefont {Hardy}\ \emph {et~al.}(1952)\citenamefont {Hardy},
  \citenamefont {Littlewood},\ and\ \citenamefont
  {P{\'o}lya}}]{hardy1952inequalities}%
  \BibitemOpen
  \bibfield  {author} {\bibinfo {author} {\bibfnamefont {G.~H.}\ \bibnamefont
  {Hardy}}, \bibinfo {author} {\bibfnamefont {J.~E.}\ \bibnamefont
  {Littlewood}},\ and\ \bibinfo {author} {\bibfnamefont {G.}~\bibnamefont
  {P{\'o}lya}},\ }\href@noop {} {\emph {\bibinfo {title} {Inequalities. By GH
  Hardy, JE Littlewood, G. P{\'o}lya..}}}\ (\bibinfo  {publisher} {University
  Press},\ \bibinfo {year} {1952})\BibitemShut {NoStop}%
\bibitem [{\citenamefont {Muirhead}(1902)}]{muirhead1902some}%
  \BibitemOpen
  \bibfield  {author} {\bibinfo {author} {\bibfnamefont {R.~F.}\ \bibnamefont
  {Muirhead}},\ }\bibfield  {title} {\bibinfo {title} {Some methods applicable
  to identities and inequalities of symmetric algebraic functions of n
  letters},\ }\href@noop {} {\bibfield  {journal} {\bibinfo  {journal}
  {Proceedings of the Edinburgh Mathematical Society}\ }\textbf {\bibinfo
  {volume} {21}},\ \bibinfo {pages} {144} (\bibinfo {year} {1902})}\BibitemShut
  {NoStop}%
\bibitem [{\citenamefont {Marshall}\ \emph {et~al.}(1979)\citenamefont
  {Marshall}, \citenamefont {Olkin},\ and\ \citenamefont
  {Arnold}}]{marshall1979inequalities}%
  \BibitemOpen
  \bibfield  {author} {\bibinfo {author} {\bibfnamefont {A.~W.}\ \bibnamefont
  {Marshall}}, \bibinfo {author} {\bibfnamefont {I.}~\bibnamefont {Olkin}},\
  and\ \bibinfo {author} {\bibfnamefont {B.~C.}\ \bibnamefont {Arnold}},\
  }\href@noop {} {\emph {\bibinfo {title} {Inequalities: theory of majorization
  and its applications}}},\ Vol.\ \bibinfo {volume} {143}\ (\bibinfo
  {publisher} {Springer},\ \bibinfo {year} {1979})\BibitemShut {NoStop}%
\bibitem [{\citenamefont {Bhatia}(2013)}]{bhatia2013matrix}%
  \BibitemOpen
  \bibfield  {author} {\bibinfo {author} {\bibfnamefont {R.}~\bibnamefont
  {Bhatia}},\ }\href@noop {} {\emph {\bibinfo {title} {Matrix analysis}}},\
  Vol.\ \bibinfo {volume} {169}\ (\bibinfo  {publisher} {Springer Science \&
  Business Media},\ \bibinfo {year} {2013})\BibitemShut {NoStop}%
\bibitem [{\citenamefont {Cvetkovski}(2012)}]{cvetkovski2012schurs}%
  \BibitemOpen
  \bibfield  {author} {\bibinfo {author} {\bibfnamefont {Z.}~\bibnamefont
  {Cvetkovski}},\ }\bibfield  {title} {\bibinfo {title} {Schur’s inequality,
  muirhead’s inequality and karamata’s inequality},\ }\href
  {https://doi.org/10.1007/978-3-642-23792-8_12} {\bibfield  {journal}
  {\bibinfo  {journal} {Inequalities}\ ,\ \bibinfo {pages} {121–132}}
  (\bibinfo {year} {2012})}\BibitemShut {NoStop}%
\bibitem [{\citenamefont {Kim}\ \emph {et~al.}(1989)\citenamefont {Kim},
  \citenamefont {de~Oliveira},\ and\ \citenamefont
  {Knight}}]{kim1989properties}%
  \BibitemOpen
  \bibfield  {author} {\bibinfo {author} {\bibfnamefont {M.~S.}\ \bibnamefont
  {Kim}}, \bibinfo {author} {\bibfnamefont {F.~A.~M.}\ \bibnamefont
  {de~Oliveira}},\ and\ \bibinfo {author} {\bibfnamefont {P.~L.}\ \bibnamefont
  {Knight}},\ }\bibfield  {title} {\bibinfo {title} {Properties of squeezed
  number states and squeezed thermal states},\ }\href
  {https://doi.org/10.1103/PhysRevA.40.2494} {\bibfield  {journal} {\bibinfo
  {journal} {Physical Review A}\ }\textbf {\bibinfo {volume} {40}},\ \bibinfo
  {pages} {2494–2503} (\bibinfo {year} {1989})}\BibitemShut {NoStop}%
\bibitem [{\citenamefont {Agarwal}\ and\ \citenamefont
  {Tara}(1991)}]{agarwal1991nonclassical}%
  \BibitemOpen
  \bibfield  {author} {\bibinfo {author} {\bibfnamefont {G.~S.}\ \bibnamefont
  {Agarwal}}\ and\ \bibinfo {author} {\bibfnamefont {K.}~\bibnamefont {Tara}},\
  }\bibfield  {title} {\bibinfo {title} {Nonclassical properties of states
  generated by the excitations on a coherent state},\ }\href
  {https://doi.org/10.1103/PhysRevA.43.492} {\bibfield  {journal} {\bibinfo
  {journal} {Physical Review A}\ }\textbf {\bibinfo {volume} {43}},\ \bibinfo
  {pages} {492–497} (\bibinfo {year} {1991})}\BibitemShut {NoStop}%
\bibitem [{\citenamefont {Domínguez-Serna}\ \emph {et~al.}(2016)\citenamefont
  {Domínguez-Serna}, \citenamefont {Mendieta-Jimenez},\ and\ \citenamefont
  {Rojas}}]{dominguez-serna2016entangled}%
  \BibitemOpen
  \bibfield  {author} {\bibinfo {author} {\bibfnamefont {F.~A.}\ \bibnamefont
  {Domínguez-Serna}}, \bibinfo {author} {\bibfnamefont {F.~J.}\ \bibnamefont
  {Mendieta-Jimenez}},\ and\ \bibinfo {author} {\bibfnamefont {F.}~\bibnamefont
  {Rojas}},\ }\bibfield  {title} {\bibinfo {title} {Entangled photon-added
  coherent states},\ }\href {https://doi.org/10.1007/s11128-016-1325-9}
  {\bibfield  {journal} {\bibinfo  {journal} {Quantum Information Processing}\
  }\textbf {\bibinfo {volume} {15}},\ \bibinfo {pages} {3121–3136} (\bibinfo
  {year} {2016})}\BibitemShut {NoStop}%
\bibitem [{\citenamefont {Gard}\ \emph {et~al.}(2016)\citenamefont {Gard},
  \citenamefont {Li}, \citenamefont {You}, \citenamefont {Seshadreesan},
  \citenamefont {Birrittella}, \citenamefont {Luine}, \citenamefont
  {Rafsanjani}, \citenamefont {Mirhosseini}, \citenamefont {Magana-Loaiza},
  \citenamefont {Koltenbah} \emph {et~al.}}]{gard2016photon}%
  \BibitemOpen
  \bibfield  {author} {\bibinfo {author} {\bibfnamefont {B.~T.}\ \bibnamefont
  {Gard}}, \bibinfo {author} {\bibfnamefont {D.}~\bibnamefont {Li}}, \bibinfo
  {author} {\bibfnamefont {C.}~\bibnamefont {You}}, \bibinfo {author}
  {\bibfnamefont {K.~P.}\ \bibnamefont {Seshadreesan}}, \bibinfo {author}
  {\bibfnamefont {R.}~\bibnamefont {Birrittella}}, \bibinfo {author}
  {\bibfnamefont {J.}~\bibnamefont {Luine}}, \bibinfo {author} {\bibfnamefont
  {S.~M.~H.}\ \bibnamefont {Rafsanjani}}, \bibinfo {author} {\bibfnamefont
  {M.}~\bibnamefont {Mirhosseini}}, \bibinfo {author} {\bibfnamefont {O.~S.}\
  \bibnamefont {Magana-Loaiza}}, \bibinfo {author} {\bibfnamefont {B.~E.}\
  \bibnamefont {Koltenbah}}, \emph {et~al.},\ }\bibfield  {title} {\bibinfo
  {title} {Photon added coherent states: nondeterministic, noiseless
  amplification in quantum metrology},\ }\href@noop {} {\bibfield  {journal}
  {\bibinfo  {journal} {arXiv preprint arXiv:1606.09598}\ } (\bibinfo {year}
  {2016})}\BibitemShut {NoStop}%
\bibitem [{\citenamefont {Bu{\v{z}}ek}\ \emph {et~al.}(1991)\citenamefont
  {Bu{\v{z}}ek}, \citenamefont {Lai},\ and\ \citenamefont
  {Knight}}]{buvzek1991displaced}%
  \BibitemOpen
  \bibfield  {author} {\bibinfo {author} {\bibfnamefont {V.}~\bibnamefont
  {Bu{\v{z}}ek}}, \bibinfo {author} {\bibfnamefont {W.}~\bibnamefont {Lai}},\
  and\ \bibinfo {author} {\bibfnamefont {P.}~\bibnamefont {Knight}},\
  }\bibfield  {title} {\bibinfo {title} {Displaced number states},\ }in\
  \href@noop {} {\emph {\bibinfo {booktitle} {Quantum Aspects of Optical
  Communications}}}\ (\bibinfo  {publisher} {Springer},\ \bibinfo {year}
  {1991})\ pp.\ \bibinfo {pages} {295--304}\BibitemShut {NoStop}%
\bibitem [{\citenamefont {Lachman}\ \emph {et~al.}(2016)\citenamefont
  {Lachman}, \citenamefont {Slodička},\ and\ \citenamefont
  {Filip}}]{lachman2016nonclassical}%
  \BibitemOpen
  \bibfield  {author} {\bibinfo {author} {\bibfnamefont {L.}~\bibnamefont
  {Lachman}}, \bibinfo {author} {\bibfnamefont {L.}~\bibnamefont {Slodička}},\
  and\ \bibinfo {author} {\bibfnamefont {R.}~\bibnamefont {Filip}},\ }\bibfield
   {title} {\bibinfo {title} {Nonclassical light from a large number of
  independent single-photon emitters},\ }\href
  {https://doi.org/10.1038/srep19760} {\bibfield  {journal} {\bibinfo
  {journal} {Scientific Reports}\ }\textbf {\bibinfo {volume} {6}},\ \bibinfo
  {pages} {1} (\bibinfo {year} {2016})}\BibitemShut {NoStop}%
\bibitem [{\citenamefont {Hug}\ and\ \citenamefont
  {Weil}(2020)}]{hug2020lectures}%
  \BibitemOpen
  \bibfield  {author} {\bibinfo {author} {\bibfnamefont {D.}~\bibnamefont
  {Hug}}\ and\ \bibinfo {author} {\bibfnamefont {W.}~\bibnamefont {Weil}},\
  }\bibfield  {title} {\bibinfo {title} {Lectures on convex geometry},\
  }\bibfield  {journal} {\bibinfo  {journal} {Graduate Texts in Mathematics}\
  }\href {https://doi.org/10.1007/978-3-030-50180-8}
  {10.1007/978-3-030-50180-8} (\bibinfo {year} {2020})\BibitemShut {NoStop}%
\bibitem [{\citenamefont {Schneider}(2009)}]{schneider2009convex}%
  \BibitemOpen
  \bibfield  {author} {\bibinfo {author} {\bibfnamefont {R.}~\bibnamefont
  {Schneider}},\ }\href {https://doi.org/10.1017/CBO9781139003858} {\emph
  {\bibinfo {title} {Convex Bodies: The Brunn-Minkowski Theory}}}\ (\bibinfo
  {publisher} {Cambridge University Press},\ \bibinfo {year}
  {2009})\BibitemShut {NoStop}%
\bibitem [{\citenamefont {Zavatta}\ \emph {et~al.}(2005)\citenamefont
  {Zavatta}, \citenamefont {Viciani},\ and\ \citenamefont
  {Bellini}}]{zavatta2005singlephoton}%
  \BibitemOpen
  \bibfield  {author} {\bibinfo {author} {\bibfnamefont {A.}~\bibnamefont
  {Zavatta}}, \bibinfo {author} {\bibfnamefont {S.}~\bibnamefont {Viciani}},\
  and\ \bibinfo {author} {\bibfnamefont {M.}~\bibnamefont {Bellini}},\
  }\bibfield  {title} {\bibinfo {title} {Single-photon excitation of a coherent
  state: Catching the elementary step of stimulated light emission},\ }\href
  {https://doi.org/10.1103/PhysRevA.72.023820} {\bibfield  {journal} {\bibinfo
  {journal} {Physical Review A}\ }\textbf {\bibinfo {volume} {72}},\ \bibinfo
  {pages} {023820} (\bibinfo {year} {2005})}\BibitemShut {NoStop}%
\end{thebibliography}%


%

\clearpage
\newpage

\begin{appendices}

\section{Generalized Klyshko's inequalities}
\label{app:sec:generalized_klyshkos}
\parTitle{Section outline}
In this section we show how to generalize Klyshko's criteria.
For the purpose, we will start by reviewing and providing a proof for the original result presented in~\cite{klyshko_observable_1996}, and then show how this is a special case of a more general class of criteria.

\parTitle{Proof of generalized criteria}
Let $\NN\ni k\ge 1$. We want to prove that, for any classical state,
\begin{equation}
    Q_k^2\le Q_{k-1}Q_{k+1},
    \label{app:eq:klyshko_with_Qs}
\end{equation}
where $Q_k\equiv k! P_k$. The proof we present here follows the same ideas given in~\cite{klyshko_observable_1996}.
For any classical state, the probabilities $P_k$ have the form
\begin{equation}
    P_k = \sum_\lambda p_\lambda e^{-\lambda} \frac{\lambda^k}{k!},
\end{equation}
and thus $Q_k=\sum_\lambda p_\lambda e^{-\lambda} \lambda^k$. The sum symbol can be replaced with an integral if needed without affecting the calculations.
\Cref{app:eq:klyshko_with_Qs} is equivalent to, bringing all the terms on the left hand side,
\begin{equation}
\begin{aligned}
    \sum_{\lambda\mu} p_\lambda p_\mu
    e^{-\lambda-\mu} [
        (\lambda\mu)^k - \lambda^{k+1} \mu^{k-1}
    ] \\ =
    \sum_{\lambda\mu} p_\lambda p_\mu e^{-\lambda-\mu} (\lambda\mu)^{k-1}
    (\lambda\mu - \lambda^2) \\ = -
    \sum_{\lambda<\mu} p_\lambda p_\mu
    e^{-\lambda-\mu} (\lambda\mu)^{k-1} (\lambda-\mu)^2,
\end{aligned}
\label{app:eq:klyshkos_basic_mechanism}
\end{equation}
where in the last step we used the fact that, for any symmetric tensor $f_{ij}=f_{ji}$ such that $f_{ii}=0$, we have
$\sum_{ij} f_{ij}=2\sum_{i<j} f_{ij}$.
The conclusion then follows from the observation that $(\lambda-\mu)^2\ge0$ for all $\lambda,\mu$, and that all the sums are extended over only positive $\lambda,\mu\ge0$, by the definition of classical states.

\Cref{app:eq:klyshko_with_Qs} can be further generalized.
Let $I,J$ be arbitrary multi-indices with $s$ elements for some positive integer $s$: $I\equiv(I_1,...,I_s)$ and $J\equiv(J_1,...,J_s)$, and suppose $|I|=|J|$, where $|I|\equiv\sum_k I_k$.
Suppose $I\preceq J$, that is, that $I$ is \textit{majorized} by $J$.
This means that, for all $1\le k\le s$, the sum of the first $k$ greatest elements of $I$ is smaller than that of the first $k$ greatest elements of $J$.
We will then prove that, for all classical states, we must have
\begin{equation}
    \prod_{i=1}^s Q_{I_i} \le \prod_{i=1}^s Q_{J_i}.
    \label{app:eq:super_general_klyshkos}
\end{equation}
We start by remembering the general equality for products of sums:
\begin{equation}
    \prod_{i=1}^n \sum_{j=1}^m a_{ij} = \sum_J \prod_{i=1}^n a_{i J_i},
\end{equation}
where the last sum is over all multi-indices $J$ of length $n$ with $J_i\in\{1,\dots,m\}$ for all $i=1,...,n$.
Applying this to $Q$ we have
\begin{equation}
\begin{aligned}
    \prod_i Q_{I_i} = \prod_i \sum_\lambda p_\lambda e^{-\lambda} \lambda^{I_i} =
    \sum_{\bs\lambda} p_{\bs\lambda} e^{-|\bs\lambda|} \bs\lambda^{I},
\end{aligned}
\end{equation}
where we used the shorthand notation
$p_{\bs\lambda}\equiv \prod_i p_{\lambda_i}$,
$|\bs\lambda|\equiv\sum_i \lambda_i$, and
$\bs\lambda^I\equiv \prod_i \lambda_i^{I_i}$,
and the sum is over all tuples of possible values of $\lambda$.
We thus see that
\begin{equation}
    \prod_i Q_{I_i} -\prod_i Q_{J_i} =
    \sum_{\bs\lambda} p_{\bs\lambda} e^{-|\bs\lambda|} (\bs\lambda^I-\bs\lambda^J).
    \label{app:eq:proving_generalized_klyshkos_nastyexpr}
\end{equation}
The conclusion now follows from Miurhead's inequalities~\cite{hardy1952inequalities,cvetkovski2012schurs}.
To see this, we first notice that $q_{\bslambda}\equiv p_{\bslambda}e^{-|\bslambda|}$ is symmetric upon permutations of $\bslambda$.
This means that we can separate the sum $\sum_{\bslambda}$ by summing first over ordered tuples, and then for each ordered tuple sum over all of its possible permutations.
For example, if $\bslambda$ has length $2$, $\bslambda=(\lambda_1,\lambda_2)$ with each $\lambda_i$ taking values in $\{1,2,3\}$, we write
\begin{equation}
    \sum_{\bslambda} =
    \sum_{\bsgamma} \sum_{\sigma\in S_\bsgamma},
\end{equation}
where
$
    \bsgamma \in \{
        (11), (12), (13),
        (22), (23), (33)
    \}
$
and $S_\bsgamma$ denotes the set of permutations of the tuple $\bsgamma$.
\Cref{app:eq:proving_generalized_klyshkos_nastyexpr} then becomes
\begin{equation}
    \sum_{\bsgamma} q_{\bsgamma} \sum_{\sigma\in S_\bsgamma} 
    (\bsgamma_\sigma^I - \bsgamma_\sigma^J),
\end{equation}
where $\bsgamma_\sigma$ is the tuple obtained by permuting $\bsgamma$ according to $\sigma$.
For each $\bsgamma$, Muirhead's inequalities then tell us that, if $I\preceq J$, and provided $\gamma_i\ge0$, then
\begin{equation}
    \sum_{\sigma\in S_\bsgamma}\bsgamma_\sigma^I \le
    \sum_{\sigma\in S_\bsgamma}\bsgamma_\sigma^J.
\end{equation}
We conclude that~\cref{app:eq:super_general_klyshkos} holds whenever $I\preceq J$.

\subsection{Examples of inequalities}
We show here a few special cases of the criteria~\eqref{app:eq:super_general_klyshkos}.

In the $N=2$ case, the only significant majorization relation is $(1,1)\prec(2,0)$,
corresponding to the well-known Klyshkos criterion $Q_1^2\le Q_0 Q_2$.

Fo $N=3$, we have the chain of relations:
\begin{equation}
	(1,1,1)\prec(2,1,0)\prec(3,0,0),
\end{equation}
corresponding to
$Q_1^3 \le Q_0 Q_1 Q_2 \le Q_0^2 Q_3$,
in which we can recognize the well-known Klyshko criterion $Q_1^2\le Q_0 Q_2$ as one of the constraints satisfied by the probabilities.
For $N=4$, we get the following sequence of inequalities:
\begin{equation}
    Q_1^4 \le Q_2 Q_1^2 Q_0 \le Q_2^2 Q_0^2 \le Q_3 Q_1 Q_0^2 \le Q_4 Q_0^3.
\end{equation}
Similar chains of criteria are obtained for higher-dimensional spaces.
It is interesting to note that it is not however true that all criteria always fall into a single chain of inequalities, like is the case for $N=2, 3, 4$. Indeed, for $N\ge6$ some tuples that are not related via majorization --- for example, $(3,1,1,1)$ and $(2,2,2,0)$ are neither majorized by the other.

\section{Characterization of classical sets in different probability spaces}
\label{sec:characterization_classical_sets}

\subsection{Two-dimensional slices}
\label{subsec:2d_slices}

\begin{figure}[tb]
	\centering
	\includegraphics[width=0.9\linewidth]{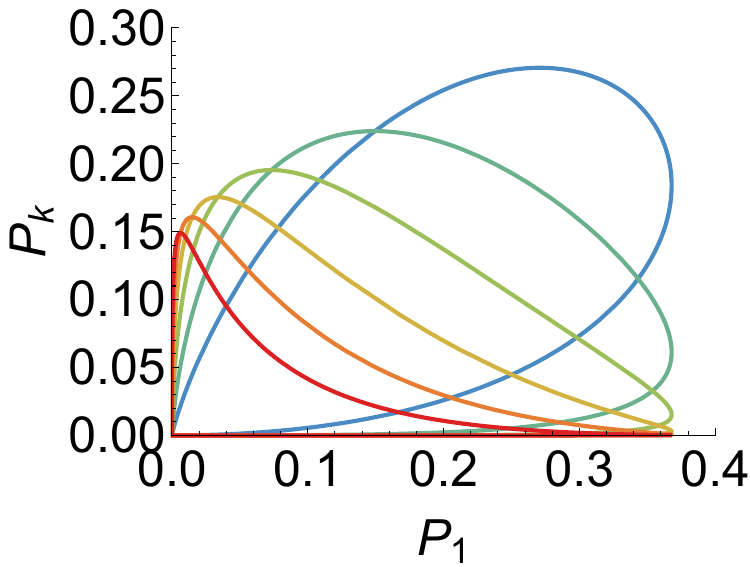}
	\caption{Classical sets in two-dimensional slices of the form $(P_1, P_k)$ with $k=2,3,4,5,6,7$ (from colder/blue to warmer/red colours).}
	\label{app:fig:twodimslices_P1Pk}
\end{figure}

\begin{figure}[tb]
    \centering
    \hspace{-25pt}\includegraphics[width=1.1\linewidth]{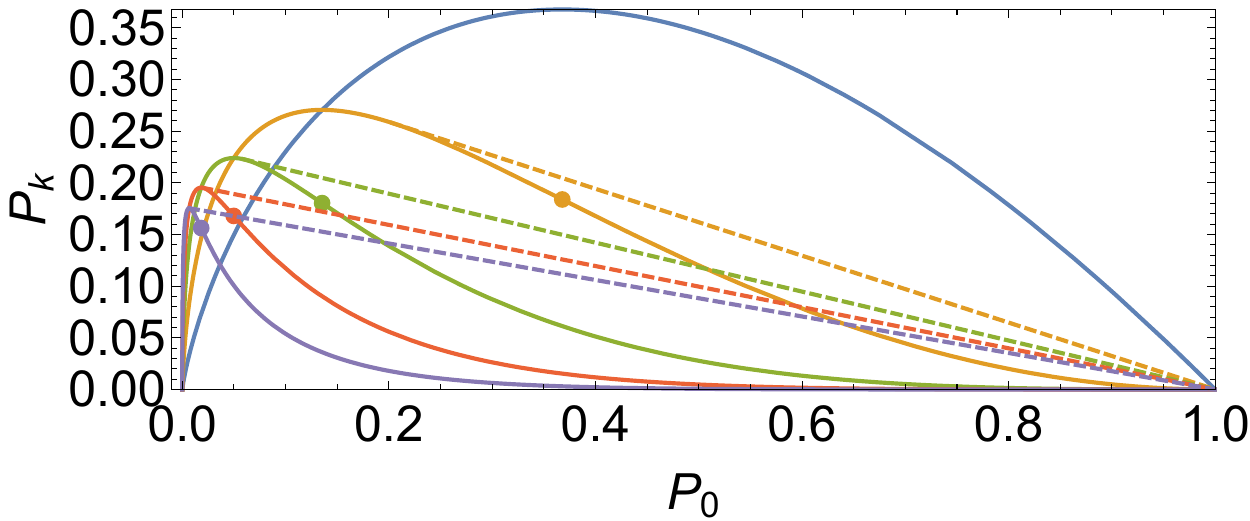}
    \caption{
        Curves drawn by the coherent states in the $(P_0,P_k)$ reduced probability spaces, for $k=1,2,3,4,5$.
        A dot marks the inflection point for each curve. Note how the blue (top right) one, corresponding to $k=1$, is the only curve without an inflection point, consistently with it being the only convex one.
        The dashed line is obtained by closing each curve with a line joining the $(1,0)$ point with the point on that curve that is such that the slope of the corresponding interpolating segment equals the tangent of the curve in that same point. This is the region described by the system given in~\cref{app:eq:system_describing_P0k}.
    }
    \label{app:fig:2dslices_P0Pk}
\end{figure}

\parTitle{Section summary}
We study in this section the projection of the set of classical states in two-dimensional probability spaces of the form $(P_n, P_m)$.
\Cref{app:fig:twodimslices_P1Pk,app:fig:2dslices_P0Pk} give examples of such sets in $(P_1,P_k)$ and $(P_0,P_k)$ for various $k$.
We seek an algebraic criterion that certifies whether a state is inside or outside such classicality regions.

\parTitle{Direct approach}
We first of all note that an algebraic relation characterizing the \emph{boundary} of any such region can be derived with ease: assuming $n<m$, we know that
$Q_n = e^{-\mu}\mu^n$, and thus
$(Q_m/Q_n)^{1/(m-n)}=\mu$, from which we find
\begin{equation}
	Q_n^{\frac{m}{m-n}} = e^{-(Q_m/Q_n)^{1/(m-n)}} Q_m^{\frac{n}{m-n}}.
	\label{app:eq:algebraic_char_2dslieces}
\end{equation}
For example, in $(P_0, P_k)$, this gives
$Q_k = Q_0 [\ln(1/Q_0)]^k$.
In $(P_1, P_k)$, we instead get the relations
\begin{equation}
	Q_1^{\frac{k}{k-1}} =
	e^{-\left(
		Q_k / Q_1
	\right)^{\frac{1}{k-1}}}
	Q_k^{\frac{1}{k-1}},
\end{equation}
which correspond to the shapes given in~\cref{app:fig:twodimslices_P1Pk}.
These two examples show the existence of two possible scenarios: in some cases, coherent states are projected onto a \emph{convex} curve, as is the case for $(P_0, P_1)$. However, in other cases, the coherent states project onto non-convex curves, as seen in~\cref{app:fig:2dslices_P0Pk,app:fig:twodimslices_P1Pk}.
In the convex instances, making~\cref{app:eq:algebraic_char_2dslieces} into an inequality is sufficient to get us a tight nonclassicality criterion.
However, when this curves is non-convex, this would not work, as the boundary of the classical set is the \emph{convex closure} of the curve, which differs from the curve itself.
To deal with these cases, we have to resort to a different technique, discussed in~\cref{app:sec:characterization_via_tangent_planes}.

\parTitle{Criteria via Klyshko-like bounds}
This method does not however generalize to spaces of dimension larger than two, in which no individual algebraic relation describes the probabilities corresponding to coherent states.
Let us then revisit the problem from a different perspective.
Focus on the $(P_0,P_1)$ space.
From~\cref{app:sec:generalized_klyshkos} we know that for all $s\ge2$, classical states satisfy
$Q_s \ge Q_1^s / Q_0^{s-1}$.
Combining these with the constraint $\sum_k P_k=1$ we have
\begin{equation}
	P_0 + P_1 + Q_0 \sum_{s=2}^\infty \frac{1}{s!} \left(\frac{Q_1}{Q_0}\right)^s \le1,
\end{equation}
which simplifies to $P_0 e^{P_1/P_0} \le 1$.
Remarkably, we recover the same bound given explicitly in~\cref{app:eq:algebraic_char_2dslieces}.
This idea of combining bounds given by Klyshko-like inequalities on the unknown probabilities has the advantages of carrying over to higher dimensions.

\parTitle{A case-study: $(P_0,P_2)$}
To highlight a complication sometimes arising with this approach, let us consider the $(P_0,P_2)$ case. This case reveals to be laborious, with the direct approach, due to its non-convexity.
Using the inequalities derived in~\cref{app:sec:generalized_klyshkos}, we have for all classical states and integers $s\ge3$ the lower bound
$Q_s^2 \ge Q_2^s / Q_0^{s-2}$.
However, for $Q_1$, we get an \emph{upper} bound: $Q_1^2 \le Q_0 Q_2$.
This is a problem, as we now cannot simply go from $\sum_k P_k=1$ to an inequality in terms of only $P_0$ and $P_2$. While we \emph{could} use the trivial lower bound $Q_1 \ge 0$, and this would give us a valid nonclassicality criterion, this bound is not tight and the corresponding criterion thus not optimal.
To solve this problem we would need to know the optimal lower bound on $Q_1$. More precisely, we would need to know the minimum value of $\rho_{11}$ when $\rho$ varies over all possible classical states such that $\rho_{00}=Q_0$ and $\rho_{22}=Q_2/2$.
We will obtain this quantity by studying the nonclassicality in $(P_0,P_1,P_2)$, but the nontriviality of this process highlight the complexity that sometimes arises when characterizing the classical region in different probability spaces.

\parTitle{Why some cases are harder to analyze}
To gain some physical intuition on the source of such complexity, observe that Klyshko-like criteria characterize sections of the classicality boundary corresponding to (possibly rescaled) coherent states. Therefore, whenever the boundary also contains \emph{non-pure} states, such criteria will not be able to account for that.

\subsection{Nonclassicality in \texorpdfstring{$(P_0,...,P_n)$}{(P0,...,Pn)}}

\parTitle{Derivation of $\calK_\infty$ criterion}
In $(P_0,...,P_n)$,
we have multiple Klyshko-like inequalities, corresponding to different pairs of tuples related via majorization.
Some of these inequalities can be structured hierarchically. Let us consider for example the following chain of inequalities, which is satisfied by classical states:
\begin{equation}
	\frac{Q_n}{Q_{n-1}} \ge 
	\frac{Q_{n-1}}{Q_{n-2}} \ge
	...
	\ge \frac{Q_2}{Q_1}
	\ge \frac{Q_1}{Q_0}.
	\label{app:eq:sequence_of_klyshkos}
\end{equation}
From~\cref{app:eq:sequence_of_klyshkos} we see that, if $Q_1 Q_{n-1}= Q_0 Q_n$, then the whole sequence of inequalities collapses to the same numerical value.
We also have, for all $s>n$, the inequalities:
\begin{equation}
	Q_s \ge \frac{Q_{n-1}^n}{Q_n^{n-1}} \left(\frac{Q_n}{Q_{n-1}}\right)^s.
	\label{app:eq:lower_bound_Qs}
\end{equation}
These follow from the relations
\begin{equation}
	(\underbrace{N,...,N}_{s-N+1}) \prec (s,\underbrace{N-1,...,N-1}_{s-N}).
	\label{app:eq:majorization_corresponding_to_naive_Koo}
\end{equation}
Putting~\cref{app:eq:lower_bound_Qs} together with $\sum_k P_k=1$ we get
\begin{equation}
\scalebox{0.94}{$\displaystyle
	\sum_{k=0}^{n-2} P_k +
	\frac{Q_{n-1}^n}{Q_n^{n-1}} \left[
		e^{\frac{Q_n}{Q_{n-1}}} - \sum_{k=0}^{n-2} \frac{(Q_n/Q_{n-1})^k}{k!}
	\right]\le1.$}
	\label{app:eq:klyshko_infinite_inequality_N}
\end{equation}
For ease of notation, let us introduce the quantities $K_n\equiv Q_1 Q_{n-1}-Q_0 Q_n$, and $K_\infty$ equal to the left hand-side of~\cref{app:eq:klyshko_infinite_inequality_N}, so that the nonclassicality criteria take the form
$K_n > 0$ and $K_\infty > 1$.

\parTitle{Equality implies coherent state}
We will show here $K_n=K_\infty-1=0$ implies that the corresponding probability vector is compatible with a coherent state.
We know that $K_n=0$ implies
\begin{equation}
	\frac{Q_0}{Q_1} =
	\frac{Q_1}{Q_2} =
	... =
	\frac{Q_{n-2}}{Q_{n-1}} =
	\frac{Q_{n-1}}{Q_n}.
	\label{app:eq:sequence_klyshko_equality}
\end{equation}
For all $s\ge2$, these imply $Q_s=Q_0 (Q_1/Q_0)^{s}$.
Moreover, using~\cref{app:eq:sequence_klyshko_equality} in the definition of $K_\infty$ gives
$
	K_\infty =
	P_0 e^{P_1 / P_0} = 1.
$ 
Consider a coherent state with average boson number $\mu\equiv P_1/P_0$.
Then, $Q_s=Q_0 \mu^s$. Moreover, $K_\infty=1$ translates into $P_0=e^{-P_1/P_0}=e^{-\mu}$,
allowing us to conclude that $Q_s = e^{-\mu}\mu^s$, implying that $\bs P$ is indeed compatible with a coherent state $\ket{\alpha}$ such that $|\alpha|^2=\mu$.

\section{Applications}
\label{app:applications}

\subsection{Attenuated Fock states}
\label{app:attenuated_fock_states}

\parTitle{Definitions}
Let $\mathcal E_T$ denote the channel corresponding to attenuation through a beamsplitter with transmittivity $T$. Thus, in particular, we have
\begin{equation}
    \rho_{k,T}\equiv \mathcal E_T(\ketbra k) = 
    \sum_{j=0}^k \binom{n}{k} T^k (1-T)^{n-k} \ketbra j,
\end{equation}
for any Fock state $\ket k$, $k\ge0$. The corresponding boson-number probabilities are therefore $P_{j}^{(k,T)} = \binom{n}{j}T^k(1-T)^{n-k}$, for all $j=0,...,k$.

\begin{figure}
    \centering
    \includegraphics[width=\linewidth]{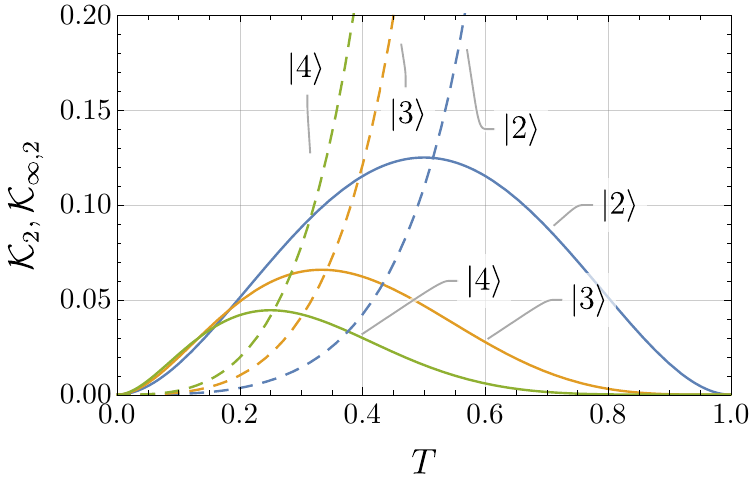}
    \caption{Violation of $\calK_1$ (sollid lines) and $\calK_{\infty,2}$ (dashed lines) criteria for attenuated Fock states $\ket2,\ket3$ and $\ket4$, for different tranmissivities $T\in[0,1]$. The violation is quantified via $K_1(\bs P)$ and $K_{\infty,2}(\bs P)-1$, defined in the previous section, so that positive values represent nonclassicality certified by the corresponding criterion. As shown here, the nonclassicality of attenuated Fock states can be certified for all $T$ with both criteria. However, remarkably, it becomes much harder to detect Klyshko's criterion $\calK_2$ for small attenuations, $T\sim 1$. On the other hand, the $K_{\infty,2}$ diverges for $T\to 1^-$, making it much easier to use with more limited statistics.}
    \label{fig:attenuated_Fock_states_K2Kinf}
\end{figure}

\parTitle{Nonclassicality results}
To investigate the nonclassicality of $\rho_{k,T}$, we observe that, for $k\ge1$,
\begin{equation}
\begin{gathered}
    K_1(P^{(k,T)}) \equiv (P^{(k,T)}_1)^2 - 2P^{(k,T)}_0 P^{(k,T)}_2 \\
    = k(1-T)^{2(k-1)} T^2
\end{gathered}
\end{equation}
This shows that $K_1(P^{(k,1)})=0$ for all $k\ge2$, implying all Fock states $\ket k$ for $k\ge2$ are not detected as nonclassical by $\calK_1$, and that small attenuations of such states still make the nonclassicality hard to detect.
Simiarly, one can verify that $\calK_{\infty,2}$ also always detects the nonclassicality of these states, but the violation is significantly easier to detect for small attenuations, as $K_{\infty,2}(P^{(k,T)})$ goes to $+\infty$ when $T\to1^-$. Such behaviour can be clearly seen plotting the amount with which the two different nonclassicality criteria are violated as a function of $T$ and $k$. The results for $k=2,3,4$ are reported in~\cref{fig:attenuated_Fock_states_K2Kinf}.
The $k=1$ case is, on the other hand, well detected as nonclassical by $\calK_1$, but not by $\calK_{\infty,2}$.

\subsection{Noisy Fock states}
\label{app:noisy_fock_states}

\parTitle{Definition}
Displaced Fock states are defined as $\ket{\alpha;k}=D(\alpha)\ket k$.
Averaging over $\alpha$, these correspond to Fock states with added Poissonian noise, which also share the same Fock-state probabilities.

\begin{figure}[tb]
    \centering
    \includegraphics[width=\linewidth]{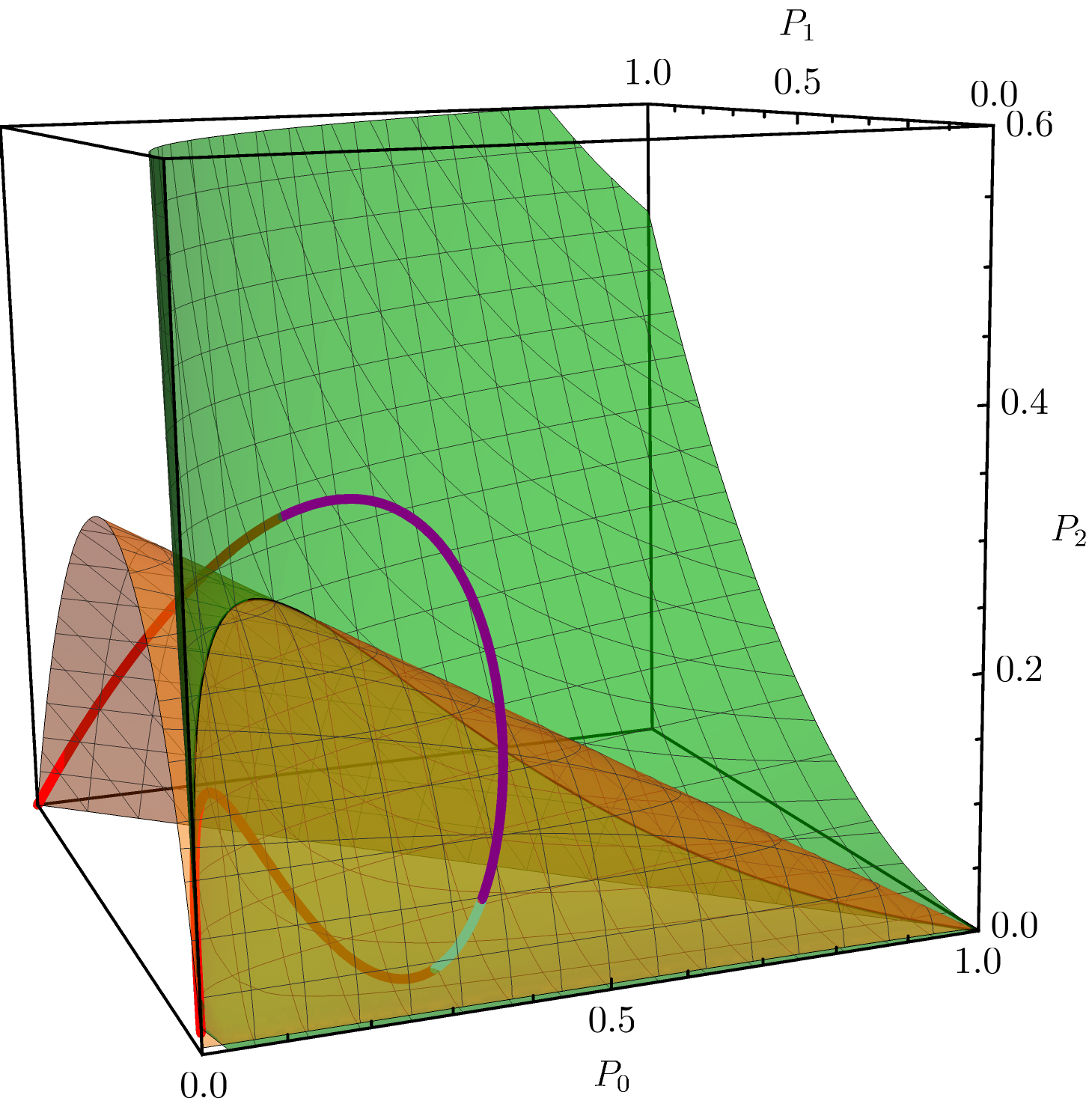}
    \caption{Nonclassicality of noisy single Fock states with probability distribution $P^{(1;\mu)}_j$, for different $\mu$. The coloured curve gives the probabilities $(P_0,P_1,P_2)$ corresponding to $\ket{\alpha;1}$ for different $\alpha$. The green and orange surfaces correspond to the two nonclassicality boundaries~\cref{eq:P012_criterion_K1,eq:P012_criterion_Kinf}, respectively. A state is certifiably nonclassical whenever it is outside of the closed region defined by these two surfaces. In particular, the red sections of the curve correspond to values of $\mu$ for which the state is detected as nonclassical by~\cref{eq:P012_criterion_K1}. On the other hand, the \emph{purple} section of the curve corresponds to $\mu$ for which the state is detected as nonclassical by~\cref{eq:P012_criterion_Kinf}. Finally, the cyan section of the curve corresponds to $\mu$ for which the state lies within the two surfaces, and is therefore compatible with a classical state.
    }
    \label{fig:displaced_fock1_P012}
\end{figure}

For $k=1$, the Fock-state probabilities are
\begin{equation}
    P^{(1;\mu)}_j \equiv \braket{j}{\alpha;1} = \frac{e^{-\mu}}{j!}\mu^{j-1}(j-\mu)^2,
\end{equation}
where $\mu\equiv|\alpha|^2$.
The corresponding probability vectors in $(P_0,P_1,P_2)$ go from $(0,1,0)$ to $(0,0,0)$ for $\mu$ going from $0$ to $\infty$, as shown in~\cref{fig:displaced_fock1_P012}.

\begin{figure}[tb]
    \centering
    \includegraphics[width=\linewidth]{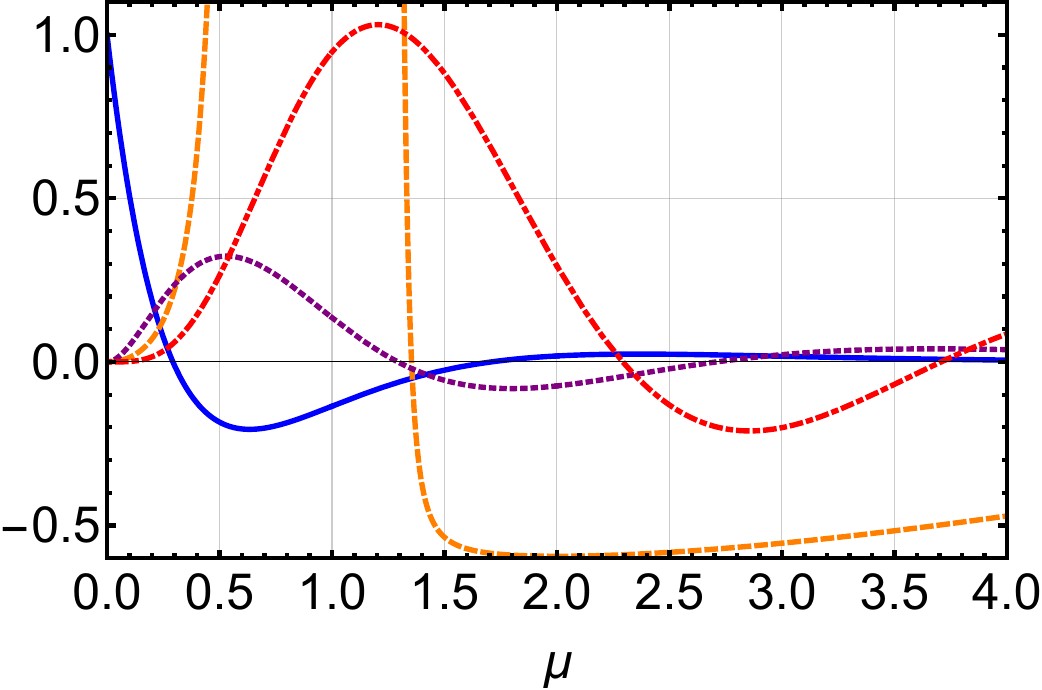}
    \caption{%
        Degrees of violation of nonclassicality criteria for noisy single-boson Fock states with probability distribution $P^{(1;\mu)}_j$, for different $\mu$.
        Blue (continuous) line: $K_1$;
        orange (dashed) line: $K_{\infty,2}-1$;
        purple (dotted) line: $K_2$;
        red (dotdashed) line: $K_3$.
        For each $\mu$, nonclassicality is certified by the \emph{positivity} of at least one of the lines.
        We stress how which criteria are available to certify nonclassicality depends on the number of accessible Fock-state probabilities. For example, if only $P_0,P_1,P_2$ are known, the only states that are recognisable as nonclassical are the ones revealed by the blue and orange curves.
    }
    \label{fig:Ki_displacedFock1_vs_mu}
\end{figure}

\begin{figure}[tb]
	\centering
	\includegraphics[width=\linewidth]{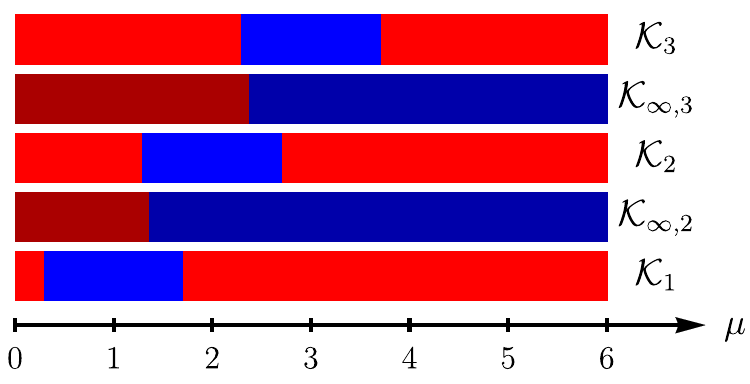}
	\caption{
		\emph{Nonclassicality of noisy Fock states statistics $P^{(1;\mu)}$}.
		Light (dark) red indicates nonclassicality according to a corresponding $\calK_i$ ($\calK_{\infty,i}$) criterion.
		The horizontal axis corresponds to the average boson number $\mu$.
		For each criterion $\calK_i,\calK_{\infty,j}$, we show in red values of $\mu$ that are certified as nonclassical, and in blue values that are not.
		Note that $\calK_{\infty,i}$ uses probabilities up to $i$, whereas $\calK_i$ uses probabilities up to $i+1$.
		We find that noisy Fock states are always certified as nonclassical by Klyshko's criteria \emph{provided that $P_0, P_1, P_2, P_3, P_4$ are all known}.
		However, if only the first \emph{three} Fock-state probabilities are known, only $\calK_1$ and $\calK_2$ can be used, and there are values of $\mu$ for which these criteria are not sufficient (\emph{e.g.} $\mu\simeq 1.5$).
		Nonetheless, $\calK_{\infty,3}$, which does not require the knowledge of $P_4$, can be used to certify nonclassicality in these regions.
		We do not show higher-order criteria here because as can be seen, knowing the first five probabilities is sufficient for a full characterization of the nonclassicality.
	}
	\label{fig:noisyFock1_nonclassicality_rectangles}
\end{figure}

\begin{figure}[tb]
	\centering
	\includegraphics[width=\linewidth]{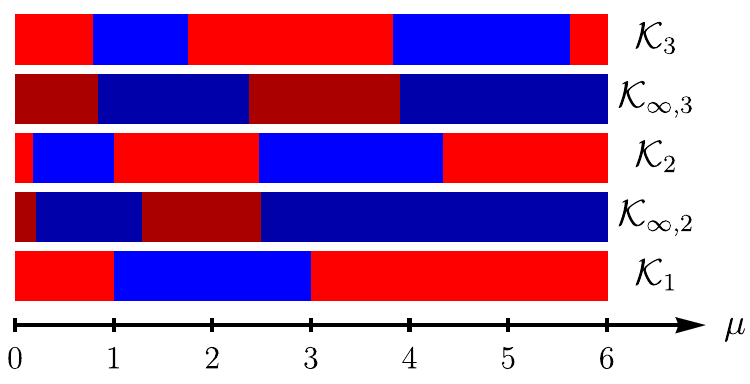}
	\caption{
		\emph{Nonclassicality of noisy Fock state statistics $P^{(2;\mu)}$}.
		Notation is as in~\cref{fig:noisyFock1_nonclassicality_rectangles}.
		We notice how, as in~\cref{fig:noisyFock1_nonclassicality_rectangles}, $\calK_{\infty,3}$ allows to fully capture the nonclassicality of the states without knowing $P_4$, as was necessary using only Klyshko's criteria.
	}
	\label{fig:noisyFock2_nonclassicality_rectangles}
\end{figure}

\parTitle{Nonclassicality of noisy Fock states}
Explicitly, we have
\begin{equation}
    (P_1^{(1,\mu)})^2 - 2 P_0^{(1,\mu)} P_2^{(1,\mu)}
    = e^{-2\mu} (2\mu^2 - 4\mu + 1).
\end{equation}
This quantity is non-positive for $\mu\in[\mu_-,\mu_+]\approx[0.29, 1.71]$ where $\mu_\pm\equiv 1\pm1/\sqrt2$, implying that these states are detected as nonclassical by $\calK_1$ for small and large Poissonian noise: $\mu<\mu_-$ and $\mu> \mu_+$, but not for $\mu\in[\mu_-,\mu_+]$.
On the other hand, one can verify numerically that $K_{\infty,2}>1$ for $\mu \lesssim 1.35 < \mu_+$.
This means that, in the range $\mu\in[0.29, 1.35]$, the nonclassicality of the states can only be certified by $\calK_{\infty,2}$.
At the same time, in the range $\mu\in[1.35, 1.71]$, for which $K_1,K_{\infty,2}-1<0$, we are ensured that the observed probabilities $(P_0, P_1, P_2)$ are compatible with classical states.
In~\cref{fig:Ki_displacedFock1_vs_mu} we give the degrees of violation of the different criteria for different values of $\mu$.
In~\cref{fig:noisyFock1_nonclassicality_rectangles,fig:noisyFock2_nonclassicality_rectangles} we show for various values of $\mu$ how nonclassicality is detected by the different criteria, for noisy Fock state statistics $P^{(1;\mu)}$ and $P^{(2;\mu)}$, respectively.

\subsection{Boson-added coherent states}
\label{app:boson_added_coherent_states}

\parTitle{Definition}
Boson-added coherent states $\ket{\alpha,\ell}$ are defined as
$
	\ket{\alpha,\ell}\equiv C_{\alpha,\ell}a^{\dagger \ell}\ket\alpha,
$
with $C_{\alpha,\ell}$ normalisation constants.
These correspond to the same Fock-state probabilities as photon- and phonon-added Poissonian noise.
Denoting with $\mu\equiv|\alpha|^2$ the average boson number of the coherent state, the corresponding Fock-state probabilities $P^{(\mu,\ell)}$ are related to the Poissonian distribution $k! P^\mu_k\equiv e^{-\mu}\mu^k$, by
\begin{equation}
	P^{(\mu,\ell)}_k = C_{\mu,\ell} P^\mu_k \frac{(k^{\underline\ell})^2}{\mu^\ell},
\end{equation}
where $k^{\underline\ell}\equiv k(k-1)\cdots (k-\ell+1)$.
In particular
\begin{equation}
\begin{gathered}
	P_k^{(\mu,1)} = \frac{1}{\mu+1}\frac{e^{-\mu}\mu^k}{k!} \frac{k^2}{\mu}, \\
	P_k^{(\mu,2)} = \frac{1}{\mu^2+4\mu+2}\frac{e^{-\mu}\mu^k}{k!} \frac{[k(k-1)]^2}{\mu^2}.
\end{gathered}
\end{equation}
The corresponding first Fock-state probabilities are
\begin{equation}
\begin{gathered}
	\frac{e^{-\mu}}{\mu+1}(0,1,2\mu,3\mu^2/2) \quad\,\,\,\,\text{ for } \ell=1, \\
	\frac{2e^{-\mu}}{\mu^2+4\mu+2}(0,0,1,3\mu) \quad\text{ for } \ell=2.
\end{gathered}	
\end{equation}
\parTitle{Nonclassicality of boson-added coherent states}
As pointed out in the main text,
the nonclassicality of $\ell=1$ follows from $\calK_1$.
On the other hand, for $\ell=2$ we have $P_0=P_1=0$, and thus $\calK_1$ is inconclusive.
If $P_3$ is known, $\calK_2$ could be used, but this is not the case when only the first three Fock-state probabilities are known.
On the other hand, $\calK_{\infty,2}$ allows to certify nonclassicality even without the knowledge of $P_3$.

\parTitle{Probabilistic boson addition case}
Consider now the nonclassicality of states of the form $p\, a^\dagger \ketbra\alpha a+(1-p)\ketbra\alpha$, corresponding to the probabilistic boson addition to a coherent state.
These correspond to the Fock-state probabilities
\begin{equation}
	\tilde P_k^{(\mu,\ell,p)} = \frac{e^{-\mu}\mu^k}{k!}
	\left[
		p \frac{k^2}{\mu(\mu+1)} + (1-p)
	\right].
\end{equation}
As discussed in the main text, we again find that having access to more Fock-state probabilities we can certify nonclassicality for more values of $\mu$, and that the use of the $\calK_\infty$ criteria increases our predictive power.

\subsection{Thermally averaged Fock states}
\parTitle{Definition}
Thermally averaged displaced Fock states model realistic Fock states prepared in trapped-ion and superconducting-circuit experiments.
These are obtained by computing thermal averages of displaced Fock states. The thermal average corresponding to the displaced Fock state $\ket{\alpha;1}$ gives
\begin{equation}\scalebox{0.86}{$\displaystyle
    P_k^{(1;\mu;\on{th})}
        = \int_0^\infty \!d\lambda \frac{e^{-\lambda/\mu}}{\mu} P_k^{(1;\lambda)}
        = \left( 1 + \frac{1}{\mu}\right)^{-k} \frac{k+\mu^2}{\mu(1+\mu)^2}.
$}\end{equation}
\parTitle{Nonclassicality of thermally averaged Fock states}
Testing the criteria on these probabilities we find that $P_k^{(1;\mu;\on{th})}\in\calK_1^>$ for $0\le \mu < \sqrt{-1+\sqrt2}\simeq0.64$, thus certifying the nonclassicality in this region. On the other hand, for $\mu\ge \sqrt{-1 + \sqrt2}$, we have $P_k^{(1;\mu;\on{th})}\in\calK_1^\le\cap\calK_\infty^\le$, and the statistics are therefore compatible with classical states.

\subsection{Boson-added thermal states}
\label{app:boson_added_thermal_states}

\parTitle{Definition and nonclassicality}
A \emph{thermal state} $\rho^{(\on{th};\mu)}$ corresponds to Fock-state probabilities
\begin{equation}
	P^{(\on{th};\mu)} = \frac{\mu^k}{(1+\mu)^{k+1}}.
\end{equation}
These states can be written as convex combinations of coherent states and are thus known to be classical.
Let us consider instead \emph{boson-added thermal states}, that is, states of the form $a^\dagger \rho^{(\on{th};\mu)} a$, up to normalisation.
These correspond to the Fock-state probabilities
\begin{equation}
	P^{(\on{th};1,\mu)}_k = \frac{k P^{(\on{th;\mu})}_{k-1}}{\mu+1} =
	\frac{k\mu^{k-1}}{(1+\mu)^{k+1}}.
\end{equation}
These probabilities are certified as nonclassical by $\calK_1$ for all $\mu$,
thus in this case there is no need for stronger criteria.

\parTitle{Probabilistic boson addition}
Mixing this state with the corresponding thermal state with probability $p$ gives the Fock-state probabilities
\begin{equation}
	\left[(1-p)\frac{k}{\mu} + p\right] P_k^{(\on{th};\mu)}.
\end{equation}
Depending on the value of $p$, we get nonclassical state for some ranges of $\mu$.
For example, for $p=0.5$, we find that nonclassicality is only detected for $\mu\lesssim 0.4$ by $\calK_1$. The $\calK_\infty$ criteria do not appear to provide further predictive power in these cases.

\subsection{Squeezed thermal states}
\begin{figure}[tb]
	\centering
	\includegraphics[width=0.95\linewidth]{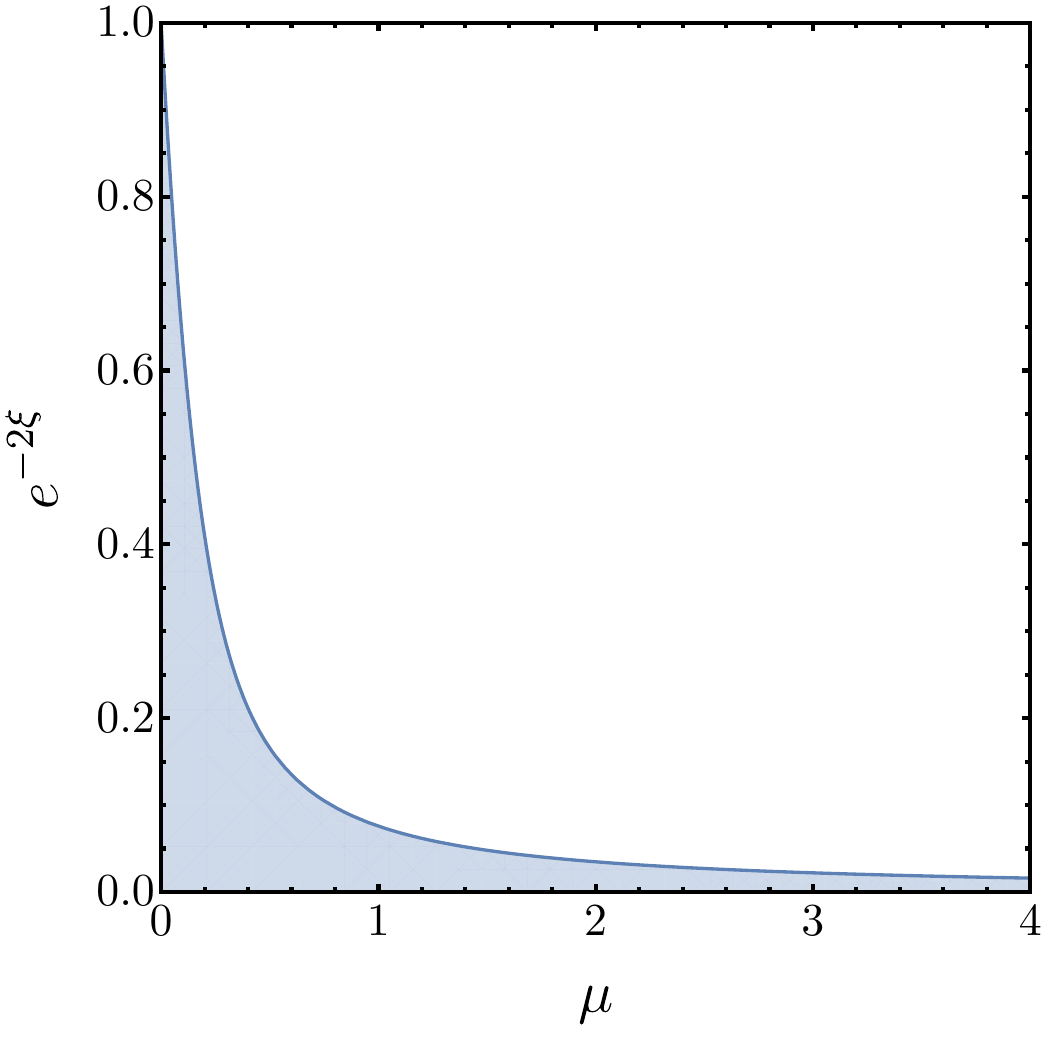}
	\caption{Nonclassicality of squeezed thermal state with average boson number $\mu$ and squeezing parameter $\xi$. The blue region identifies values of the parameters for which the $\calK_{\infty,2}$ criterion identifies the state as nonclassical. Klyshko's $\calK_2$ never identifies nonclassicality in these cases.}
	\label{fig:squeezedThermalStateNonclassicalityP012}
\end{figure}

\parTitle{Definition}
Squeezed thermal states are defined as $S(\xi)\rho^{(\on{th};\mu)}S(\xi)^\dagger$, where
$S(\xi)\equiv \exp[\frac12(\xi a^{\dagger 2} - \xi^* a^2)]$ is the squeezing operator and $\rho^{(\on{th};\mu)}$ is a thermal state.
The $\calK_1$ criterion does not work for these states. At the same time, 
as shown in~\cref{fig:squeezedThermalStateNonclassicalityP012}, 
$\calK_{\infty,2}$ successfully certifies nonclassicality for some values of $\mu$ and $\xi$.

\section{Characterizing convex hulls via tangent planes}
\parTitle{Section outline}
In this section we outline a general method to compute the convex hull of arbitrary compact regions in $\mathbb R^N$, which generalizes the linear functional techniques of~\cite{lachman2016nonclassical,lachman2019criteria,lachman2019faithful}.
We will then showcase the use of this method to recover some of the results presented in the main text. The mathematical background underlying these methods to handle convex Euclidean spaces can be found \textit{e.g.} in~\cite{hug2020lectures,schneider2009convex}.

\subsection{General procedure}
\label{app:sec:characterization_via_tangent_planes}

\parTitle{Convex hulls with tangent planes}
Let $A\subset\RR^n$ be some bounded region. 
The convex hull $\calC_A$ of $A$ is the smallest convex region containing $A$.
Any such $\calC_A$ can be characterized by the set of its tangent planes.
More precisely, given a unit vector $\bs n_\theta\in\RR^n$, the boundary of $\partial\calC_A$ is characterized by the tangent planes $T_\theta$, defined as
\begin{equation}
    T_\theta\equiv 
    \{\bs x\in \RR^n : \langle \bs n_\theta, \bs x\rangle = \max_{\bs a\in A} \langle\bs n_\theta, \bs a\rangle \},
\end{equation}
where $\langle \cdot,\cdot\rangle$ denotes the inner product in the space.
Define for future convenience
\begin{equation}
	F_{\max}(\theta)\equiv F_{\max}(\bs n_\theta)\equiv\max_{\bs a\in A}\langle\bs n_\theta,\bs a\rangle.
\end{equation}
Each such plane $T_\theta$ separates $\RR^n$ into two half-planes. Let us denote with $T_\theta^\le$ the set of $\bs x$ such that
$\langle \bs n_\theta,\bs x\rangle\le F_{\rm max}(\theta)$.
Then 
	$\calC_A=\bigcap_\theta T_\theta^\le.$

\parTitle{How to retrieve convex hull description}
This suggests the following general procedure to recover the convex hull of a region $A$: compute the value of $F_{\max}(\theta)$ for each direction $\theta$ by solving the associated maximisation problem. The region of the boundary of $\calC_A$ tangent to this direction will then equal the convex hull of the elements $\bs a\in A$ such that $F_{\max}(\theta)=\langle \bs n_\theta,\bs a\rangle$. This observation will generally be enough to reconstruct the convex hull of the regions we are interested in.

\subsection{An example}
\label{app:sec:example_convexreconstruction}

Suppose $A\equiv\{(t,0):t\in[0,1]\}\cup\{(0,t):t\in[0,1]\}$, and we are interested in finding the convex hull of this set. Clearly, $A$ is the union of two segments, whose convex hull is trivially the triangle touching the points $(0,0), (1,0), (0,1)$. Let us see how this convex hull is retrieved by means of the general procedure given in~\cref{app:sec:characterization_via_tangent_planes}.
Given an arbitrary unit vector $\bs n_\theta$, we need to find $F_{\max}(\theta)\equiv \max_{\bs a\in A}\langle \bs n_\theta,\bs a\rangle$.
In polar coordinates,
$ \langle \bs n_\theta,\bs a\rangle = \cos\theta a_x + \sin\theta a_y$.
The maximum thus equals
\begin{equation}
    F_{\max}(\theta) =
    \begin{cases}
        \cos\theta, & \theta\in[-\pi/2,\pi/4], \\
        \sin\theta, & \theta\in[\pi/4, \pi], \\
        0, & \theta\in[\pi,3\pi/2].
    \end{cases}
\end{equation}
This tells us that, for $\theta\in(-\pi/2,\pi/4)$, the boundary corresponds to the single point $\bs a=(1,0)$.
For $\theta=\pi/4$, both $\bs a=(1,0)$ and $\bs a=(0,1)$ achieve $F_{\max}(\pi/4)=1/\sqrt2$, and thus the boundary of the region orthogonal to this direction is the set of points in the convex hull of $(1,0)$ and $(0,1)$, that is, $\{(t,1-t): t\in[0,1]\}$.
For $\theta\in(\pi/4,\pi)$ we again have $\bs a=(0,1)$. For $\theta=\pi$ we have $\bs a=(0,1)$ and $\bs a=(0,0)$, corresponding to $\{(0,t): t\in[0,1]\}$. Finally, for $\theta\in(\pi,3\pi/2)$ we have $\bs a=(0,0)$ and for $\theta=3\pi/2=-\pi/2$ we have $\bs a=(0,0)$ and $\bs a=(1,0)$, and thus $\{(t,0): t\in[0,1]\}$.

Overall, this is telling us that the convex hull of $A$ is bounded by the triangle with vertices $(0,0)$, $(1,0)$ and $(0,1)$, as expected from a direct geometric analysis.

\section{2D projections}
\label{sec:2d_projections}

\parTitle{Section outline}
In this section we apply the method discussed in~\cref{app:sec:characterization_via_tangent_planes} to retrieve the classicality regions for two- and three-dimensional slices of the probability space.
\Cref{app:sec:2dslices_naivederivation} focuses on two-dimensional slices of the form $(P_0,P_k)$, and solves the convex hull via direct geometric analysis. The same problem is tackled in~\cref{app:sec:2dspaces_via_tangent_lines}, this time using the procedure given in~\cref{app:sec:characterization_via_tangent_planes}, to illustrate the difference between the two methods.
Finally,~\cref{app:sec:3dslices_from_tangentplanes} applies the same method to find the classicality region in the $(P_0,P_1,P_2)$ space. We thus recover the same results reported in the main text with a completely different approach, that is more methodical and general, but also significantly more convoluted.

\subsection{Direct method for \texorpdfstring{$(P_0,P_k)$}{(P0,Pk)}}
\label{app:sec:2dslices_naivederivation}

\parTitle{Nonclassicality in two-dimensional spaces}
As we discussed in~\cref{subsec:2d_slices}, the characterization of nonclassicality in two-dimensional slices is in many cases complicated by the non-convexity of the associated boundary.
An explicit example of this is $(P_0,P_k)$ with $k>1$, as follows from the non-convexity of $P_0\mapsto P_0 (-\ln P_0)^k$, which has an inflection point at $P_0=e^{1-k}$.

\parTitle{Closing non-convex curves}
In these cases, the convex hull is obtained by ``closing'' the region with a line connecting $(1,0)$ and the point $\bs P_t\equiv (P_{0t},f(P_{0t}))$ such that $f'(P_{0t})=-f(P_{0t})/(1-P_{0t})$, where $f(P_0)\equiv P_0(-\ln P_0)^k/k!$.
More explicitly, the convex hull of the set of coherent states in the $(P_0,P_k)$ space is therefore defined algebraically by the equations:
\begin{equation}
\begin{cases}
    0 \le P_k \le f(P_0), & \text{for } P_0\in[0, P_{0t}], \\
    0 \le P_k \le \frac{P_0 - 1}{P_{0t} - 1} f(P_{0t}), & \text{for } P_0\in[ P_{0t}, 1].
\end{cases}
\label{app:eq:system_describing_P0k}
\end{equation}
The region corresponding to this convex closure can be understood from direct geometric intruition of the corresponding curves, given in~\cref{app:fig:2dslices_P0Pk}.

\subsection{Convex hull in 2D spaces}
\label{app:sec:2dspaces_via_tangent_lines}
\parTitle{Section outline}
In this section we work out the convex hulls of the classical set in the probability subspaces $(P_0,P_1)$ and $(P_0,P_2)$ using the method of~\cref{app:sec:characterization_via_tangent_planes}.

\parTitle{Convex hull in $(P_0,P_1)$}
Define $F_\rho(\theta) = \cos\theta P_0 + \sin\theta P_1$.
Then, for all classical states $\rho$, $F_\rho(\theta)\le F_{\max}(\theta)$,
where
\begin{equation}
    F_{\max}(\theta)\equiv\max_{\lambda\in[0,\infty]} e^{-\lambda}[\cos\theta + \sin\theta \lambda].
\end{equation}
We need to find, for each $\theta$, the value of $F_{\max}(\theta)$.
There are three distinct regions to consider:
\begin{enumerate}
    \item When $\theta\in[-\pi/2,\pi/4]$ the maximum is achieved for $\lambda=0$, and equals $F_{\max}(\theta)=\cos\theta$.
    \item When $\theta\in[\pi/4, \pi]$ the maximum is achieved within the interval, for $\lambda=1-\cot\theta$, and is $F_{\max}(\theta) = e^{-1+\cot\theta}\sin\theta$.
    \item When $\theta\in[\pi,3\pi/2]$, the maximum is achieved for $\lambda=\infty$, with $F_{\max}(\theta)=0$.
\end{enumerate}
The boundary of the convex hull is then obtained by joining the points
\begin{itemize}
	\item $(1,0)$,
	\item $(P_0(\lambda),P_1(\lambda))$ for $\lambda=\lambda(\theta)=1-\cot\theta$ and $\theta\in[\pi/4,\pi]$, which corresponds to $\lambda\in[0,\infty]$,
	\item and $(0,0)$ corresponding to the interval $[\pi,3\pi/2]$.
\end{itemize}
This is consistent with what can be seen in~\cref{app:fig:2dslices_P0Pk}.
Notice how the convexity of the associated curve is reflected into this formalism into the continuity of the first derivative of $\theta\mapsto F_{\max}(\theta)$.

\begin{figure}[tb]
	\centering\hspace{-20pt}
	\includegraphics[width=\linewidth]{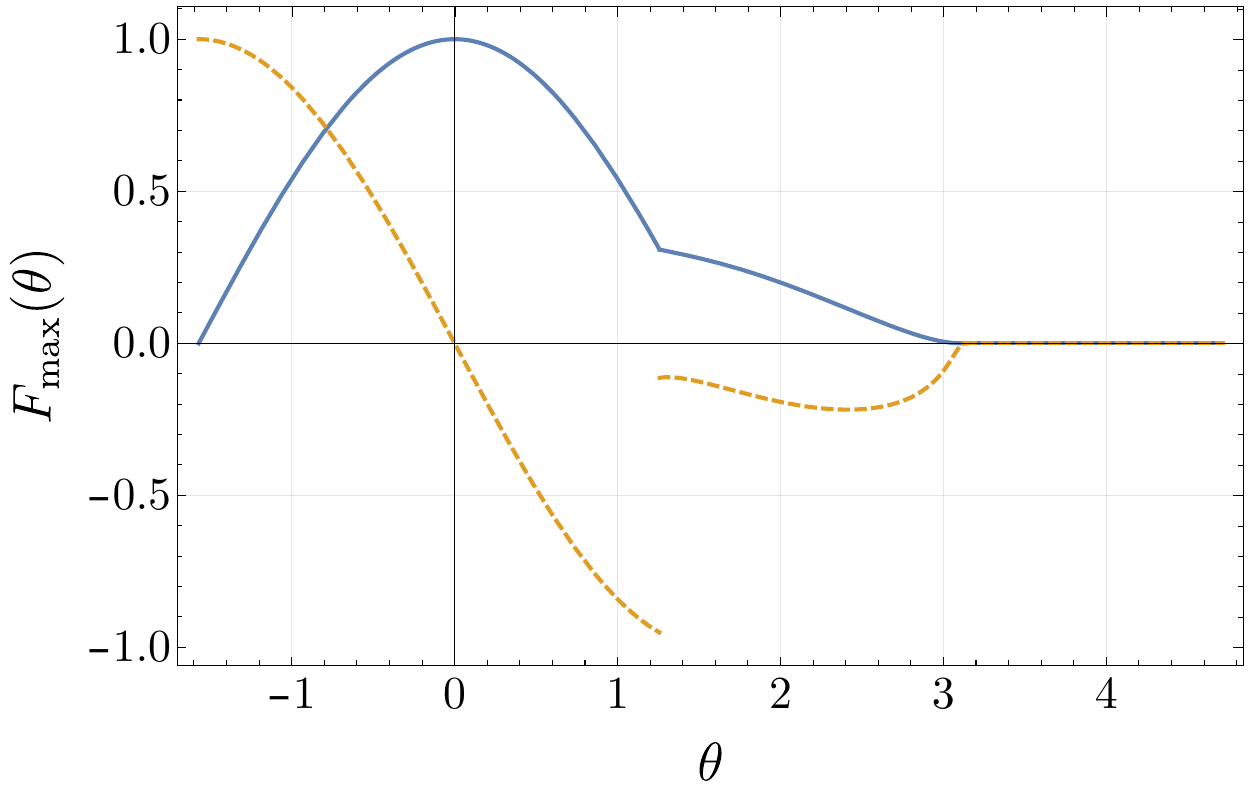}
	\caption{Shape of $F_{\max}(\theta)$ as given in~\cref{eq:Fmax} (continuous blue line) and its derivative (dashed orange line) as a function of $\theta$, computed in the $(P_0,P_2)$ space. The discontinuity in the first derivative corresponds to the non-convexity of the associated generating function.}
	\label{fig:P02_Fmax_curve}
\end{figure}

\parTitle{Convex hull in $(P_0,P_2)$}
In $(P_0,P_2)$ we have
\begin{equation}
    F_{\max}(\theta)\equiv\max_{\lambda\in[0,\infty]} e^{-\lambda}[\cos\theta + \sin\theta \lambda^2/2].
\end{equation}
Define $f_\theta(\lambda)=e^{-\lambda}(c+s \lambda^2/2)$, using the shorthand notation $c\equiv\cos\theta$ and $s\equiv\sin\theta$.
The stationary points are the solutions of
$-c-s\lambda^2/2 + s\lambda=0$
and thus
$\lambda^2 -2\lambda + 2\cot\theta=0$.
This gives $\lambda_\pm = 1\pm \sqrt{1-2\cot\theta}$
for $\cot\theta\le1/2$, that is,
$\theta\in[\arccot(1/2),\pi]\cup[\pi+\arccot(\pi/2),2\pi]$.
Moreover, whereas $\lambda_+$ is always positive (where well-defined), $\lambda_-$ is only positive in the subset
$\theta\in[\arccot(1/2),\pi/2]\cup[\pi+\arccot(1/2),3\pi/2]$.
We thus need to consider these cases separately.
\begin{itemize}
    \item For $\theta\in[0,\arccot(1/2)]$ there are no viable stationary points to check. The values at the extreme of the $\lambda$ intervals are $0$ for $\lambda\to\infty$ and $\cos\theta$ for $\lambda=0$.
        Because $\cos\theta\ge0$ for the angles considered, the maximum is achieved at the point of contact $\lambda=0$ (which is the $(1,0)$ in the considered probability space), and $F_{\max}(\theta)=\cos\theta$.
    \item When $\theta\in[\arccot(1/2),\pi/2]$, it can be checked that $\lambda_+\ge\lambda_-\ge0$. We therefore only need to test whether $\lambda_+$ or $\lambda=0$ give the biggest value for $F_\rho(\theta)$.
        Studying the functions, we find that there are two cases.
        Defining $\theta_0$ as the unique solution in the interval of the equation
        \begin{equation}
            e^{-1-\sqrt{1-2\cot\theta}}(1+\sqrt{1-2\cot\theta})=\cot\theta,
            \label{app:eq:equation_for_theta0}
        \end{equation}
        for $\theta\le\theta_0$ the maximum still corresponds to $\lambda=0$, but for $\theta\ge\theta_0$ it is achieved by $\lambda=\lambda_+$. Numerically, we find that $\theta_0\sim 1.26$. Note that this $\theta_0$ corresponds to the line tangent to the classical set at the point of contact between the dashed and the continuous orange line in~\cref{app:fig:2dslices_P0Pk}.
        In summary, we found that
        \begin{equation}
        \begin{cases}
            F_{\max}(\theta) = \cos\theta, &\theta<\theta_0, \\
            F_{\max}(\theta) = f_\theta(\lambda_+), &\theta>\theta_0.
        \end{cases}
        \end{equation}
    \item For $\theta\in[\pi/2,\pi]$ the optimal solution remains the one corresponding to $\lambda=\lambda_+$, and thus $F_{\max}(\theta)=f_\theta(\lambda_+)$.
    \item For $\theta\in[\pi,\pi+\arccot(1/2)]$ there are no local stationary points and $\cos\theta<0$, and therefore the maximum is achieved by $\lambda=\infty$, and $F_{\max}(\theta)=0$.
    \item For $\theta\in[\pi+\arccot(1/2),3\pi/2]$ there are again the stationary points $\lambda=\lambda_\pm$ to consider, but these both satisfy $\lambda_\pm<0$ in this interval, and so are not viable solutions. We also have $\cos\theta<0$ in this interval, and therefore the maximum remains the $F_{\max}(\theta)=0$ corresponding to $\lambda=\infty$.
    \item Finally, for $\theta\in[3\pi/2,2\pi]$ we the stationary point $\lambda_-$ is negative and therefore non-physical, while $\lambda_+$ gives $f_\theta(\lambda_+)<0$. Because in this region $\cos\theta\ge0$, we can conclude that the maximum is achieved by $\lambda=0$ and $F_{\max}(\theta)=\cos\theta$.
\end{itemize}
We thus showed that
\begin{equation}
    F_{\max}(\theta) = \begin{cases}
        \cos\theta, & -\pi/2 \le \theta \le \theta_0, \\
        f_\theta(\lambda_+(\theta)), & \theta_0\le \theta \le \pi, \\
        0, & \pi \le \theta \le 3\pi /2
    \end{cases},
    \label{eq:Fmax}
\end{equation}
where $\theta_0$ is defined as the solution of~\cref{app:eq:equation_for_theta0},
$\lambda_+(\theta)\equiv 1+\sqrt{1-2\cot\theta}$, and
$f_\theta(\lambda)\equiv e^{-\lambda}(\cos\theta+\sin\theta \lambda^2/2)$.
This can be again be seen to be consistent with~\cref{app:eq:system_describing_P0k}.
The shape of $F_{\max}(\theta)$ together with its first derivative is given in~\cref{fig:P02_Fmax_curve}.

Completely analogous reasoning can be used to derive the convex hull for other two-dimensional subspaces.

\subsection{Recovering Klyshko's inequalities in \texorpdfstring{$(P_0,P_1,P_2)$}{(P0,P1,P2)}}
\label{app:sec:3dslices_from_tangentplanes}

In $(P_0,P_1,P_2)$ we have, for every $\bshatn\equiv (a,b,c)$ with $a^2+b^2+c^2=1$,
\begin{equation}
    F_{\max}(\bshatn) = \max_{\lambda\in[0,\infty]} e^{-\lambda} (a + b \lambda + c \lambda^2/2).
    \label{app:eq:P012_fmax_definition}
\end{equation}
Define the function $f(\lambda,\bshatn)\equiv e^{-\lambda}(a+b \lambda+c\lambda^2/2)$, so that $F_{\max}(\bshatn)=\max_{\lambda\ge0} f(\lambda,\bshatn)$.
The stationary points are the solutions of
\begin{equation}
    -c/2 \lambda^2 + (c-b)\lambda + (b-a) = 0,
    \label{app:eq:quadratic_eq_when_recovering_P012_condition}
\end{equation}
which gives
\begin{equation}
    \lambda_\pm = \frac{1}{c} \left[ (c-b) \pm \sqrt{c^2 + b^2 - 2ac} \right].
    \label{eq:app:P012_lambdapm_def}
\end{equation}
From the second derivative we find that, when both solutions exist, only $\lambda_+$ is a local maximum (that is, corresponds to a negative second derivative), and therefore of interest to us.
\Cref{app:eq:quadratic_eq_when_recovering_P012_condition} has no real solution if and only if
\begin{equation}
    c^2 + b^2 - 2ac < 0.
\end{equation}
Using the spherical coordinates
\begin{equation}
    a = \cos\theta, \quad
    b = \sin\theta\cos\varphi, \quad
    c = \sin\theta\sin\varphi,
\end{equation}
this condition reads $\sin\theta(\sin\theta-2\cos\theta\sin\varphi)<0$.
For these angles, the maximum $F_{\max}$ is achieved by either $\lambda=0$ or $\lambda=\infty$, corresponding to $F_{\max}=a$ and $F_{\max}=0$, respectively (as there are no possible stationary points in these directions). The former case corresponds to points on the surface whose tangent plane touches the vacuum state (that is, the point $(1,0,0)$ in the reduced probability space), whereas the latter corresponds to point with tangent plane passing throgh the origin.
We should therefore expect the associated sets of points to match with the surfaces $\calK_\infty$ and $\calK_1$, respectively.

These solutions do not, however, give useful information about the nontrivial sectors of the classicality boundary, as they all correspond to flat sections of it.
This means that one has to consider directions for which~\cref{app:eq:P012_fmax_definition} also admits stationary points. This is, however, a rather lengthy procedure: for a given direction $\bshatn$ one has to check the value of $F_{\max}$ corresponding to $\lambda=\lambda_\pm$ and $\lambda=0,\infty$, finding which one of these corresponds to the largest value \emph{and} is positive (which is not ensured for $\lambda_\pm$).
We can simplify this procedure by taking a hint from the results about the $(P_0,P_1,P_2)$ space obtained in the main text.
Indeed, we know that the two sections of the classicality boundary are ``partially flat'' surfaces, that is, surfaces obtained by connecting a fixed point (either $(0,0,0)$ or $(1,0,0)$) with the points corresponding to coherent states. In terms of the linear functional approach considered here, this means that the directions corresponding to these surfaces must be such that~\cref{app:eq:P012_fmax_definition} has a maximum corresponding to \emph{two} values of $\lambda$: some $\lambda\in(0,\infty)$, and then either $\lambda=0$ or $\lambda=\infty$. By focusing on these directions we can avoid considering the other cases.

Consider then in particular the section of the boundary corresponding to $P_1^2=2P_0P_2$.
The points on this boundary are convex combinations of $(0,0,0)$ with coherent states, that is, points of the form $(e^{-\lambda}, e^{-\lambda}\lambda, e^{-\lambda}\lambda^2/2)$.
Consider the directions for which~\cref{app:eq:P012_fmax_definition} has a maximum for $\lambda=\infty$ (that is, for which $F_{\max}=0$) that is \emph{also} achieved as a local maximum for some finite $0<\lambda<\infty$. This amounts to the condition
\begin{equation}
    a+  b \lambda_+ + c \lambda_+^2/2 = 0,
\end{equation}
with $\lambda_+$ given by~\cref{eq:app:P012_lambdapm_def}.
Explicitly, this is equivalent to $c+\sqrt{b^2+c^2-2ac}=0$, which using spherical coordinates then translates into $\theta\in[\pi/2,\pi]$ and $\varphi$ being related to $\theta$ via one of the following two conditions:
\begin{equation}
    \varphi = 2\pi - \arcsin\cot(\theta/2), \,\,
    \varphi =  \pi + \arcsin\cot(\theta/2).
\end{equation}
The second one corresponds to $\lambda_+<0$, and is thus not physical.
We conclude that the angles $\theta\in[\pi/2,\pi]$ and $\varphi=\varphi(\theta)=2\pi-\arcsin\cot(\theta/2)$ correspond to points of the boundary for which the tangent plane touches both $(0,0,0)$ and the coherent state $\lambda_+(\theta,\varphi)$.
The associated convex hull must then be given, for these angles, by the set of convex combinations $p\ketbra0+(1-p)\ketbra\lambda$, for all coherent states with
$\lambda\in\{\lambda_+(\theta,\varphi): \theta\in[\pi/2,\pi], \, \varphi=\varphi(\theta)\}$. Moreover, $\lambda_+(\theta,\varphi)=\sqrt{-\cos\theta}\sec(\theta/2)$, and therefore this set equals the full interval $[0,\infty]$.
The section of the classicality boundary we just found corresponds to~\cref{eq:P012_criterion_K1}, thus giving further confirmation of the results in the main text.

\begin{figure}[tb]
	\centering
    \hspace{-20pt}\begin{tikzpicture}
		\node[anchor=south west] (A) at (0, 0)%
			{\includegraphics[width=\linewidth]{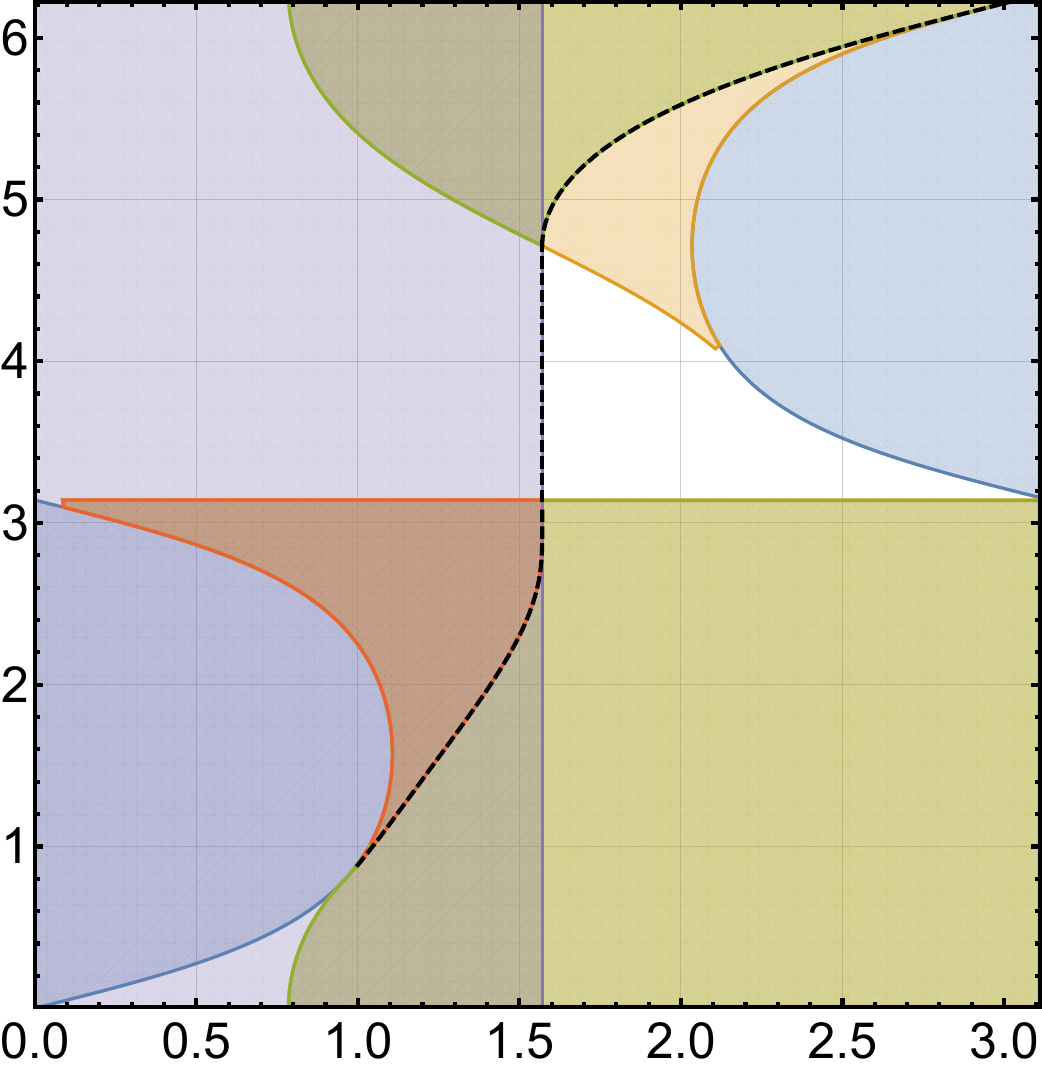}};
		\node[above] at (4.5, -.3) {$\theta$};
		\node[above] at (-.2, 4.4) {$\varphi$};
		\node at (2.4, 9.2) {$\overbrace{\hspace{116pt}}^{\displaystyle \,\,f(0)\ge f(\infty)}$};
		\node at (1.7, 2.6) {no $\lambda_\pm$};
		\node at (7.4, 6.7) {no $\lambda_\pm$};
		\node at (4.4, 5.5) {\Large$\lambda_+ < 0$};
		\node at (3.5, 4.2) {\small$f(0)\ge f(\lambda_+)$};
		\node at (3.6, 3.8) {\small$\lambda_+\ge0$};
		\node at (6.5, 2.7) {\large$f(\lambda_+)\ge0$};
		\node at (6.5, 2.1) {\large$\lambda_+\ge0$};
		\node at (4.2, 8.2) {$f(\lambda_+)\ge0$};
		\node at (4.39, 7.8) {$\lambda_+\ge0$};
		\node at (5.16, 6.9) {\small$\lambda_+\ge0$};
	\end{tikzpicture}
	\caption{%
        Summary of behaviour of $f_\bshatn(\lambda)\equiv f(\lambda,\bshatn)$ for different values of $\theta\in[0,\pi]$ and $\varphi\in[0,2\pi]$. Each coloured region represents the angles corresponding to a certain relation between $f_{\theta,\varphi}(0), f_{\theta,\varphi}(\lambda_+)$, and $f_{\theta,\varphi}(\infty)$. The slightly darker left half of the plot (where $\theta\le\pi/2$) corresponds to $f_{\theta,\varphi}(0)\ge f_{\theta,\varphi}(\infty)=0$. In the blue regions on the upper right and lower left $f_{\theta,\varphi}$ does not have local stationary points. The white regions (including the slightly darker white region in the left half-plane) corresponds to $\lambda_+$ existing but being negative (which means that it cannot possibly correspond to a maximum). Green, orange, and red regions all correspond to $\lambda_+\ge0$. Green and red regions correspond to $f_{\theta,\varphi}(\lambda_+)\ge0$. The red region corresponds to $f_{\theta,\varphi}(0)\ge f_{\theta,\varphi}(\lambda_+)$. The black dashed line corresponds to the points for which the maximum $F_{\max}$ is achieved by two distinct values of $\lambda$. These are thus the angles corresponding to the boundary of the classicality region.
    }
    \label{fig:P012_linearfunctional_landscape}
\end{figure}
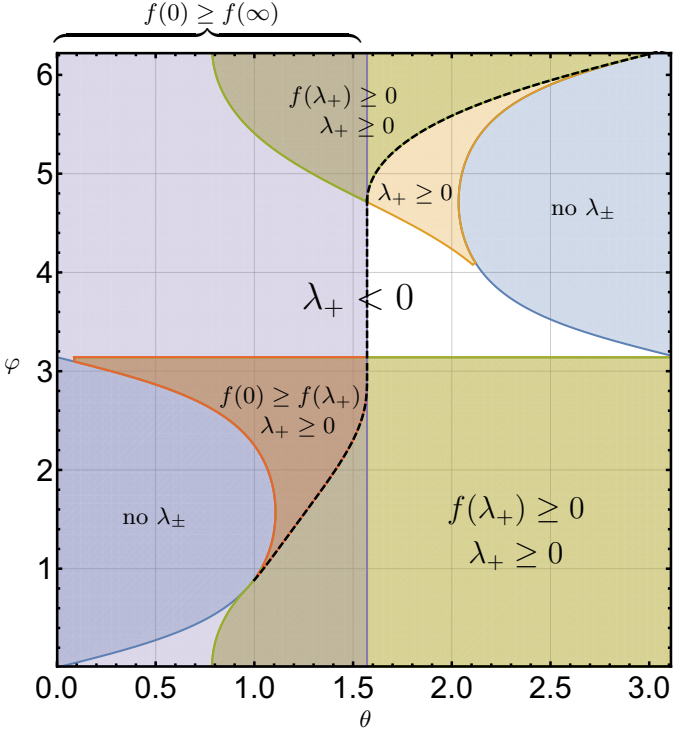

For the boundary corresponding to~\cref{eq:P012_criterion_Kinf}, we instead look for some $\lambda$ such that $F_{\max}$ is achieved both by a stationary point \emph{and} $\lambda=\infty$. This amounts to the condition
\begin{equation}
    e^{-\lambda_+}(a + b \lambda_+ + c \lambda_+^2/2) = a.
\end{equation}
In this case, however, there seems to be no nice algebraic relation between $\theta$ and $\varphi$.
To get a better idea of the overall behaviour we then resort to resolving numerically the $\lambda$s for which $f_{\bshatn}(\lambda)=F_{\max}(\bshatn)$ for the different directions $\bshatn\equiv(\theta,\varphi)$. The result of this is shown in~\cref{fig:P012_linearfunctional_landscape}. In particular, the directions $(\theta,\varphi)$ corresponding to the black dashed line, which are the values for which $F_{\max}$ is achieved by two distinct values of $\lambda$ at the same time, are the ones that give the most information about the structure of the boundary.
The black dashed curve between green and orange regions in the upper right of the figure corresponds to the directions for which $f_\bshatn(\lambda_+)=f_\bshatn(\infty)=0$. In each such direction, the boundary is therefore given by the segment joining $\lambda_+$ and the origin. This corresponds to the part of the boundary covered by Klyshko's criterion, as per~\cref{eq:P012_criterion_K1}.
The black dashed curve between red and green regions in the lower half of the figure corresponds to directions for which $f_\bshatn(\lambda_+)=f_\bshatn(0)$. Each such direction corresponds to a piece of boundary that is a segment joining $\lambda_+$ and $\lambda=0$ (that is, the point $(1,0,0)$). These therefore draw the part of the boundary corresponding to the $\calK_\infty$ criterion:~\cref{eq:P012_criterion_Kinf}.
Finally, the black dashed line between white and darker white regions, which joins the other two black dashed curves, corresponds to directions for which $f_\bshatn(0)=f_\bshatn(\infty)>f_\bshatn(\lambda_+)$. All of these directions thus correspond to the segment joining the origin with $(1,0,0)$.

The above provides us with a complete characterization of the boundary of the classical region in this reduced space: the boundary of the classical set is drawn by segments joining the various coherent states with either the origin or $(1,0,0)$, consistently with the results reported in the main text.

\section{Comparison with Wigner and \texorpdfstring{$P$}{P}-function criteria}

In this section we will present an analysis of the classes of states discussed in the main text using the standard methods to certify nonclassicality via the non-positivity of Wigner and $P$-functions.
While the results of such methods are not comparable with those obtained with the techniques we discussed in the main text, in that they rely on completely different assumptions on what information is needed to apply the methods, they can be of interest to compare how restricting the available information can change the capacity of predicting the nonclassicality of a given state.

\subsection{Wigner of Fock and coherent states}

In this section we will report the standard expressions for the Wigner functions of Fock and coherent states, for future reference in the later sections.
The Wigner function of a Fock state $\ket k$ is
\begin{equation}
\begin{aligned}
    W_k(x,p) &= \frac{(-1)^k}{\pi} e^{-x^2-p^2} L_k(2(x^2+p^2)) \\ 
             &= \frac{(-1)^k}{\pi} e^{-|z|^2} L_k(2|z|^2),
\end{aligned}
\end{equation}
where $z\equiv x+ip$ and $L_k$ are the Laguerre polynomials.

If $\ket\alpha$, $\alpha\in\CC$, is a coherent state, its Wigner function reads
\begin{equation}
    W_\alpha(z) = \frac{1}{\pi} e^{-|z-\sqrt2\alpha|^2}.
\end{equation}
If instead of a coherent state we consider the corresponding fully depolarized state, \textit{i.e.} what we referred to as \textit{Poissonian noise state}, we get a state $\rho_\mu$ of the form
\begin{equation}
    \rho_\mu = \sum_{k=0}^\infty \frac{e^{-\mu}\mu^k}{k!},
\end{equation}
whose Wigner is
\begin{equation}
    W_\mu(z) = \sum_{k=0}^\infty \frac{e^{-\mu}\mu^k}{k!}
           \frac{(-1)^k}{\pi} e^{-|z|^2} L_k(2|z|^2).
\end{equation}

\subsection{Wigner of boson-added Poissonian noise states}
\label{subsec:wigner_bosonaddedPoissonianNoise}
\begin{figure}
    \centering
    \hspace{-20pt}\begin{tikzpicture}
        \node at (0, 0) {\includegraphics[width=\linewidth]{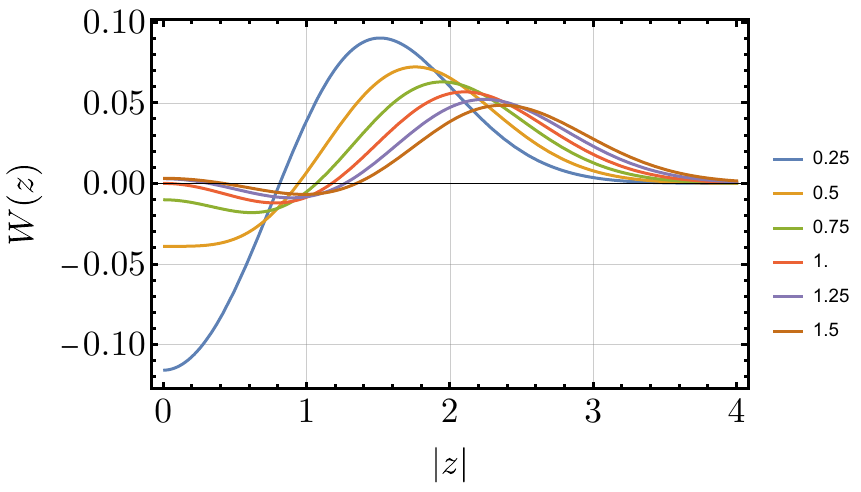}};
        \node[overlay] at (3.65, 1.36) {$\mu$};
    \end{tikzpicture}
    \caption{Each plot shows the value of the Wigner function $W_{\mu;1}$ corresponding to boson-added Poissonian noise states. Different colors correspond to different values of $\mu$. The Wigner function is found to be negative for all tested values of $\mu$, albeit the negativity becomes harder to detect for larger $\mu$.fi}
    \label{fig:wignerDecoheredBosonAddedCoherent}
\end{figure}

Consider now \textit{boson-added} coherent states. To compare with the results in Fig. 2 in the main text, we consider the case of a single boson added.
A \textit{boson-added coherent state} has the form
\begin{equation}
    \frac{1}{\sqrt{1+|\alpha|^2}}a^\dagger\ket\alpha,
\end{equation}
and its Wigner, as shown \textit{e.g.} in~\cite{zavatta2005singlephoton}, is
\begin{equation} 
    W_{\alpha,1}(z) = \frac{2 |z-\alpha/\sqrt2|^2-1}{\pi(1+|\alpha|^2)} e^{-|z-\sqrt2\alpha|^2}.
\end{equation}
Let $\rho_\mu$ denote the states obtained by dephasing a coherent state with average boson number $\mu$. The corresponding boson-added Poissonian noise state is then
\begin{equation}
    \rho_{\mu;1}\equiv (a^\dagger \rho_\mu a)/(1+\mu).
\end{equation}
These are diagonal in the Fock basis, so that the corresponding Wigner reads
\begin{equation}
    W_{\mu;1}(z) = \sum_{k=0}^\infty P_k^{(\mu,1)} W_k(z),
\end{equation}
where $P_k^{(\mu,1)}$ are the Fock probabilities of the Boson added noise, which are
\begin{equation}
    P_k^{(\mu,1)} = \frac{1}{\mu+1} \frac{e^{-\mu}\mu^k}{k!} \frac{k^2}{\mu}.
\end{equation}
As shown numerically in~\cref{fig:wignerDecoheredBosonAddedCoherent}, we find that the Wigner function of boson-added Poissonian noise states takes negative values for all values of $\mu$.

\begin{figure}
    \centering
    \hspace{-20pt}\begin{tikzpicture}
        \node at (0, 0) {\includegraphics[width=\linewidth]{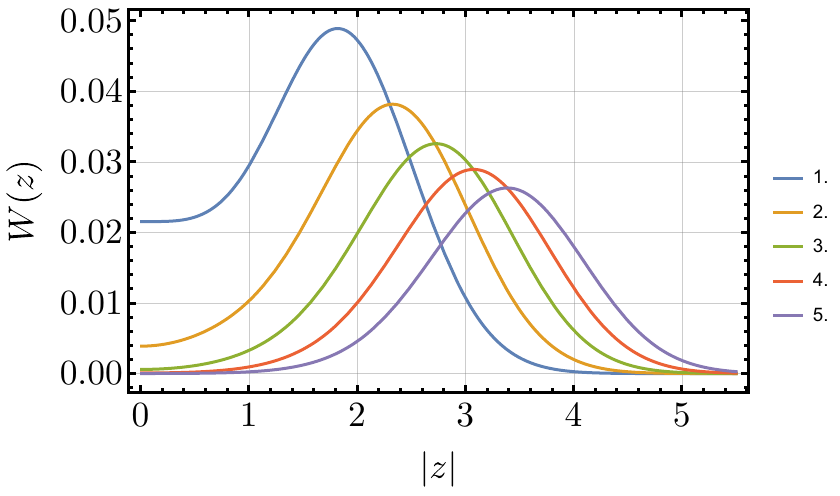}};
        \node[overlay] at (3.75, 1.2) {$\mu$};
    \end{tikzpicture}
    \caption{As~\cref{fig:wignerDecoheredBosonAddedCoherent} but now we consider boson-added Poissonian noise states, mixed with the corresponding non-boson-added coherent noise state. The mixing ratio is here chosen to be $p=1/2$.}
    \label{fig:wignerNegativityDecoherentBosonAddedMixedNoise}
\end{figure}

\begin{figure}
    \centering
    \hspace{-10pt}\begin{tikzpicture}
        \node at (0, 0) {\includegraphics[width=\linewidth]{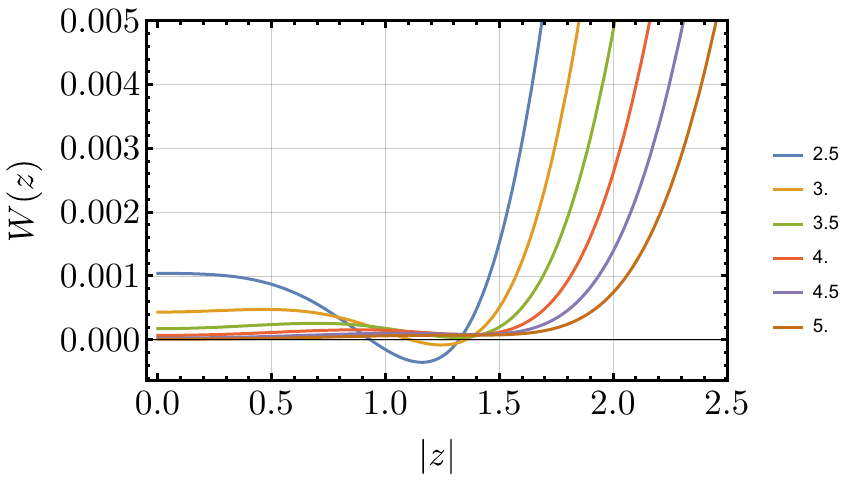}};
        \node[overlay] at (3.7, 1.4) {$\mu$};
    \end{tikzpicture}
    \caption{Same as~\cref{fig:wignerDecoheredBosonAddedCoherent}, but considering a mixing ratio of $p=0.9$, so that the state is a boson-added Poissonian noise state with larger probability.}
    \label{fig:wignerNegativityDecoherentBosonAddedMixedNoisep0.9}
\end{figure}

To consider the same case studied in the main text, we consider now states of the form $(\rho_{\mu;1}+\rho_\mu)/2$. As shown in~\cref{fig:wignerNegativityDecoherentBosonAddedMixedNoise}, we find that the Wigner function stays positive for all $\mu$.
Changing the mixing ratios affects the nonclassicality detected by the Wigner function. Considering mixtures of the form
$p\rho_{\mu;1}+(1-p)\rho_\mu$, the Wigner starts being nonpositive at around $p\simeq0.9$, as shown in~\cref{fig:wignerNegativityDecoherentBosonAddedMixedNoisep0.9}.

\subsection{Thermal states}
\begin{figure}
    \centering
    \hspace{-15pt}\begin{tikzpicture}
        \node at (0, 0) {\includegraphics[width=\linewidth]{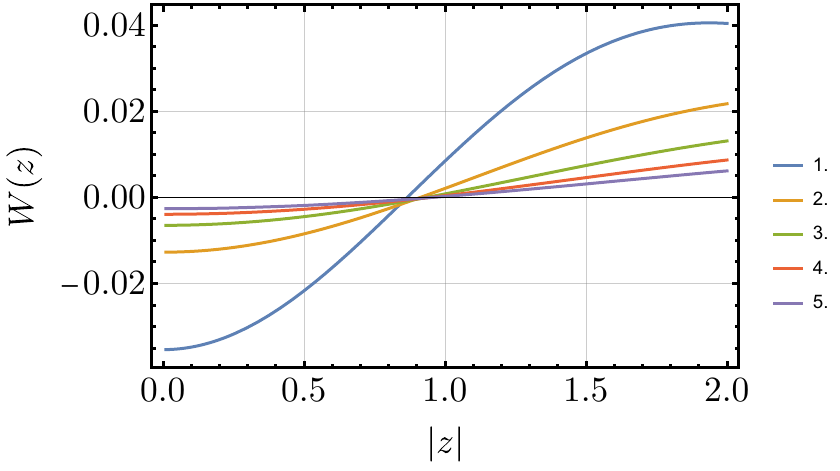}};
        \node[overlay] at (3.82, 1.25) {$\mu$};
    \end{tikzpicture}
    \caption{Wigner function of boson-added thermal states. The non-positivity again certifies nonclassicality of these states.}
    \label{fig:wignerBosonAddedThermal}
\end{figure}

The Wigner function of a thermal state with average boson number $\mu$ is
\begin{equation}
    W^{\rm th}_{\text{th}}(z) = \frac{1}{\pi(1+2\mu)}e^{-|z|^2/(1+2\mu)},
\end{equation}
where $z\equiv x+ip$.
Boson-added thermal states have the form
\begin{equation}
    N\, a^\dagger \rho_{\on{th}}a =
    \sum_{k=0}^\infty \frac{k\mu^{k-1}}{(1+\mu)^{k+1}} \PP_k
\end{equation}
where $N$ is a normalisation constant. The corresponding Wigner function is then
\begin{equation}
\begin{aligned}
    W^{\rm th}_{\on{th};1}(z) &=
    \sum_{k=0}^\infty 
    \frac{k\mu^{k-1}}{(1+\mu)^{k+1}} W_k(z) \\
    &= \sum_{k=0}^\infty 
        \frac{k\mu^{k-1}}{(1+\mu)^{k+1}}
        \frac{(-1)^k}{\pi} e^{-|z|^2} L_k(2|z|^2).
\end{aligned}
\end{equation}
As shown in~\cref{fig:wignerBosonAddedThermal}, the corresponding Wigner function is negative for all $\mu$. For comparison, as was shown in the previous sections, Klyshko's criteria are also capable of detecting the nonclassicality of these states from their first three Fock probabilities.

\subsection{Noisy Fock states}
\textit{Noisy Fock states} have states of the form $D(\alpha)\ket m$, averaged over the phase of $\alpha$. Let us denote these states with $\rho^{(m,\mu)}$, where $\mu\equiv|\alpha|^2$.
To obtain the corresponding expressions, we start from the matrix components of $D(\alpha)$ in the Fock basis:
\begin{equation}
\begin{gathered}
	\mel{k}{D(\alpha)}{n} = \frac{1}{\sqrt{n!k!}} e^{-\mu/2}
	\alpha^{k-n} c_{k,n}, \\
	c_{k,n}\equiv \sum_{i=0}^n \frac{(-\mu)^i}{i!} \frac{k!n!}{(n-i)!(k-n+i)!}.
\end{gathered}
\end{equation}

\begin{figure}
    \centering
    \hspace{-15pt}\begin{tikzpicture}
        \node at (0, 0) {\includegraphics[width=\linewidth]{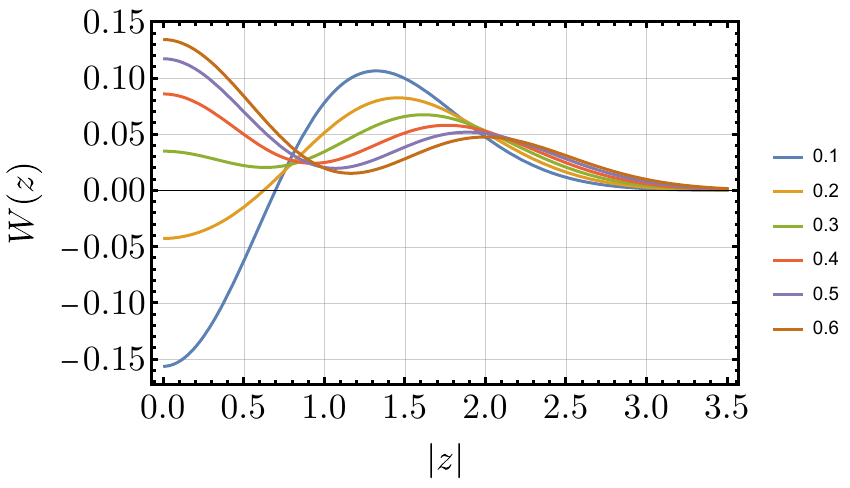}};
        \node[overlay] at (3.75, 1.3) {$\mu$};
    \end{tikzpicture}
    \caption{Wigner function of noisy Fock states $\rho^{(m,\mu)}$ with $m=1$, for different values of the average boson number $\mu$. As expected, the Wigner function is non-positive for small $\mu$, but the nonclassicality becomes undetectable for larger $\mu$s.}
    \label{fig:wignerNoisyFockStates}
\end{figure}

As discussed previously, for $n=1$, these correspond to the Fock probabilities
\begin{equation}
    P_k^{(1;\mu)} = \frac{e^{-\mu}\mu^k}{k!} \frac{(k-\mu)^2}{\mu}.
\end{equation}
As show in~\cref{fig:wignerNoisyFockStates}, the corresponding Wigner functions are non-negative, and thus cannot be used to certify nonclassicality.

\begin{figure}
    \centering
    \hspace{-15pt}\begin{tikzpicture}
        \node at (0, 0) {\includegraphics[width=\linewidth]{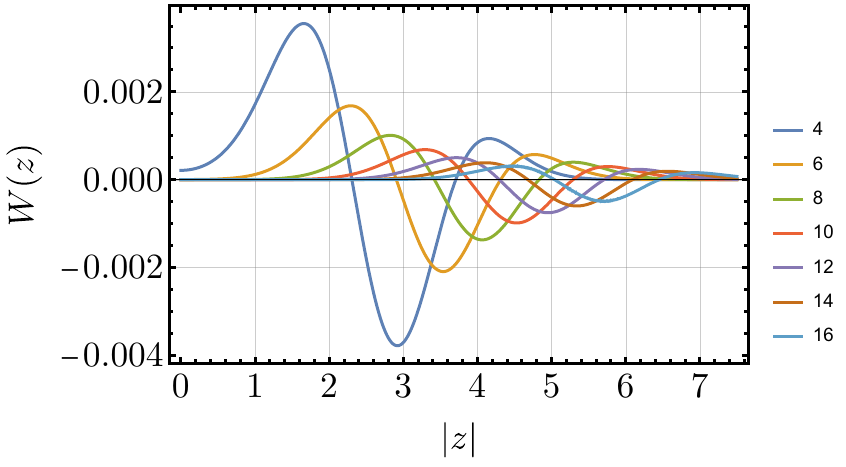}};
        \node[overlay] at (3.75, 1.44) {$\mu$};
    \end{tikzpicture}
    \caption{Wigner function of noisy Fock states $\rho^{(m,\mu)}$ with $m=2$, for different values of the average boson number $\mu$. The Wigner function is observed to remain non-negative for all tested values of $\mu$.}
    \label{fig:wignerNoisyFock2States}
\end{figure}

For $n=2$, we get
\begin{equation}
	P_k^{(2;\mu)} =
	\frac{e^{-\mu}\mu^k}{k!} \frac{(k(k-1)-2k\mu + \mu^2)^2}{2\mu^2}.
\end{equation}
In this case, as shown in~\cref{fig:wignerNoisyFock2States}, the Wigner function is non-positive, thus certifying nonclassicality.

\subsection{\texorpdfstring{$P$}{P}-function}
Another way to characterize a state is via its $P$-function. This, however, can be highly singular, and thus difficult to manipulate.
To explore the $P$-function of highly nonclassical states such as Fock states, we thus consider the limit of the $P$-functions of boson-added thermal states $\rho^{\rm th}_{\bar n;k}$ in the limit of the average boson number going to zero. More precisely, we consider the states
\begin{equation}
    \rho^{\rm th}_{\bar n;k} \propto \left(a^{\dagger}\right)^k \rho^{\on{th}}_{\bar n} a^k,
    \label{BEFock}
\end{equation}
where $\bar{n}$ is the average number of photons and $\rho^{\on{th}}_{\bar n}$ the corresponding thermal state. Note that $\rho^{\on{th}}_{\bar n}\to \ketbra 0$ for $\bar n\to 0$, and thus also
$\rho^{\rm th}_{\bar n;k}\to\ketbra k$ for $\bar{n}\rightarrow 0$.
We can thus study the $P$-function of $\ketbra k$ via that of $\rho^{\rm th}_{\bar n;k}$ in the limit of $\bar n\to 0$, effectively regularizing the singular $P$-functions.
The $P$-function of the thermal state $\rho_{\bar n}^{\on{th}}$ reads
\begin{equation}
    \pi P^{\on{th}}_{\bar n}(\alpha) = \frac{1}{\bar n} e^{-|\alpha|^2/\bar n}.
\end{equation}
The $P$-function of $\rho^{\rm th}_{\bar n;k}$ is then
\begin{equation}
    \pi P_{\bar{n};k}^{\rm th}(\alpha) =
    N_k \frac{1}{\bar n}
    e^{\vert \alpha \vert} \frac{\partial^{2k}}{\partial^k_{\alpha}\partial^k_{\alpha^*}}e^{-\vert \alpha \vert^2 (\frac{1+\bar{n}}{\bar{n}})},
    \label{intPhase}
\end{equation}
where
\begin{equation}
    N_k = \left[\Tr\left(a^{\dagger k}x^N a^k\right)\right]^{-1},
    \quad x\equiv \frac{\bar n}{\bar n+1}.
\end{equation}
For example, the $P$-function of $\rho^{\rm th}_{\bar n;1}$ is
\begin{equation}
    P^{\rm th}_{\bar{n};1}(\alpha) =
    \frac{1}{\bar n^3}
    (\lvert\alpha\rvert^2(1+\bar n)-\bar n)
    e^{-\frac{\vert \alpha \vert^2}{\bar{n}}},
\end{equation}
which is manifestly negative for $\lvert\alpha\rvert^2< \bar n/(\bar n+1)$.
This implies that the $P$ function is always negative for $\alpha=0$ when $\bar{n}>0$, consistently with the fact that the $P$-function loses its regularity in the limit $\bar n\to0$. This again certifies the nonclassicality of the Fock state $\ket 1$.

To verify the negativity of \textit{noisy} Fock states, obtained applying the displacement operator $D(\sqrt\mu e^{i\phi})$ on a Fock state and then averaging over $\phi$, we notice that the $P$-function becomes
\begin{equation}
    P^{\rm th}_{\bar n;1;\mu}(\alpha) =
    \int_{0}^{2\pi} \frac{d\phi}{2\pi}
    P_{\bar{n};1}(\alpha-\sqrt\mu e^{i\phi}).
    \label{PDisplacement}
\end{equation}
This is again negative for small $\bar n$, as expected.
To compare again with the results in Fig. 2 in the main text, as done in~\cref{subsec:wigner_bosonaddedPoissonianNoise}, we now consider mixtures of the form
\begin{equation}
    p \rho^{\rm th}_{\bar{n};1;\mu} +
    (1-p) \rho^{\rm th}_{\bar{n};0;\mu}.
    \label{P:noisySt}
\end{equation}
These are characterized by the $P$-function $P_{\bar{n},p,\mu}=p P^{\rm th}_{\bar{n};1;\mu}+(1-p)P^{\rm th}_{\bar{n};0;\mu}$, where $P^{\rm th}_{\bar{n};0;\mu}$ is the $P$-function of a thermal state displaced with a random phase.
To observe the negativity for every $p \in (0, 1)$, we need to choose sufficiently small $\bar{n}$, since the $P$-function always drops at $\vert \alpha \vert^2=\mu$ without any saturation for decreasing $\bar{n}$.
However, very small values of $\bar{n}$ cause failure in numerical evaluation of the integral (\ref{intPhase}). This limits the numerics in proving the negativity for large $\mu$ and small $p$.
We verified that $P_{\bar{n},p,\mu}$ is negative for $p=0.05$ and $\mu$ up to $19$. However, if $p=0.01$ the negativity could only be verified for $\mu$ up to $1.3$.
Predictably, we find that the $P$-function is able to capture nonclassicality better than the Wigner function, which instead was found to certify nonclassicality via non-positivity only for $p\ge0.9$.

\parTitle{Boson-added Poissonian noise states}
Consider now the boson-added state $\frac{1}{1+\mu}a^{\dagger}\rho_{\mu} a$, with $\rho_{\mu}$ the Poissonian noise state with average boson number $\mu$. Using the commutation relations between creation and displacement operators we get
\begin{equation}\scalebox{0.92}{$\displaystyle
    a^{\dagger}\rho_{\mu} a\propto
    \int_{0}^{2\pi} d\phi
    D(\sqrt{\mu}\exp(i \phi))\vert q \rangle \langle q \vert D^{\dagger}(\sqrt{\mu}\exp(i \phi)),
$}\end{equation}
where $\vert q \rangle \propto \vert 1 \rangle-\sqrt{\mu}\exp(-i \phi)\vert 0 \rangle$.
Using again thermal states to approximate the Fock states in the $\bar n\to0$ limit, we get
\begin{equation}
    \begin{aligned}
        P_{\bar n;\mu} &\propto e^{\vert \alpha \vert^2} \frac{\partial^{2}}{\partial_{\alpha}\partial_{\alpha^*}}e^{-\vert \alpha \vert^2 \frac{1+\bar{n}}{\bar{n}}}+\mu e^{-\frac{\vert\alpha \vert^2}{\bar{n}}}\\
        &-e^{\vert\alpha \vert^2\frac{\bar{n}-1}{2\bar{n}}}\sqrt{\mu}\left(e^{-i\phi}\frac{\partial}{\partial \alpha}+e^{i\phi}\frac{\partial}{\partial \alpha^*}\right)e^{-\vert \alpha \vert^2 \frac{1+\bar{n}}{2\bar{n}}}.
    \end{aligned}
    \label{Pquibit}
\end{equation}
To complete the calculation, we use the transfomation rule~\cref{PDisplacement} on the $P$-function in~\cref{Pquibit}. 

If the boson-addition happens with probability $p$, the density matrix reads
\begin{equation}
    \frac{p}{1+\mu}a^{\dagger} \rho_{\mu} a+(1-p) \rho_{\mu}.
    \label{P:bosonAddSt}
\end{equation}
Analysing the corresponding $P$-function presents difficulties similar to the ones for noisy Fock states, since the negativity appears only for small $\bar{n}$ when $p$ is small. The numerics allowed us to observe the negativity only for $\mu$ up to $1.35$ when we considered $p=0.05$.
It is worth stressing once more that the observed limits for the negativities in the $P$-functions of the states \eqref{P:noisySt} and \eqref{P:bosonAddSt} are only due to numerical precision, and bear no physical meaning. This is in contrast with the negativities of the Wigner function, which are sensitive to the noise contributions and efficiency of the processes giving the prepared states.

\end{appendices}

\end{document}